\documentclass[a4paper,10pt]{article}
\newtheorem{lemma}{Lemma}

\newtheorem{nc}{Necessary condition}

\newtheorem{cor}{Corollary}

\newtheorem{theorem}{Theorem}
\usepackage{graphicx}
\usepackage{graphicx,amssymb}
\usepackage{hyperref}
\usepackage{epsfig, color, graphicx}
\usepackage[margin=1.0in]{geometry}
\usepackage{amssymb}

 \usepackage{amsmath}
\usepackage{algorithm2e}

\newcount\stepdepth
\newcounter{stepi}
\newcounter{stepii}

\newenvironment{StepI}{
  \begin{list}{Step \arabic{stepi}.}{
    \usecounter{stepi}
    \setlength{\listparindent}{0pt}
    \setlength{\rightmargin}{0pt}
}}{\end{list}}

\newenvironment{StepII}{
  \begin{list}{Step \arabic{stepi}\alph{stepii}.}{
    \usecounter{stepii}
    \setlength{\listparindent}{0pt}
    \setlength{\rightmargin}{0pt}
}}{\end{list}}

\def\stepstart{
  \advance\stepdepth by 1
  \ifnum\stepdepth=1
    \begin{StepI}
  \else
  \ifnum\stepdepth=2
    \begin{StepII}
  \fi
  \fi
}
\def\stepend{
  \ifnum\stepdepth=1
    \end{StepI}
  \else
  \ifnum\stepdepth=2
    \end{StepII}
  \fi
  \fi
  \advance\stepdepth by -1
}
\newenvironment{step}{
  \stepstart
}{
  \stepend
}

\title{Four-connected triangulations of planar point sets}
\author{
 }

\begin{document}
\date{\vspace{-5ex}}
\maketitle
  
\vskip 0.2 in
\begin{center}
\mbox{\begin{minipage} [b] {3in}
\centerline{Ajit Arvind Diwan}
\centerline{ Department of Computer Science and Engineering}
\centerline{ Indian Institute of Technology Bombay}
\centerline{Mumbai 400076, India}
\centerline{aad@cse.iitb.ac.in}
\end{minipage}}\hspace{0.1in}
\mbox{\begin{minipage} [b] {3in}
\centerline{Subir Kumar Ghosh
}
\centerline{School of Technology and Computer Science}
\centerline{Tata Institute of Fundamental Research}
\centerline{Mumbai 400005, India}
\centerline{ghosh@tifr.res.in}
\end{minipage}}\hspace{0.1in}
\mbox{\begin{minipage} [b] {3in}
\centerline{Bodhayan Roy}
\centerline{ School of Technology and Computer Science}
\centerline{ Tata Institute of Fundamental Research}
\centerline{Mumbai 400005, India}
\centerline{bodhayan@tifr.res.in}
\end{minipage}}

\end{center}
 \centerline{\date{Dated: \today}}
\vskip 0.3 in
\begin{abstract}
$ \\$
In this paper, we consider the problem of determining in polynomial time whether a given
 planar point set $P$ of $n$ points admits 4-connected triangulation. We propose 
 a necessary and sufficient condition for recognizing $P$, 
%
and present an $O(n^3)$ algorithm
of constructing a 4-connected triangulation of $P$.
Thus, our algorithm solves a longstanding open problem in computational geometry and geometric graph theory.
We also provide a simple method for constructing a noncomplex triangulation of $P$
which requires $O(n^2)$ steps. 
This method provides a new insight to
the structure of 4-connected triangulation of point sets.
\end{abstract}  
\section{Introduction}

Let $P =\{p_1, p_2, \ldots , p_n \}$ be a set of points on the plane.
We assume that no three points of $P$ are collinear.
Consider the problem of constructing a planar graph of maximum connectivity by connecting points of $P$ with straight edges
or line segments. 
A graph is said to be \emph{k-connected} if 
the graph has at least $k+1$ vertices and
there does not exist a set of $k-1$ vertices whose removal disconnects the graph.
A planar point set $P$ is \emph{k-connectible} if there exists a $k$-connected plane graph $G$ with vertex set $P$ and all edges as
line segments. 
If $G$ is not a triangulated graph, edges can be added to $G$ preserving planarity such that the resultant graph $G'$ is a 
triangulated graph. Observe that since $G$ is $k$-connected, $G'$ is also a $k$-connected graph.
Henceforth, we consider $k$-connected plane graphs of $P$ as triangulated plane graphs. 
Observe that $k$ can be at most $5$ due to Euler's Formula for planar graphs.
$ \\ \\ $
In this paper, we consider the problem of determining in polynomial time whether a given
 planar point set $P$ of $n$ points admits a 4-connected triangulation. We propose 
 a necessary and sufficient condition for recognizing $P$, 
%
and present an $O(n^3)$ algorithm
of constructing a 4-connected triangulation of $P$.
Thus, our algorithm solves a longstanding open problem in computational
geometry and geometric graph theory \cite{tr-hc-1995, huth-2013}.
$ \\ \\$
A \emph{triangulation} of $P$ is a plane graph $T$ with vertex set $P$ such that 
all edges are line segments, the boundary of the outer face of $T$ is the boundary of the convex hull
of $P$ (denoted as $CH(P)$), and all faces of $T$ (with the 
possible exception of the exterior face) are bounded by triangles \cite{ps-cg-85}.
It can be seen that $G$ corresponds to a triangulation $T$ of $P$, where $CH(P)$ in $T$ is the outer face of $G$.
 A \emph{chord} in $T$ is an edge connecting two
nonconsecutive vertices on $CH(P)$.
 A \emph{complex triangle} of $T$ is a triangle formed by three edges of $T$, containing a point of $P$ in its interior
 and another point of $P$ in its exterior.
We have the following properties on the connectivity of $P$ and $T$ from Dey et al. \cite{tr-hc-1995} and Laumond \cite{cpt_laumond}.
 
\begin{lemma}
A triangulation $T$ of a point set $P$, $|P| \geq 3$, in general position is always 2-connected.
\end{lemma}
\begin{cor}
 A point set $P$, $|P| \geq 3$, in general position is always 2-connectible.
\end{cor}

\begin{lemma}
 A triangulation $T$ of $P$, $|P| \geq 4$, is 3-connected if and only if it does not have a chord.
\end{lemma}
\begin{cor}
 A point set $P$, $|P| \geq 4$, is 3-connectible if and only if there is at least one point in the interior of $CH(P)$.
\end{cor}
\begin{figure}[h]\label{anom}
\begin{center}
\centerline{\hbox{\psfig{figure=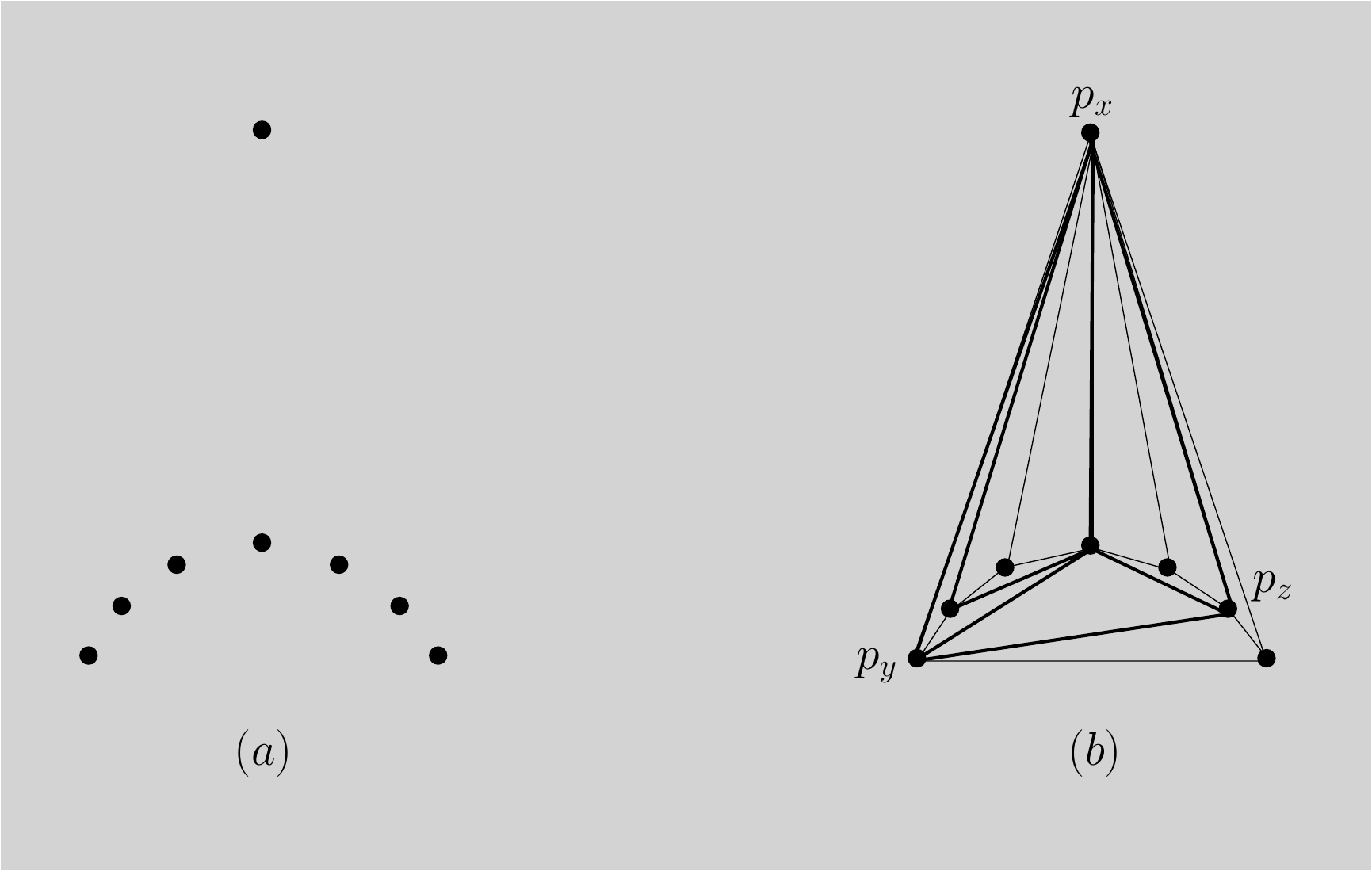,width=0.600\hsize}}}
\caption{(a) An anomalous point set $Q$. 
(b) A triangulation of the anomalous set, containing a complex triangle $p_xp_yp_z$.}
\end{center}
\end{figure}

\begin{lemma}
 A triangulation $T$ of $P$, $|P| \geq 5$, is 4-connected if and only if:
\begin{enumerate}
\item $T$ does not have a chord.
\item No point of $P$ is connected in $T$ to two non-consecutive points on $CH(P)$.
\item $T$ does not have a complex triangle.
\end{enumerate}
\end{lemma}
\begin{cor} 
 If $|P| \geq 5$ and $|CH(P)|=3$, then a triangulation $T$ of $P$ is 4-connected if and only if $T$ has no
complex triangle.
\end{cor}

$ \\ $
Let $Q$ be a set of planar points such that $|CH(Q)| =3$.
Let $p_i$ be a point of $CH(Q)$ .
Let $Q' = Q \setminus \{ p_i \}$.
If all points of $Q'$ are on $CH(Q')$, then $Q$ is called an \emph{anomalous set} (Figure \ref{anom}(a)).
It can be seen that any triangulation of $Q$ must have a complex triangle (Figure \ref{anom}(b)). 
We have the following theorems from Dey et al. \cite{tr-hc-1995}.
\begin{theorem}\label{thm1}
 $Q$ is 4-connected if and only if $Q$ is not anomalous.
\end{theorem}
\begin{theorem}\label{thm2}
 A 4-connected triangulation of $Q$ can be constructed in O($n$ log $n$) time.
\end{theorem}
A triangulation $T$ of $P$ is said to be \emph{noncomplex} if $T$ has neither chords nor complex triangles.
So, a noncomplex triangulation of $P$ may contain an interior point (say, $p_k$) connected to two non consecutive points
(say, $p_i$ and $p_j$) of $CH(P)$. 
 We refer to such a path  $(p_i,p_k,p_j)$ of length 2 
as a \emph{2-chord}. 
Dey et al. \cite{tr-hc-1995} characterized point sets that admit noncomplex triangulation and gave a 
polynomial time algorithm for constructing such a triangulation as follows. 
\begin{theorem}
 A point set $P$ 
admits a noncomplex triangulation if and only if
$P$ is not anomalous and the interior of $CH(P)$ is not empty.
%
\end{theorem}
\begin{cor}
 A noncomplex triangulation $T$ of $P$ can be constructed in $O(n \ log \ n)$ time.
\end{cor}
%
%
In the next section,
 we present an alternative, simple and short proof for constructing a noncomplex triangulation of $P$ leading to  
 an $O(n^2)$ time algorithm.
 In Section \ref{secnc},
 we present three necessary conditions for characterizing $P$ that admits 4-connected triangulation.
 We prove that if $P$ satisfies the third necessary condition, then $P$ also satisfies the first and 
 second necessary conditions. 
 Necessary Condition \ref{nc1} and  Necessary Condition \ref{nc2} are stated here to provide intuition 
 on geometric structures of point sets that allow 4-connected triangulation.
 In Section 4, we give an $O(n^2)$ time algorithm for testing the third necessary condition.
 In Sections 5, 6 and 7, we prove that if $P$ satisfies the third necessary condition, then $P$ admits a 4-connected triangulation.
We 
first find
a simple polygon $C$ containing all points of $P$ that are not in $CH(P)$
together with a
suitable triangulation of the annular region bounded by $CH(P)$ and $C$.  
Then the
interior of $C$ is triangulated to complete the triangulation by suitably modifying the
triangulation of
the annular region, if necessary.
 In Section 8, we conclude the paper with a 
 few remarks and open problems.
\section{Noncomplex triangulations}\label{sec2} 
 
\begin{figure}[h]\label{rem}
\begin{center}
\centerline{\hbox{\psfig{figure=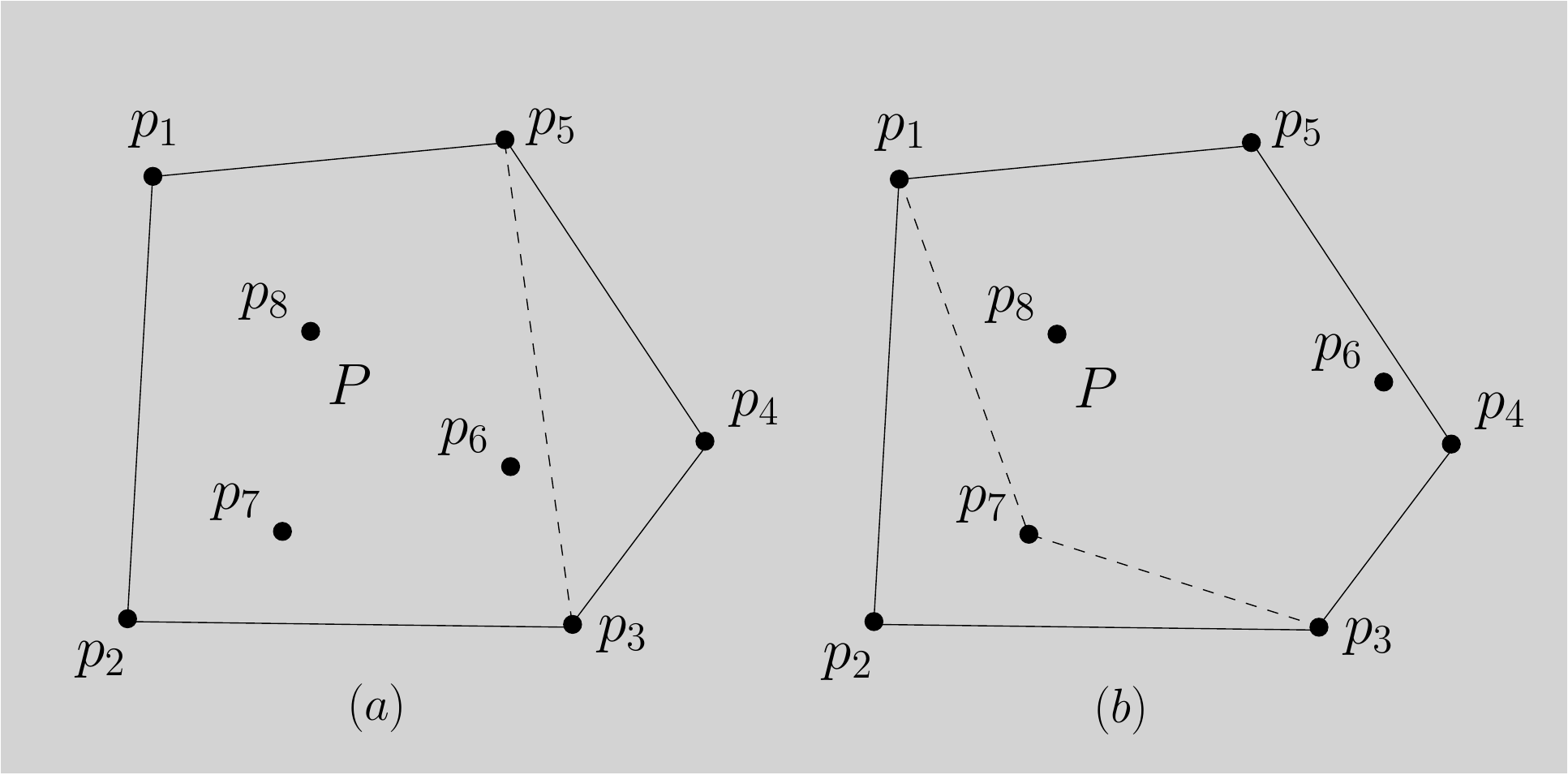,width=0.70000\hsize}}}
\caption{
(a) The triangle $p_3p_4p_5$ is  empty.
(b) The triangle $p_3p_4p_5$ is not empty.
  }
\end{center}
\end{figure}
 
\begin{figure}[h]
\begin{center}
\centerline{\hbox{\psfig{figure=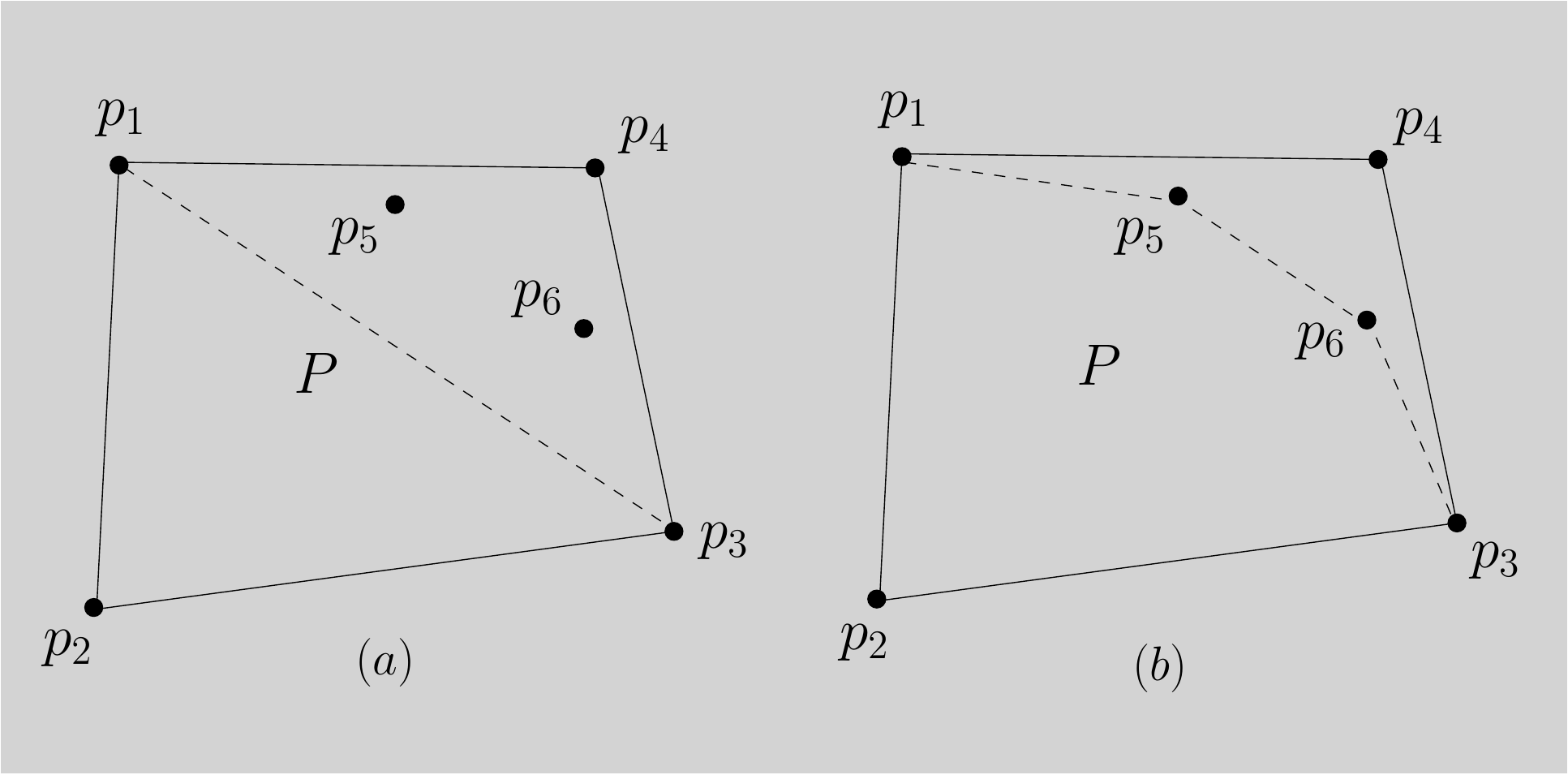,width=0.700\hsize}}}
\caption{
(a) $CH(P \setminus \{ p_2 \})$ is a triangle.
(b)  $CH(P \setminus \{ p_4 \})$ is not a triangle.
  }
\label{rem1}
\end{center}
\end{figure}

\begin{lemma}\label{pdel}
Assume that $|CH(P)| \geq 4$ and $CH(P)$ has at least two points in its interior. 
A point $p_j$ can always be located on $CH(P)$ such that $|CH(P \setminus \{ p_j \})| \geq 4$ and $CH(P\setminus \{ p_j \})$ 
has at least one point in its interior.
\end{lemma}
\textbf{Proof:}
 Consider three consecutive points $p_{i-1}$, $p_i$ and $p_{i+1}$ in the anticlockwise order on $CH(P)$.
  If the interior of the triangle   $p_{i-1} p_i p_{i+1}$ is empty (Figure \ref{rem}(a)), then delete $p_i$,
giving the required conditions for $CH(P \setminus \{ p_i \})$. Otherwise, delete any point
on $CH(P)$ except $p_{i-1}$, $p_i$ and $p_{i+1}$ (Figure \ref{rem}(b)). This method 
works for $|CH(P)| >4$, but it
may not always work if $|CH(P)|=4$ as $CH(P \setminus \{p_i\})$
can become a triangle. Let $p_1$, $p_2$, $p_3$ and $p_4$ be the vertices of $CH(P)$, and $p_5$ and $p_6$ are interior points.
Without the loss of generality, we assume that $CH(P \setminus \{p_2\})$ is a triangle (Figure \ref{rem1}(a)). 
This implies that $CH(P \setminus \{p_4\})$ cannot be a triangle. 
So, 
 $|CH(P \setminus \{p_4\})| \geq 4$. However, the interior of 
  $CH(P \setminus \{p_4\})$ may be empty (Figure \ref{rem1}(b)). 
In that case, $CH(P \setminus \{p_1\})$ or $CH(P \setminus \{p_3\})$ satisfies the required conditions.
%
%
 $\hfill{\Box}$
\begin{lemma}\label{4ch}
 If $|CH(P)| \geq 4$, then $P$ admits a noncomplex triangulation if and only if at least one point of $P$
is not on $CH(P)$.
\end{lemma}
\textbf{Proof:} The proof is by induction on the number of points. 
Let $P_i$ denote a set of $i$ points, such that $|CH(P_i)| \geq 4$ and the interior of $CH(P_i)$ is not empty. 
The base case is for all $P_i$, $i \geq 5$, such that the interior of $CH(P_i)$ contains exactly one point. 
In this case, a noncomplex triangulation can be obtained by joining the interior point to all points on $CH(P_i)$.
$ \\ \\$
Assume that $P_n$ is not a base case (Figure \ref{rem}). 
Since the number of internal points of $CH(P_n)$ is at least two, a point on $CH(P_n)$  (say, $p_j$) can always be located 
using Lemma \ref{pdel} such that
removing $p_j$ from $P_n$ gives $P_{n-1}$ whose all points do not belong to $CH(P_{n-1})$, i.e., the interior of $CH(P_{n-1})$
is not empty.
By the induction hypothesis, we assume that $P_{n-1}$ admits a noncomplex triangulation $T_{n-1}$.
We show that $P_n$ admits a noncomplex triangulation. 

\begin{figure}[h]
\begin{center}
\centerline{\hbox{\psfig{figure=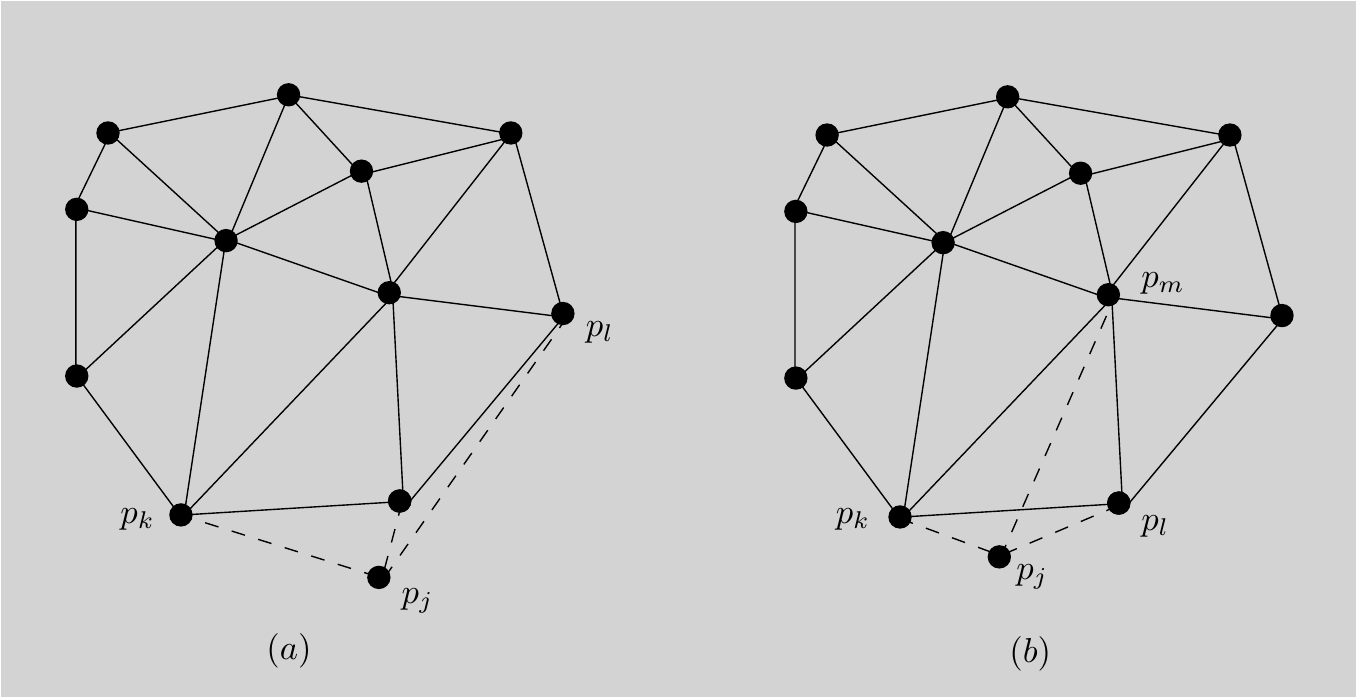,width=0.7000\hsize}}}
\caption{(a) In $T_{n-1}$, $(p_l,p_k)$ is not an edge of $CH(P_{n-1})$. 
(b) In $T_{n-1}$, $(p_l,p_k)$ is an edge of $CH(P_{n-1})$. }
  \label{fig4ch}
\end{center}
\end{figure}
$ \\$
Draw two tangents from $p_j$ to $CH(P_{n-1})$ meeting it at $p_k$ and $p_l$ (Figure \ref{fig4ch}). 
If $(p_l,p_k)$ is not an edge of $CH(P_{n-1})$ (Figure \ref{fig4ch}(a)), then draw edges from $p_j$
to all of these points of $CH(P_{n-1})$ between $p_k$ and $p_l$
that are facing $p_j$. Add these edges to $T_{n-1}$ to obtain $T_n$. 
Since there is no chord in $T_{n-1}$ by assumption, new edges from $p_j$ cannot form a complex triangle
in $T_n$. So, $T_n$ is a noncomplex triangulation of $P_n$. If $(p_l,p_k)$ is an edge of $CH(P_{n-1})$ (Figure \ref{fig4ch}(b)),
$(p_l,p_k)$ becomes a chord in $T_n$ after adding the edges $(p_l,p_j)$ and $(p_j,p_k)$ to $T_{n-1}$.
In order to obtain a noncomplex triangulation of $P_n$, $(p_k,p_l)$ is replaced by a new edge
$(p_j,p_m)$, where $(p_k,p_l,p_m)$  and $(p_k,p_l,p_j)$ are two triangles on $(p_k,p_l)$
 forming a convex quadrilateral $(p_j,p_k,p_m,p_l)$ in $T_n$.
Thus a noncomplex triangulation $T_n$ is obtained from $T_{n-1}$. $\hfill{\Box}$
 
\begin{figure}[h]
\begin{center}
\centerline{\hbox{\psfig{figure=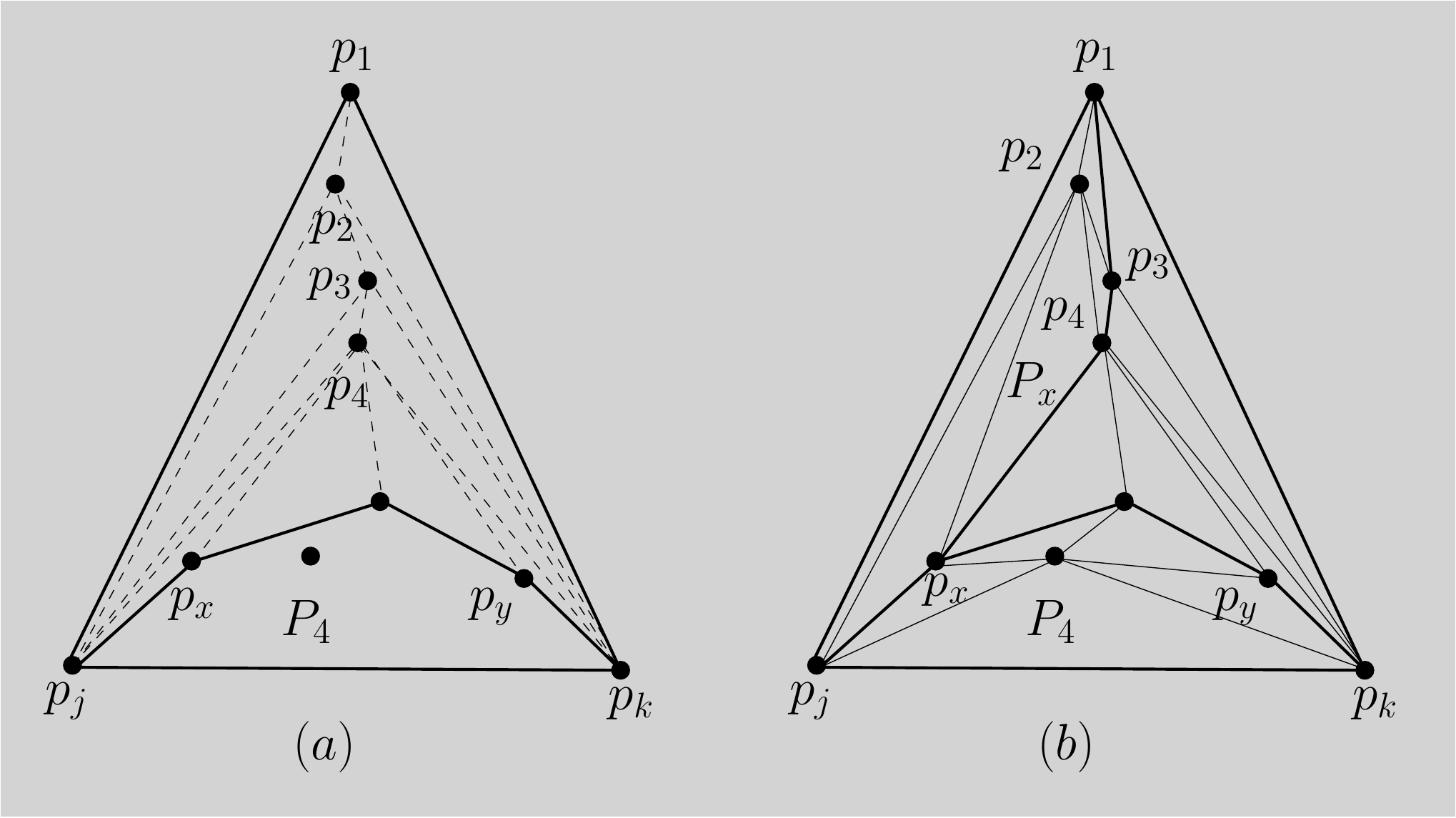,width=0.650\hsize}}}
\caption{(a) In a non-anomalous point set $P$, points $p_1$, $p_2$, $p_3$ and $p_4$ are deleted.
(b) A noncomplex triangulation of $P$, where interiors of $CH(P_i)$ and $CH(P_x)$ are not empty.
}
\label{figind5}
\end{center}
\end{figure}
\begin{lemma} \label{3ch}
 If $ | CH(P)|=3$, then $P$ admits a noncomplex triangulation
if and only if $P$ is not anomalous.
\end{lemma}
\textbf{Proof:} 
If $P$ is anomalous, then there exists a point $p_x \in CH(P)$ such that all points of $P \setminus \{ p_x \}$ are on 
$CH(P \setminus \{ p_x \})$ (Figure \ref{anom}).
So, there exists a chord $p_yp_z$ in any triangulation of $CH(P \setminus \{ p_x \})$ which forms a complex triangle $p_xp_yp_z$.
So, there is no noncomplex triangulation of $P$.
$ \\ \\$
Consider the other situation when
 $P$ is non-anomalous. Remove a convex hull point (say, $p_1$) from $P$ and let $Q_1= P \setminus \{ p_1 \}$ (Figure \ref{figind5}(a)).
If $|CH(Q_1)| \geq 4$, then triangulate $Q_1$ using Lemma \ref{4ch}, and connect $p_1$ to all points of $CH(Q_1)$
facing $p_1$ to complete the triangulation of $P$. If $|CH(Q_1)| = 3$, and the interior of $CH(Q_1)$ is not empty,
then the new convex hull point (say, $p_2$) is removed from $Q_1$ as before and let $Q_2 = Q_1 \setminus \{ p_2 \}$.
This process of deletion is repeated till the remaining point set (say, $Q_i$) forms an empty triangle or $|CH(Q_i)| \geq 4$.  
Let $Q_i' = \{ p_1, p_2, \ldots , p_i \} $.
Let $CH(Q_i) = \{ p_j, p_x, p_{x+1}, \ldots, p_y, p_k \}$, where $p_j$ and $p_k$ are two convex hull points of $P$
not deleted during the process (Figure \ref{figind5}(b)).
Let $Q_x = Q_i' \cup \{ p_x \} \cup \{ p_j \}$ and $Q_y = Q_i' \cup \{ p_y \} \cup \{ p_k \}$.
If all points of $Q_i'$ do not belong to the triangle $p_1p_jp_x$, then $|CH(Q_x)| \geq 4$ and therefore
$CH(Q_x)$ can be triangulated using Lemma \ref{4ch}. Otherwise, $|CH(Q_y)| \geq 4$ which can again be triangulated using 
Lemma \ref{4ch}. The remaining portion of $P$ between the convex hull boundaries can be triangulated arbitrarily. 
\begin{figure}[h]
\begin{center}
\centerline{\hbox{\psfig{figure=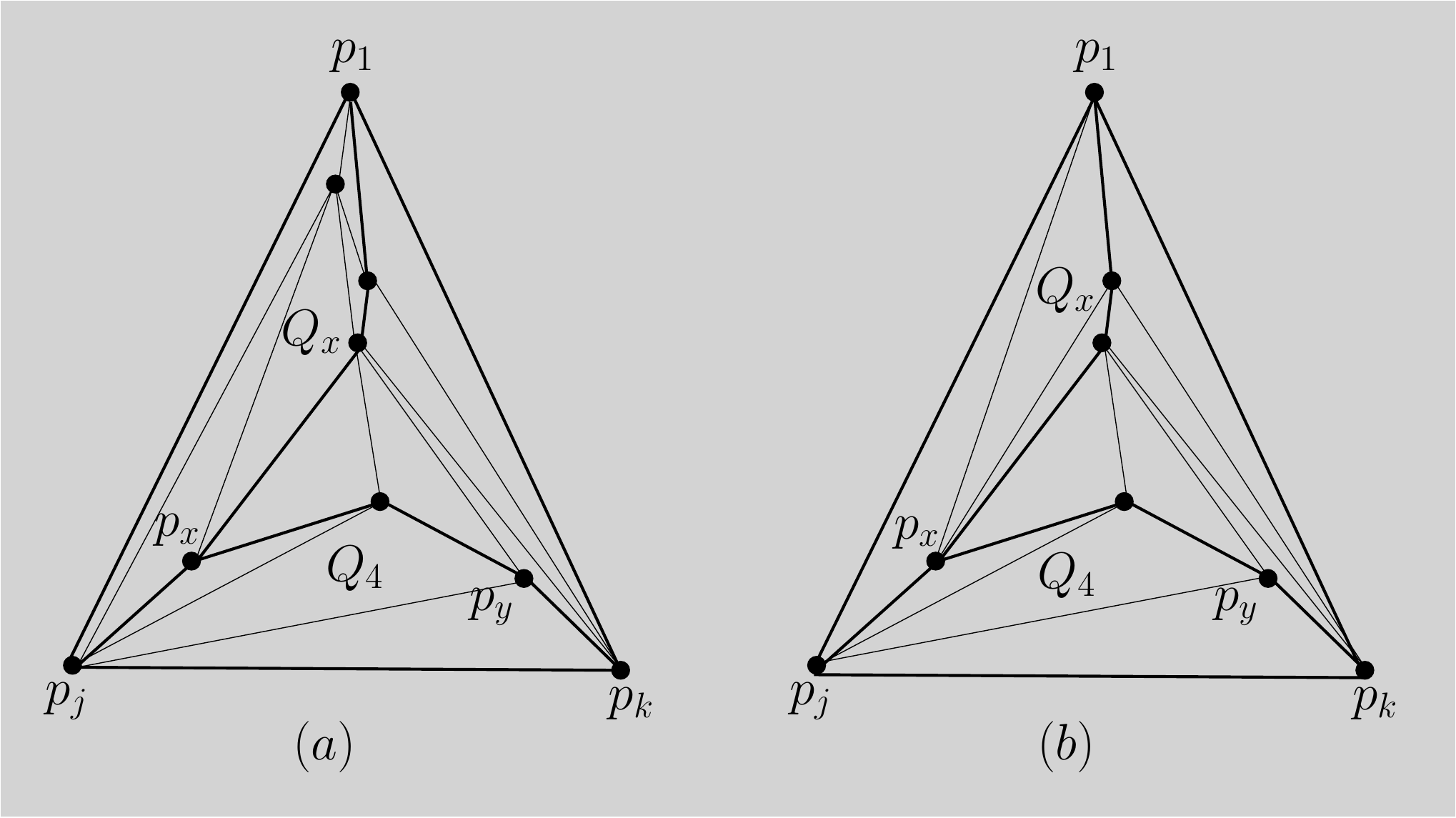,width=0.650\hsize}}}
\caption{(a) The interior of $CH(Q_i)$ is empty, but the interior of $CH(Q_x)$ is not empty. 
(b) The interiors of both $CH(Q_i$ and $CH(Q_x)$ are empty.
}
\label{ind7}
\end{center}
\end{figure}
$ \\ \\$
Observe that Lemma \ref{4ch} cannot be used to triangulate $CH(Q_i)$, $CH(Q_x)$ and $CH(Q_y)$ if their interiors
are empty. In such situations, a different method is used to triangulate $P$. Assume that the interior of $CH(Q_i)$
is empty and $Q_x$ has already been triangulated using Lemma \ref{4ch} (Figure \ref{ind7}(a)). Draw edges from $p_k$ to all points
on $CH(Q_x)$ between $p_1$ and $p_x$. 
Also, draw chords from $p_j$ to all points on $CH(Q_i)$. 
Note that these edges cannot form any complex triangle because there is no chord in the triangulation of $Q_x$.
Consider the other situation when both
$CH(Q_i)$ and $CH(Q_x)$ have empty interiors (Figure \ref{ind7}(b)). As before, draw edges from $p_j$ and $p_k$. In order to avoid forming
any complex triangle, draw edges from $p_x$ to all points on $CH(Q_x)$. The remaining portion of $P$ between
the convex hulls of $Q_i$ and $Q_x$ can be triangulated arbitrarily.
$\hfill{\Box}$
$ \\ \\$ 
Based on the above lemmas, we now present the main steps of our algorithm for constructing a noncomplex triangulation of $P$.
 \begin{step} 
 \item \label{step1} Compute the convex layers of $P$; $indicator:= false$.
\item \emph{If} $|CH(P)|=3$ \emph{then goto} Step \ref{step7}.
\item \label{step3} Locate a point $p_j \in CH(P)$ such that $|CH(P \setminus \{ p_j \})| \geq 4$ and the interior of 
$CH(P \setminus \{ p_j \})$ is not empty (see Lemma \ref{pdel}).
\item  \label{step4}Join $p_j$ with the vertices of $CH(P \setminus \{ p_j \})$ that are facing $p_j$ ; 
$P:=P \setminus \{ p_j \}$ ; Update the convex layers for $P$ (see Lemma \ref{4ch}).
\item \emph{If} $CH(P)$ has two or more interior points \emph{then goto} Step \ref{step3}.
\item \label{joinstep}  Join the interior point of $CH(P)$ to all vertices of $CH(P)$; \emph{if}
$indicator= false$ \emph{then goto} Step \ref{lastep} \emph{else goto} Step \ref{cstep}.

 \item \label{step7} $C:= \phi$. ; Let $p_i$, $p_j$, $p_k$ be the vertices of $CH(P)$; \emph{If} $P$
is anomalous \emph{then goto} Step \ref{lastep}.
\item \label{step8}  Locate a point $p_i$ on $CH(P)$; $C := C \cup \{ p_i \}$; $P:=P \setminus \{ p_i \}$;
 Update convex layers of $P$.
\item \emph{If} $P$ is a nonempty triangle \emph{then}  \emph{goto} Step \ref{step8}.
\item Let $p_x$ and $p_y$ be the next clockwise and counterclockwise point of $p_j$ and $p_k$ on $CH(P)$ respectively ;
\emph{If} $CH(P)$ and $CH(C \cup \{p_x \} \cup \{ p_j \}) $ do not overlap \emph{then} $C :=C \cup \{p_x \} \cup \{ p_j \}$ \emph{else}
 $C :=C \cup \{p_y \} \cup \{ p_k \}$ (see Lemma \ref{3ch}).
\item Triangulate the region between $CH(P)$ and $CH(C)$.
\item \emph{If} $P$ is empty then triangulate $P$; \emph{If} $C$ is empty then triangulate $C$.
\item \emph{If} $P$ is nonempty \emph{then} $indicator:= true$ \emph{and goto} Step \ref{step3}.
\item   \emph{If} \label{cstep}  $C$ is nonempty \emph{then} 
$P:=C$ \emph{and}  $indicator:= false$  \emph{and goto} Step \ref{step3}.


\item STOP. \label{lastep}
\end{step} 

 \begin{theorem}
  Noncomplex triangulation of $P$ (if it exists), can be constructed in $O(n^2)$ time.
 \end{theorem}
 \textbf{Proof:} 
Correctness of the algorithm follows from Lemmas \ref{pdel}, \ref{4ch} and \ref{3ch}. 
In Step \ref{step1}, the convex layers of $P$ can be computed recursively by computing convex hulls of $P$ which takes $O(n^2)$ time. 
In Step \ref{step4}, vertices of $CH(P \setminus \{ p_j \})$ facing $p_j$ can be obtained by drawing appropriate tangents
from its two neighbours to the next layer of $CH(P \setminus \{ p_j \})$, which can be done in $O(n)$ time.
Remaining steps of the algorithm also take $O(n)$ time. Hence, the overall time complexity of the algorithm is $O(n^2)$.

$\hfill{\Box}$
\section{Necessary conditions} \label{secnc}

\begin{figure}[h]  
\begin{center} 
\centerline{\hbox{\psfig{figure=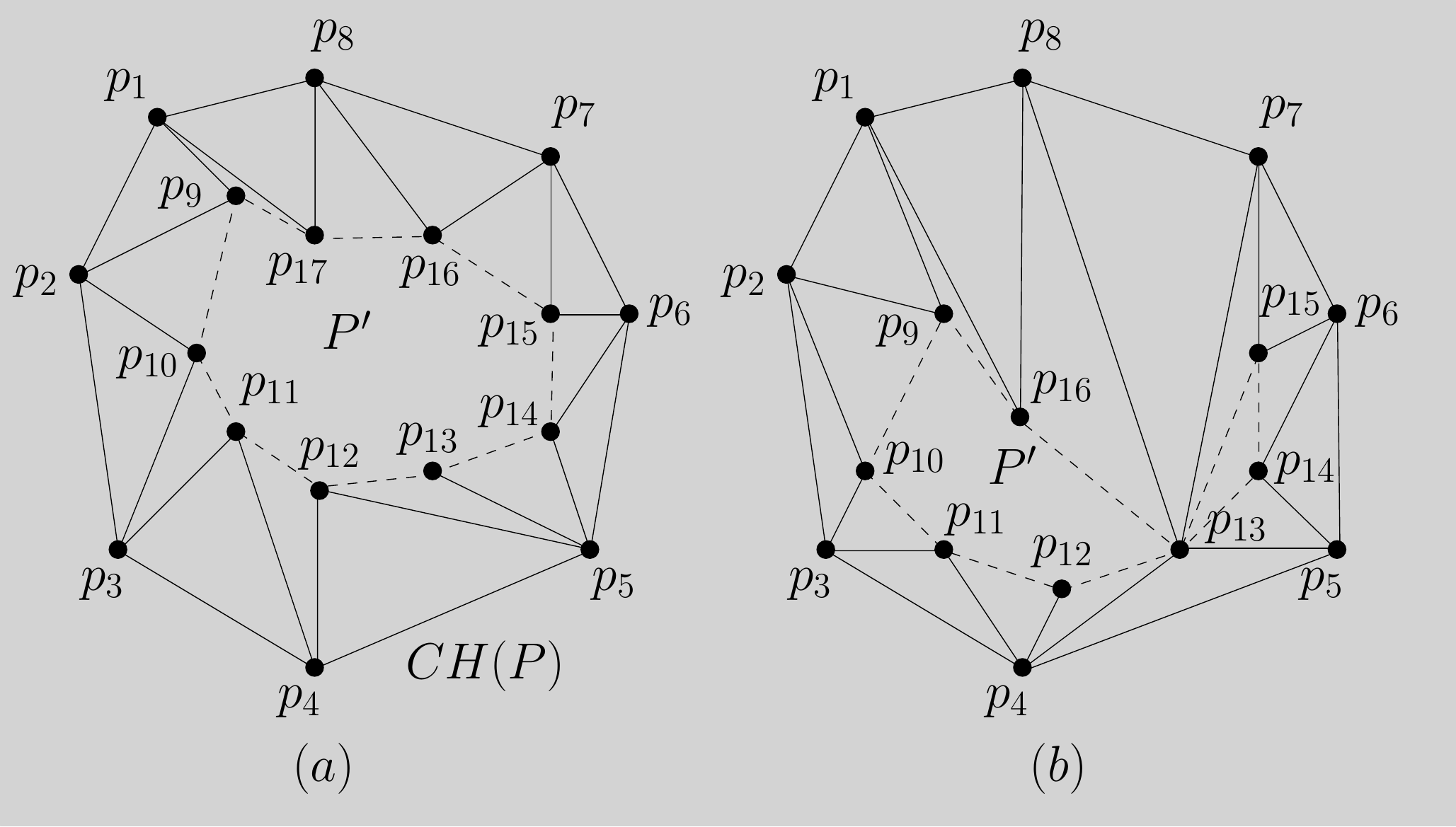,width=0.650\hsize}}}
\caption{
(a) The point set satisfies Necessary Conditions \ref{nc1} and \ref{nc2}. 
(b) The point set satisfies Necessary Condition \ref{nc1} but not Necessary Condition \ref{nc2}.
}
\label{ncpic}
\end{center}
\end{figure}
$\\$
Consider any 4-connected triangulation $T$ of $P$. A triangle of $T$ is said to be an \emph{annular triangle}
if one of its vertices belong to $CH(P)$ (see Figure \ref{ncpic}). The region covered by all annular triangles of $T$ is referred as the
 \emph{annular region} of $T$ (denoted by $A(T)$). 
Observe that $A(T)$ is a region bounded by $CH(P)$ and the \emph{inner cycle} of $A(T)$ formed by vertices of annular
triangles not belonging to $CH(P)$.
Note that all the points of $CH(P')$, where $P'$ is the set of interior points of $CH(P)$, belong to the inner cycle of $A(T)$.
In Figure \ref{ncpic}(a), the inner cycle is formed by the points
  $\{ p_9, p_{10}, p_{11}, p_{12}, p_{13}, p_{14}, p_{15}, p_{16}, p_{17} \}$. 
If exactly one vertex of an annular triangle belongs to $CH(P)$, the triangle is called an \emph{outward triangle} of $A(T)$.
Otherwise, the triangle is called an \emph{inward triangle} of $A(T)$.
For example, $p_4p_{12}p_5$ in Figure \ref{ncpic} is an inward triangle while $p_{12}p_5p_{13}$ is an outward triangle.
Note that every inward or outward triangle is empty by definition of triangulation.
The vertex of an inward triangle belonging to $P'$ is called the
\emph{inward vertex} of the triangle.
We have the following
necessary condition from Dey et al. \cite{tr-hc-1995}.
\begin{nc} \label{nc1}
 If $P$ admits a 4-connected triangulation,
 then $|P'| \geq |CH(P)|$.
\end{nc}
 \textbf{Proof:} 
Let $ab$ and $cd$ be two edges of $CH(P)$. If two inward triangles $abe$ and $cde$ share a vertex $e$,
then $(a,e,c)$ is a 2-chord, which is not permitted in a 4-connected triangulation. So, no two inward triangles
of $A(T)$ can share an inward vertex (see Figure \ref{ncpic}(a)).
Since every edge of $CH(P)$ belongs to one inward triangle, the number of points on the inner cycle of $A(T)$, 
which are   points of $P'$, is at least $|CH(P)|$.$\hfill{\Box}$
$ \\ \\$
Consider Figure \ref{ncpic}(b). The inward triangle $p_7p_8p_{13}$ 
  has touched $CH(P')$ at $p_{13}$ from the opposite side
after intersecting the edge $p_9p_{15}$ of $CH(P')$. 
 This introduces a 2-chord as any triangulation of $P$ must connect 
$p_{13}$ with $p_4$ or $p_5$. So, there cannot be any 4-connected triangulation of $P$ with
inward triangle $p_7p_8p_{13}$ as the inner cycle becomes self-intersecting.  
This observation leads to the following necessary condition.
\begin{figure}[h]  
\begin{center} 
\centerline{\hbox{\psfig{figure=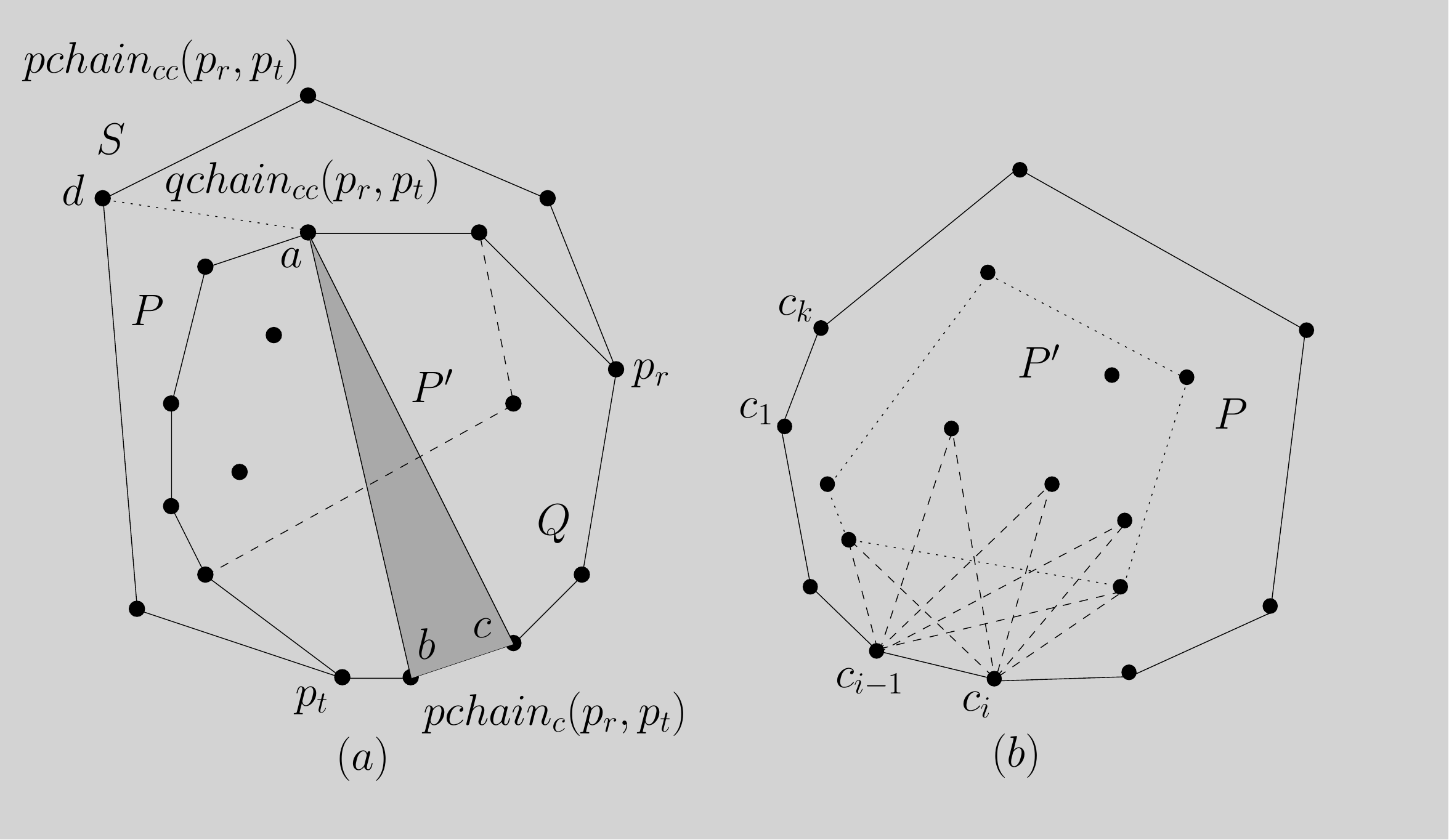,width=0.70\hsize}}}
\caption{(a) Any choice of inward triangles leads to a self-intersecting inner cycle of $A(T)$. 
(b) Candidate triangles on $c_{i-1}c_i$ in the clockwise order are stored in $L_i$.
}
\label{ncpic2}
\end{center}
\end{figure}
\begin{nc} \label{nc2}
Let $P$ be 4-connectible.  
Let $S$ be any set of consecutive points on $CH(P)$. If $S$ is deleted from $P$, the new convex hull of the set $Q$ of 
remaining points of 
$P$ must be of size at most $|P'|+1$.
\end{nc}
\textbf{Proof:} 
Let $p_r$ and $p_t$ be any two points on $CH(P)$ (Figure \ref{ncpic2}(a)). 
Let $pchain_{cc}(p_r,p_t)$ (or, $pchain_{c}(p_r,p_t)$) denote the counterclockwise
boundary (respectively, clockwise boundary) of $CH(P)$ from $p_r$ to $p_t$. For any pair of $p_r$ and $p_t$, the corresponding
$S$ is defined as all points of $pchain_{cc}(p_r,p_t)$ excluding $p_r$ and $p_t$. Similarly, $qchain_{cc}(p_r,p_t)$ 
(or, $qchain_{c}(p_r,p_t)$) is defined as the
counterclockwise boundary (respectively, clockwise boundary) of $CH(Q)$ from $p_r$ to $p_t$, 
excluding $p_r$ and $p_t$.
Note that $qchain_{c}(p_r,p_t) = pchain_{c}(p_r,p_t)$.
$ \\ \\$
Let $abc$ be an inward triangle, where $bc$ is an edge of  $pchain_{c}(p_r,p_t)$. Assume that $a \in qchain_{cc}(p_r,p_t)$.
Since any triangulation of $P$ must join $a$ with some point $a'_i$ on $pchain_{cc}(p_r,p_t)$, $(d,a,b)$ or $(d,a,c)$ become 2-chords.
So, there cannot be any such inward triangle (called \emph{forbidden triangle})
in a 4-connected triangulation of $P$. This implies that $a$ must be an interior 
point of $CH(Q)$ (i.e. $P' \setminus qchain_{cc}(p_r,p_t)$), 
and the number of interior points of $CH(Q)$ must be at least the number of edges of $pchain_c(p_r,p_t)$
(i.e. $|CH(P)| - |pchain_{cc}(p_r,p_t)| - 1 $).
In other words, 
\begin{eqnarray*}
 & &|CH(P)| - |pchain_{cc}(p_r,p_t)| - 1  \leq |P'| - |qchain_{cc}(p_r,p_t)|  \\
  &or,& |CH(P)| - |pchain_{cc}(p_r,p_t)| + |qchain_{cc}(p_r,p_t)|   \leq |P'| + 1 \\
   &or,& |CH(Q)|   \leq |P'| + 1 
\end{eqnarray*}
\begin{figure}[h]  
\begin{center} 
\centerline{\hbox{\psfig{figure=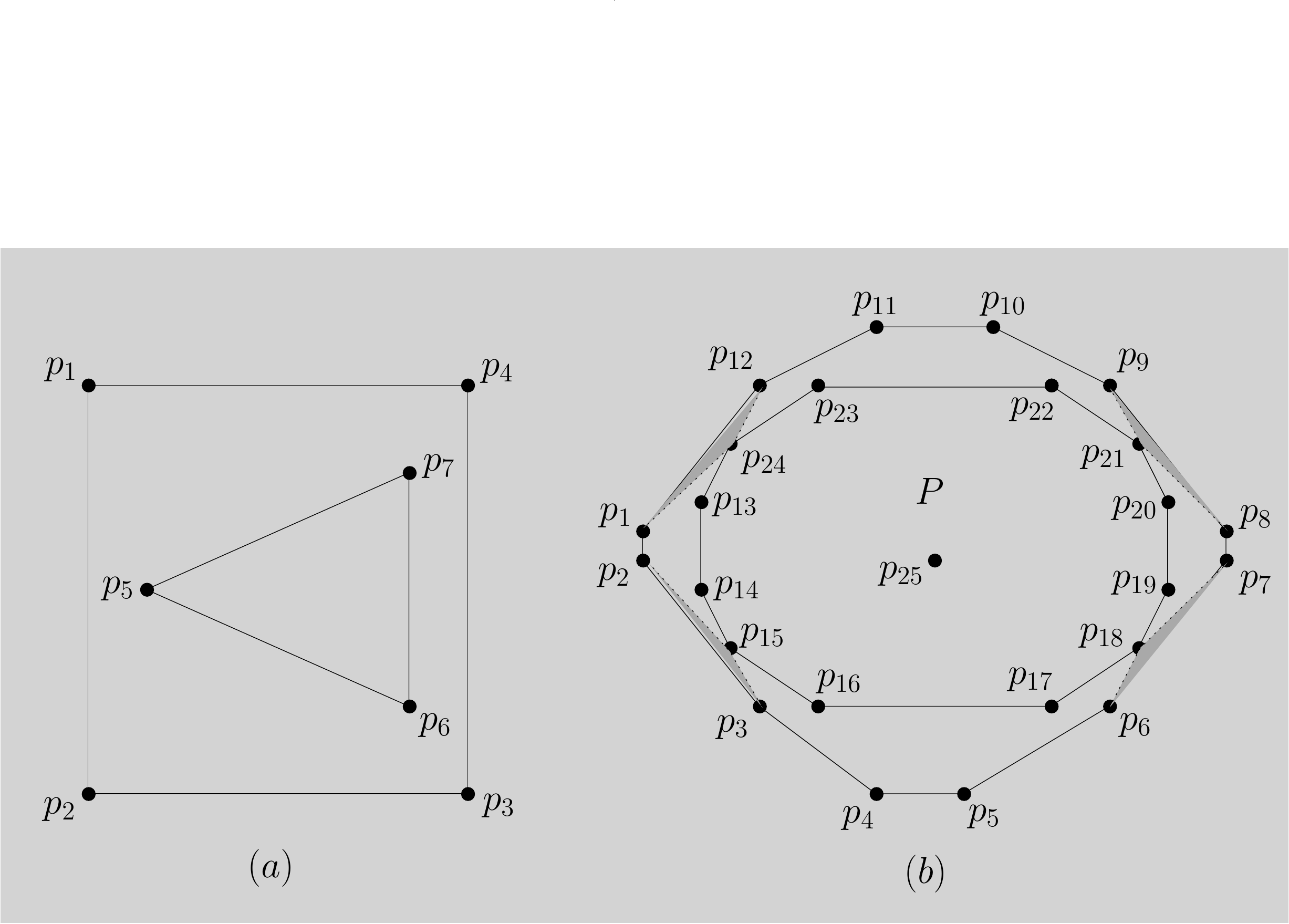,width=0.70\hsize}}}
\caption{
(a) The point set satisfies Necessary Condition \ref{nc2}  but not Necessary Condition \ref{nc1}.
 (b) The point set satisfies Necessary Condition \ref{nc1} and Necessary Condition \ref{nc2} 
 but there is no 4-connected triangulation of this point set.
}
\label{ncpic3}
\end{center}
\end{figure}
Consider Figure \ref{ncpic3}(a). Though $|P'|<|CH(P)|$,
the point set satisfies Necessary Condition \ref{nc2} for every pair of $p_r$ and $p_t$.
It may appear that if $P$ satisfies both Necessary Conditions \ref{nc1} and \ref{nc2}, then there always exists a 4-connected 
triangulation of $P$.
However, this is not true for the point set shown in Figure \ref{ncpic3}(b), though it satisfies Necessary Conditions \ref{nc1}
and \ref{nc2}.
Observe that the inward triangles $p_2p_{15}p_3$, $p_6p_{18}p_7$, $p_8p_{21}p_9$ and $p_{12}p_{24}p_1$  
 must be present in any 4-connected triangulation of $P$.
Consider the edges $p_9p_{10}$, $p_{10}p_{11}$ and $p_{11}p_{12}$ between the 
two inward triangles $p_8p_{21}p_9$ and $p_{12}p_{24}p_1$.
Since they need three inward vertices, $p_{22}$, $p_{23}$ and $p_{25}$
must be assigned  as inward vertices for these three edges.
Observe that $p_{25}$ is also required as inward vertex in addition to $p_{16}$ and $p_{17}$
for the edges $p_3p_4$, $p_4p_5$ and $p_5p_6$. Since $p_{25}$ cannot be an inward vertex of two inward triangles,
the point set does not admit a 4-connected triangulation. This leads to another necessary condition.
$ \\ \\$
A set $T_c$ of inward triangles, that are not forbidden, is said to be \emph{compatible} if no two inward triangles in $T_c$ 
 share an edge or  an inward vertex or an interior point. $T_c$ is said to be \emph{maximal} if no inward
triangle can be added to $T_c$ while keeping $T_c$ compatible.
Let $T_c' \subseteq T_c$  
be the set of all compatible inward triangles whose inward vertices 
are vertices of $CH(P')$. 
Let $max|T_c'|$ denote the maximum cardinality of $T_c'$ among all $T_c$ that are maximal.

\begin{nc} \label{nc3}
Let $P$ be 4-connectible. Then, $|P'| - |CH(P')| \geq |CH(P)| - max|T_c'|$.
\end{nc}
\textbf{Proof:} 
In a 4-connected triangulation, there are $|CH(P)|$ compatible inward triangles 
and $|T_c| = |CH(P)|$. 
Choose one such maximal $T_c$ which gives $max|T_c'|$.
In order to have $|CH(P)|$ compatible inward triangles for a 
4-connected triangulation, $|P'|- |CH(P')|$  must be of size at least 
$|T_c| - max|T_c'|$. Hence, 
 $|P'|- |CH(P')| \geq |CH(P')|- max|T_c'| $. $\hfill{\Box}$
%
%
\begin{lemma}\label{ncsat}
 If $P$ satisfies Necessary Condition \ref{nc3}, then $P$ also satisfies Necessary Conditions \ref{nc1} and \ref{nc2}.
\end{lemma}
\textbf{Proof:} 
We have $|P'|-|CH(P')| \geq |CH(P)| - max|T_c'|$. So, $|P'| \geq |CH(P)| + |CH(P')| - max|T_c'|$.
Since inward vertices of $T_c'$ are vertices of $CH(P')$, $|CH(P')| - max|T_c'| \geq 0$.
Hence, $|P'| \geq |CH(P)|$ which is Necessary Condition \ref{nc1}. 
$\\ \\$
Consider any two vertices $p_r$ and $p_t$ of $CH(P)$ and $T_c$ which gives $max|T_c'|$. 
Since an inward triangle on an edge in $pchain_{cc}(p_r,p_t)$ cannot have an inward vertex in $qchain_{cc}(p_r,p_t)$,
we have 
$max|T_c'| \leq |CH(P')| - qchain_{cc}(p_r,p_t) + pchain_{cc}(p_r,p_t) +1$. However, by Necessary Condition \ref{nc3}, 
  $max|T_c'| \geq |CH(P)| + |CH(P')| - |P'|$. This implies  
  $|CH(P)| - |P'| \leq pchain_{cc}(p_r,p_t) - qchain_{cc}(p_r,p_t) +1$, or,
$|CH(P)| + qchain_{cc}(p_r,p_t) - pchain_{cc}(p_r,p_t) + 1 = |CH(Q)|  \leq |P'| + 1$, which is Necessary Condition \ref{nc2}.
$\hfill{\Box}$

\section{An algorithm for testing necessary conditions}
For testing necessary conditions, it is enough to test Necessary Condition \ref{nc3} due to Lemma \ref{ncsat}. 
In this section, we give an $O(n^2)$ time algorithm for checking whether $P$ satisfies
Necessary Condition \ref{nc3}. 
Our algorithm first constructs a bipartite graph $G(U,V,E)$ and then computes a maximum matching $M$ in $G$.
Starting from $M$, another matching $M'$ in $G$ is constructed such that $|M| = |M'|$ and no two inward
triangles corresponding to edges in $M'$ intersect. Hence, $|M'| = max|T'_c|$.
$\\ \\$
Initially,
$U = V = E = \phi$.
Let 
$c_1, c_2, \ldots, c_k$ be the vertices of $CH(P)$ in the counterclockwise order (see Figure \ref{ncpic2}(b)).
For every edge $c_{i-1}c_i$ of $CH(P)$, add a vertex $u_i$ in $U$.  
For every vertex $a_i$ of $CH(P')$, add a vertex $v_i$ to $V$.
If $c_{i-1}a_jc_i$ is empty and is not a forbidden triangle, then add the edge $u_iv_j$ to $E$.
 Compute a maximum matching $M$ of $G$ by 
 the Hopcroft-Karp algorithm \cite{bm-85}.
 Let $T$ be the set of inward triangles of $P$ corresponding to $M$.
 We have the following lemma.

%
 \begin{figure}[h]  
\begin{center} 
\centerline{\hbox{\psfig{figure=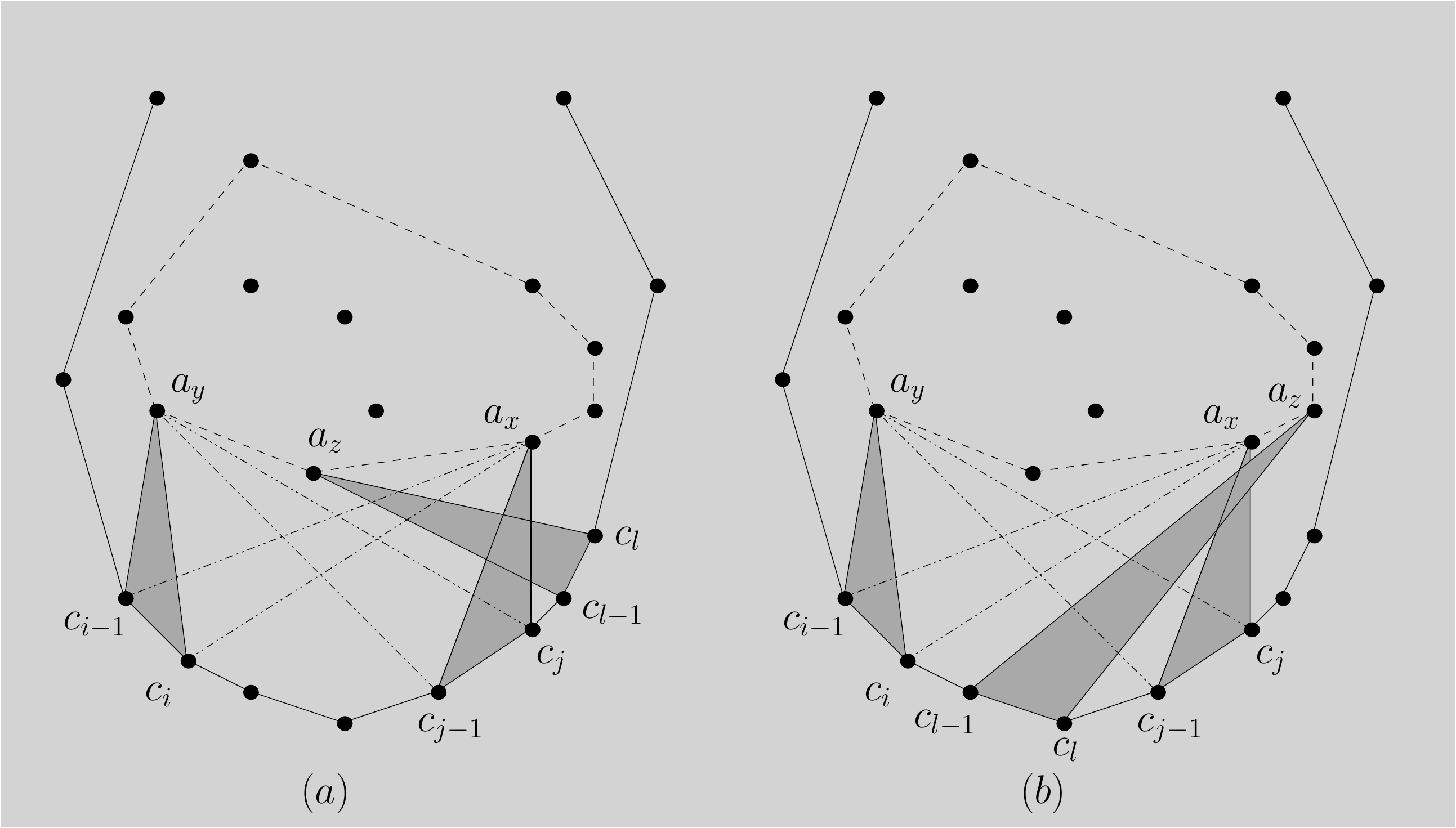,width=0.700\hsize}}}
\caption{
(a) The inward triangle $c_{l-1}a_zc_l$ intersects $c_{j-1}a_xc_j$ and $c_{i-1}a_xc_i$ from right to left.
(b) The inward triangle $c_{l-1}a_zc_l$ intersects $c_{j-1}a_xc_j$ and $c_{j-1}a_yc_j$ from left to right.
}
\label{matchfig1}
\end{center}
\end{figure}
 
 \begin{lemma}\label{checkcorrect}
  Starting from
  a maximum matching 
  $M$ in $G(U,V,E)$,
  another maximum matching $M'$ in $G$ can be constructed such that no two inward triangles
  corresponding to the edges of $M'$ intersect each other.
 \end{lemma}
\textbf{Proof:}
Let $T$ denote the inward triangles corresponding to $M$. If no two triangles in $T$ intersect then 
$M= M'$.
So, we assume that $T$ contains intersecting inward triangles.
Let $c_{i-1}a_xc_i$ and $c_{j-1}a_yc_j$ be two intersecting inward triangles in $T$.
Replace $c_{i-1}a_xc_i$ and $c_{j-1}a_yc_j$ by $c_{i-1}a_yc_i$ and $c_{j-1}a_xc_j$ 
to remove the intersection (see Figure \ref{matchfig1}). Observe that $c_{i-1}a_yc_i$ and $c_{j-1}a_xc_j$
are not forbidden triangles and hence, they are represented as edges in $G$.
So, two edges in $M$ are replaced by two other edges of $G$. 
Observe that if any inward triangle $c_{l-1}a_zc_l$ in $T$ intersects $c_{j-1}a_xc_j$, 
then the same triangle also intersects 
$c_{i-1}a_xc_i$ or $c_{j-1}a_yc_j$ as $c_{l-1}c_l$ and $a_z$ must lie on the opposite
sides of $c_{j-1}a_xc_j$ (see Figure \ref{matchfig1}). So, the number of intersecting triangles in $T$ is reduced by the 
above replacement.
Repeat this process of 
replacement of triangles
till no two inward triangles, corresponding to the modified matching, intersect.
Thus a new matching $M'$ in $G$ is constructed from $M$ with $|M| = |M'|$.
 $\hfill{\Box}$
\begin{algorithm}[H]

\textbf{testing\_necessary\_conditions(\emph{P})}

compute $CH(P)$ and $CH(P')$\;
 let the points of $CH(P)$ be $\{ c_1, c_2, \ldots, c_k \}$ in counterclockwise order\;
 \tcp{test}
 $U := \phi$, $V := \phi$, $E := \phi$\;
 $G := (U, V, E)$\;
 $i := 1$\;
 \While{$i \leq |CH(P)|$}
 {
 $U : = U \cup \{ u_i \}$\;
 $i := i + 1$\;
 }
 \tcp{creates $U$}
 $i := 1$\;
 \While{$i \leq |CH(P')|$}
 {
 $V : = V \cup \{ v_i \}$\;
 $i := i + 1$\;
 }
 \tcp{creates $V$}
 $i := 1$\;
 \While{$i \leq |CH(P)|$}
 {
 $j := 1$\;
 \While{$j \leq |CH(P')|$}
 {
 \If{$c_{i-1}a_jc_i$ is 
 empty and it is not a forbidden triangle of $P$}
 {
 $ E := E \cup \{ u_iv_j \}$\;
 }
 $j := j + 1$\;
 }
 $i := i + 1$\;
 }
 \tcp{creates $E$}
 \tcp{completes construction of $G$}
  compute a maximum matching $M$ of $G$\;
  $T : = \phi$\;
  add inward triangles of $P$ to $T$ that correspond to the edges of $M$\;
  \While{two triangles $c_{i-1}a_xc_i$ and  $c_{j-1}a_yc_j$ in $T$ intersect}
  {
  $T = (T \setminus \{ c_{i-1}a_xc_i, c_{j-1}a_yc_j \}) \cup  \{ c_{i-1}a_yc_i, c_{j-1}a_xc_j \}$\;
  $M = (M \setminus \{ u_iv_x, u_jv_y \}) \cup  \{  u_iv_y, u_jv_x \}$\;
  
  }
  \tcp{computes final $M$ and $T$ with non-intersecting and non-forbidden inward triangles}
  report $T$ \;
  \If{$|CH(P)|-|T| \leq |P'|-|CH(P')|$}
{report that $P$ satisfies Necessary Condition \ref{nc3}\;
}
\end{algorithm}

 \begin{lemma}\label{checklem}
The procedure \textbf{testing\_necessary\_conditions(\emph{P})} correctly computes $max|T'_c|$ in $O(n^3)$ time.
 \end{lemma}
\textbf{Proof:}
The correctness of the algorithm follows from Lemma \ref{checkcorrect}.
Constructing $G$ requires $O(n^2)$ time. The Hopcroft-Karp algorithm for
computing maximum matching of bipartite graphs takes $O(n^{2.5})$ time.
Since 
there can be at most $n^2$ intersections among triangles,
locating each such pair takes $O(n)$ time. 
Hence, replacement of all intersecting
pairs of triangles takes $O(n^3)$ time.
$\hfill{\Box}$

\section{Construction of initial set of inward triangles}
\begin{figure}[h]  
\begin{center} 
\centerline{\hbox{\psfig{figure=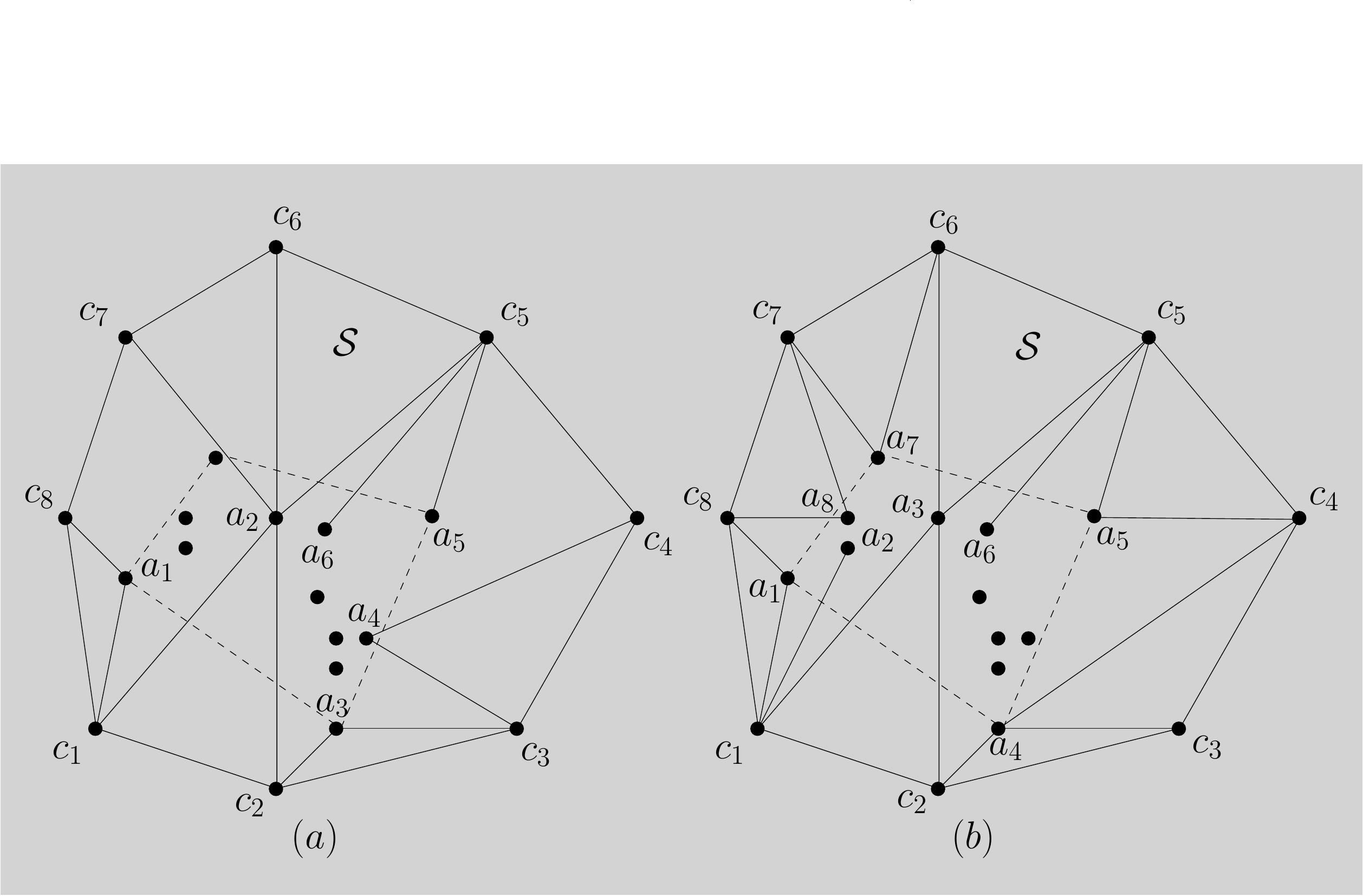,width=0.650\hsize}}}
\caption{
(a) $S =$ $\{c_8a_1c_1$, $c_1a_2c_2$, $c_2a_3c_3$, $c_3a_4c_4$, $c_5a_5$, $c_5a_6,$ $c_5a_2c_6,$ $c_6a_2c_7 \}$
is a set of inward triangles but not a good set.
(b) $S = \{c_8a_1c_1,$ $c_1a_2,$ $c_1a_2,$ $c_1a_2,$ $c_1a_3c_2,$ $c_2a_4c_3,$ $c_3a_4c_4,$
$c_4a_5c_5,$ $c_5a_6,$ $c_5a_6,$ $c_5a_6,$ $c_5a_3c_6,$ $c_6a_7c_7,$ $c_7a_8c_8 \}$ is a good set.
}
\label{13}
\end{center}
\end{figure}

\begin{figure}[h]  
\begin{center} 
\centerline{\hbox{\psfig{figure=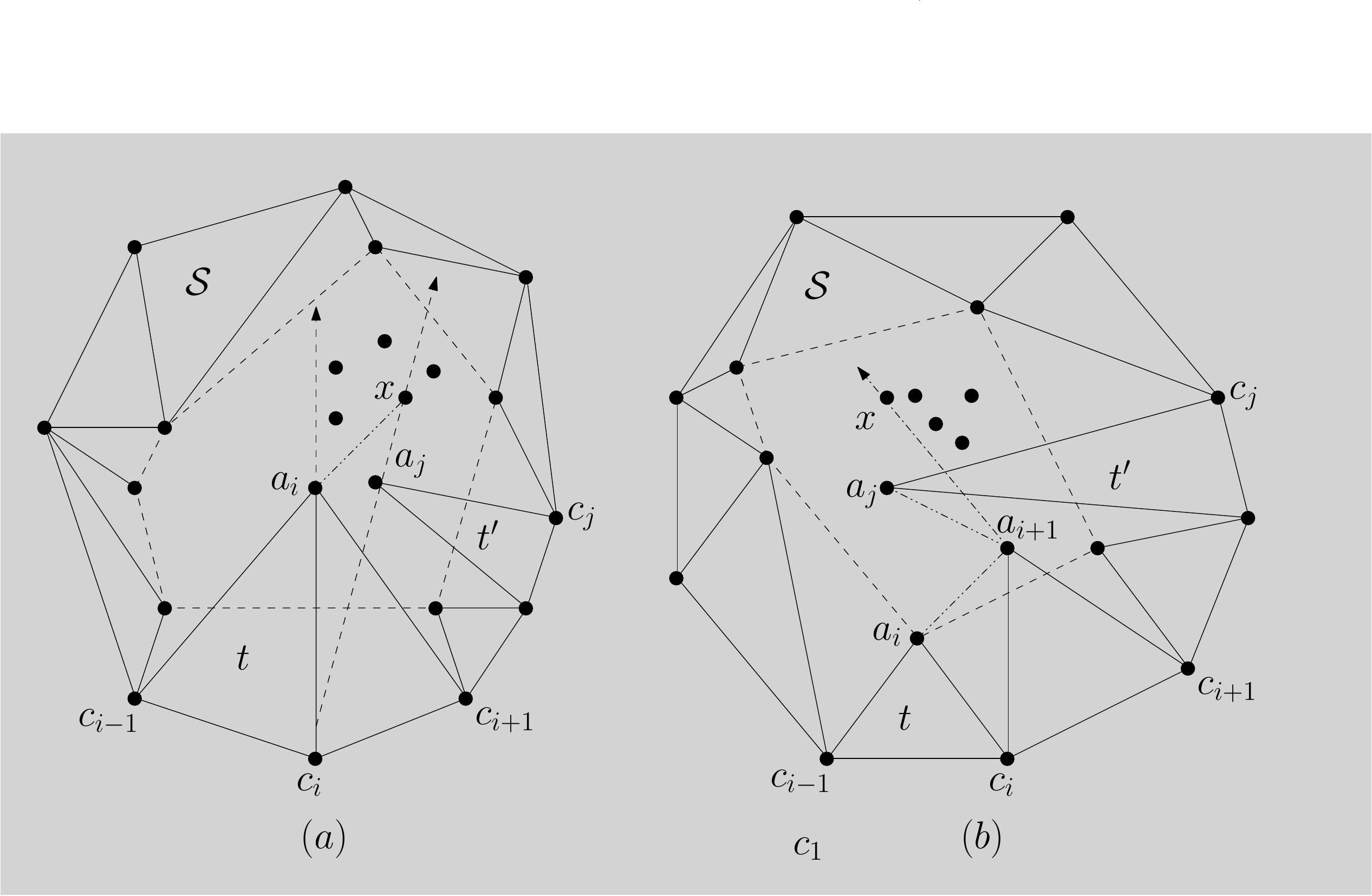,width=0.70\hsize}}}
\caption{ 
(a) $S$ satisfies Property 7 for where consecutive inward triangles $c_{i-1}a_ic_i$ and $c_{i}a_ic_{i+1}$ are
sharing the common inward vertex $a_i$.
(b) $S$ satisfies Property 8 for two consecutive inward triangles $c_{i-1}a_ic_i$ and $c_{i}a_{i+1}c_{i+1}$
for $a_i \neq a_{i+1}$.
}
\label{prop78}
\end{center}
\end{figure}

In this section, we introduce the notion of a good set $\mathcal{S}$ of inward triangles which is used as the
first step for constructing a 4-connected triangulation of $P$.
Let us start with a few definitions (see Figure \ref{13}(a)).
Two triangles  are said to be \emph{pairwise disjoint} if their interiors do not intersect. Since 
the definition permits two triangles  to share vertices or edges, pairwise disjoint triangles may not
be compatible triangles as defined earlier.
We refer to a line segment joining a vertex $c_i$ 
of $CH(P)$ to a point $u$ of $P'$ as a \emph{degenerate} inward triangle, where $u$ is the inward
vertex of this degenerate inward triangle. The line segment $c_iu$ is 
called \emph{forbidden} if $u$ is a point of $CH(P')$ and $c_iu$ intersects the interior of $CH(P')$.
In our definition of $\mathcal{S}$, we allow $\mathcal{S}$ to include degenerate inward triangles.
We allow repetition of only degenerated inward triangles in $\mathcal{S}$, making $\mathcal{S}$ a multiset.
For every vertex $c_i$, let $\mathcal{S}_i$ denote the set of inward triangles of $\mathcal{S}$ incident on $c_i$. 
For all $i$, order the triangles of $\mathcal{S}_i$ around $c_i$ in the clockwise order starting from $c_{i-1}$.
Construct a list $L(\mathcal{S})$ by concatenating $\mathcal{S}_1, \mathcal{S}_2, \ldots, \mathcal{S}_k$
in the same order, and remove the duplicate inward triangles from  $L(\mathcal{S})$.
The edge $c_1a_i$ of an inward triangle $c_{i-1}a_ic_i$ (or a degenerate inward triangle $c_ia_i$)
is referred to as the right edge of $c_{i-1}a_ic_i$.
A point 
$u$ in $P'$ is said to be \emph{free} if it is not the inward vertex of any triangle in $\mathcal{S}$. 
 Let $\overrightarrow {c_iy}$ denote the ray drawn from $c_i$ through a point 
 $y \in P'$. The segment $c_iy$ is called the \emph{left
 tangent} of $c_i$ to $P'$ if all points of $P'$ lie to the right of $\overrightarrow {c_iy}$. 
$\\ \\$
We say  $\mathcal{S}$ is \emph{good} if it satisfies the following properties (see Figure \ref{13}(b)):
\begin{enumerate}
\item
$|\mathcal{S}| = |P'|$.
\item
$\mathcal{S}$ does not contain any forbidden triangle.
\item
The triangles in $\mathcal{S}$ are pairwise disjoint.
\item
Every vertex of $CH(P')$ is the inward vertex of some triangle in $\mathcal{S}$.
\item
Every edge of $CH(P)$ has an inward triangle in $\mathcal{S}$.
\item
No line segment joining two free points intersects any triangle in $\mathcal{S}$.
\item
\label{cons}
Let $t$ be a triangle in $\mathcal{S}$ with right edge $c_ia_i$ such that the next triangle 
in $L(\mathcal{S})$ has the same inward vertex $a_i$ (i.e., the counterclockwise next triangle of $t$ in $L(\mathcal{S})$
is either $c_ia_ic_{i+1}$ or a
degenerate triangle $c_ia_i$) (see Figure \ref{prop78}(a)).
For any free point $x$, the following properties hold:
\begin{enumerate}
\item
The point $x$ lies to the right of $\overrightarrow{c_ia_i}$.
\item
If $t'$ is a triangle in $\mathcal{S}$ with right edge $c_ja_j$ intersecting the line segment $c_ix$, 
then $c_j$ lies to the right of $\overrightarrow{c_ix}$, $a_j$ lies to the left of $\overrightarrow{c_ix}$, and either $a_j = a_i$ 
or $a_j$ lies to the right of $\overrightarrow{c_ia_i}$.
\end{enumerate}
\item
\label{non-cons}
Let $t$ be a triangle in $\mathcal{S}$ with right edge $c_ia_i$ such that the next triangle  
in $L(\mathcal{S})$ has inward vertex $a_{i+1} \neq a_i$ (i.e., 
the counterclockwise next triangle of $t$ in $L(\mathcal{S})$
is either $c_ia_{i+1}c_{i+1}$ or a
degenerate triangle $c_ia_{i+1}$) (see Figure \ref{prop78}(b)). 
For any free point $x$, the following properties hold:
\begin{enumerate}
\item
The line segment $a_{i+1}x$ does not intersect $t$.
\item
If $t'$ is a triangle in $\mathcal{S}$ with right edge $c_ja_j$ intersecting the line segment $a_{i+1}x$, 
then $c_j$ lies to the right of $\overrightarrow{a_{i+1}x}$, $a_j$ lies to the left of 
$\overrightarrow{a_{i+1}x}$ and the line segment
$a_{i+1}a_j$ does not intersect $t$.
\item
There is no point of $P'$ in the interior of the triangle $a_ic_ia_{i+1}$.
\end{enumerate}
\end{enumerate}
\begin{figure}[h]  
\begin{center} 
\centerline{\hbox{\psfig{figure=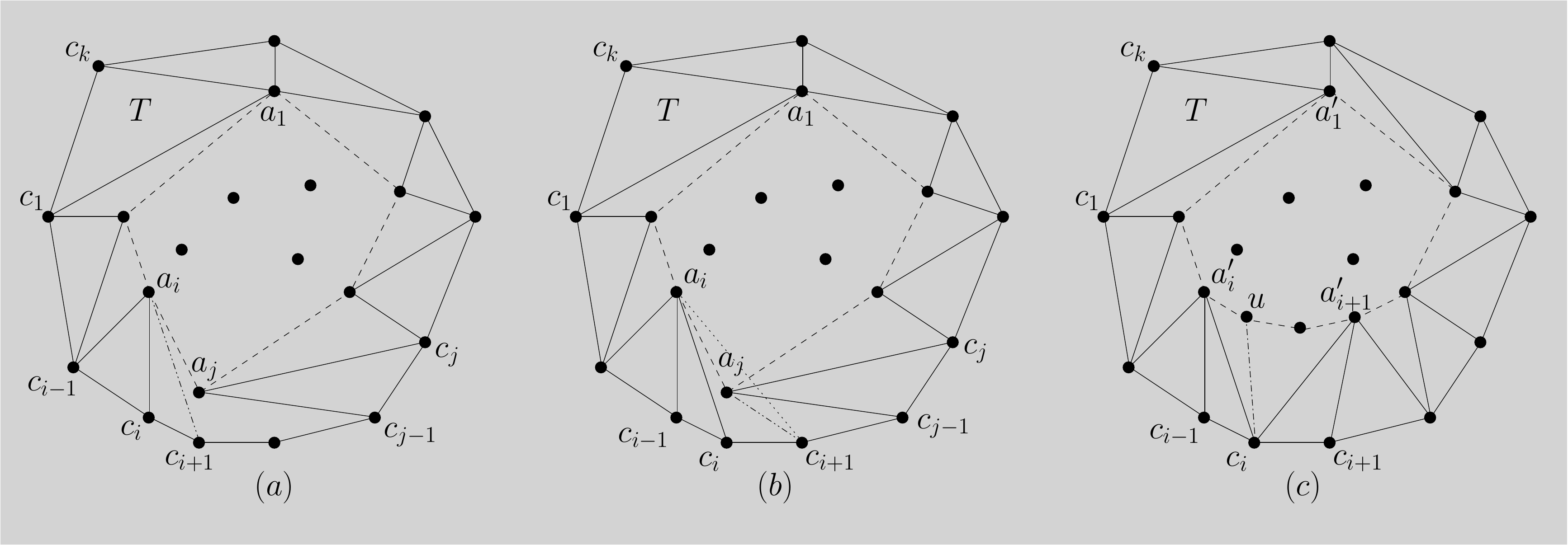,width=1.0\hsize}}}
\caption{ 
(a) The inward triangle $c_ia_ia_{i+1}$ is added to $T$.
(b) The inward triangle $c_ia_ja_{i+1}$ instead of $c_ia_ia_{i+1}$ is added to $T$.
(c) The degenerate inward triangle $c_iu$ is added to $T$.
}
\label{ttos1}
\end{center}
\end{figure}
\newpage
\begin{figure}[h]  
\begin{center} 
\centerline{\hbox{\psfig{figure=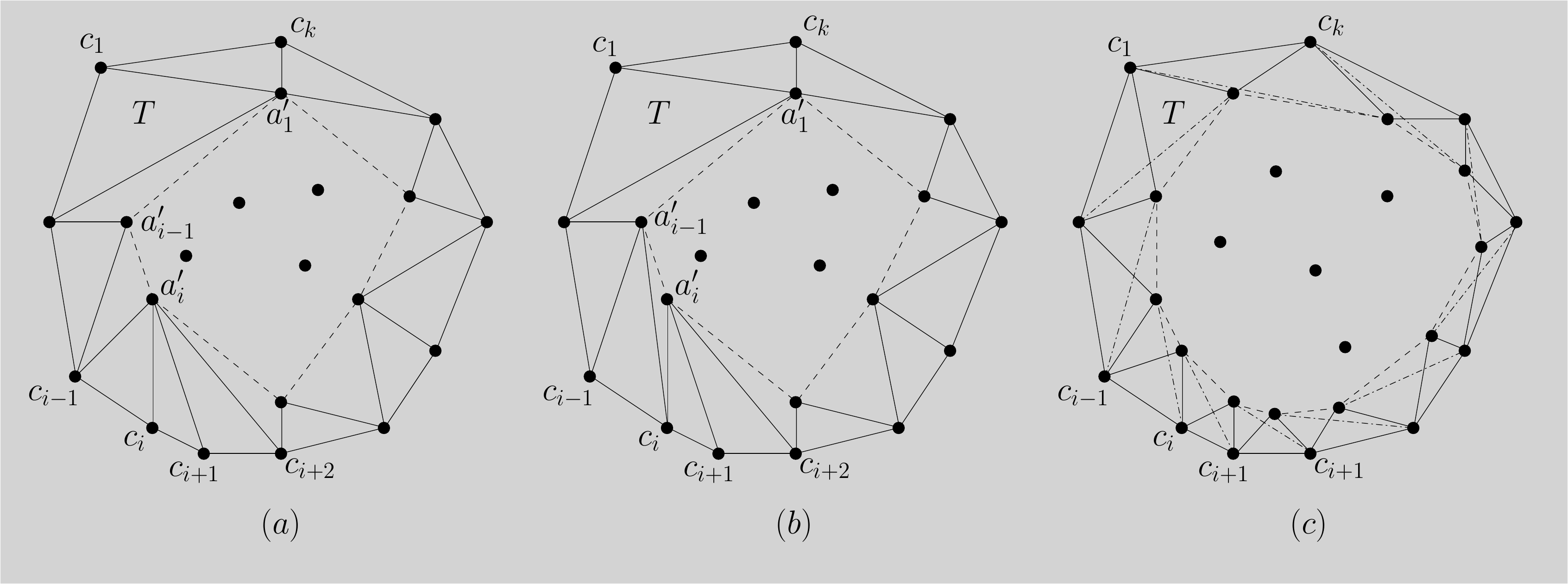,width=1.0\hsize}}}
\caption{ 
(a) The inward triangles $c_{i-1}a_ic_i$, $c_{i}a_ic_{i+1}$ and $c_{i-1}a_ic_{i+1}$ have a common inward vertex $a_i$.
(b) The triangle $c_{i-1}a_ic_i$ is replaced by $c_{i-1}a_{i-1}c_i$ in $T$.
(c) The inward vertex of each inward triangle can be shifted to the vertex in the clockwise order along $CH(P')$. 
}
\label{ttos2}
\end{center}
\end{figure}
\begin{lemma}
\label{cons1}
If $P$ satisfies Necessary Condition \ref{nc3}, then there exists a good set $\mathcal{S}$ of inward triangles.
\end{lemma}
\textbf{Proof:}
Assume that a set $P$ satisfies Necessary Condition \ref{nc3}. So, a set $T$ of inward triangles having maximum cardinality
can be computed while testing for Necessary Condition \ref{nc3}
(see Procedure {testing\_necessary\_conditions()} and Lemma \ref{checklem}).
Observe that $T$ may not satisfy all properties of a good set. We show that $T$ can be converted to a good set $\mathcal{S}$
as follows. 
$\\ \\$
We know that $T$ satisfies Properties 2, 3 and 6 of a good set. 
However, all edges of $CH(P)$ may not have inward triangles
i.e., $T= \{ c_{i-1}a_ic_i \}$ for some values of $i$. Let
$c_ic_{i+1}$ be one such edge where the inward triangle $c_{i-1}a_ic_i$ belongs to $T$. If  
$c_{i}a_ic_{i+1}$ is empty (see Figure \ref{ttos1}(a)), 
add the inward triangle $c_{i}a_ic_{i+1}$ to $T$.
Otherwise, $c_{i}a_ic_{i+1}$ contains a vertex $a_{j}$ of $CH(P')$ which forms an empty triangle
on $c_ic_{i+1}$ (see Figure \ref{ttos1}(b)). 
Add the inward triangle  $c_{i}a_{j}c_{i+1}$ to $T$.
The process is repeated so that every edge of $CH(P)$ has an inward triangle in $T$, satisfying property 5 of a good set.
If a point $u \in CH(P')$ has not been assigned as an inward vertex, then add the degenerate inward triangle 
$c_iu$ to $T$, where $u$ lies between 
two inward triangles on $c_{i-1}c_i$ and $c_{i}c_{i+1}$ with distinct inward vertices.
(see Figure \ref{ttos1}(c)). Note that $c_iu$ does not intersect 
any inward triangle in $T$. This process is repeated so that every vertex of $CH(P')$ is the inward vertex of some inward
triangle (possibly degenerate) in $T$, satisfying property 4 of a good set.
$ \\ \\$
Observe that since all vertices of $CH(P')$ are assigned as inward vertices, all free points lie in the 
interior of $CH(P')$. So, for any inward triangle $c_{i-1}a_ic_i$ and free point $x$, the line segment $a_ix$ does not intersect any 
inward triangle in $T$, satisfying Property 8(a). 
Let $a_{i+1}$ be the next counterclockwise vertex of $a_i$ on $CH(P')$.
Since $a_{i+1}$ is also an inward vertex,
  $a_ic_ia_{i+1}$ does not contain any point, satisfying Properties 8(b) and 8(c).
$\\ \\$
Let $a'_i$ be the inward vertex for two or more triangles in $T$, say, $\{ c_{i-1}a'_ic_i, c_{i}a'_ic_{i+1}, \ldots,
c_{i+j-1}a'_{i}c_{i+j}\}$. 
Let $a'_{i-1}$ be the next clockwise vertex of $a'_i$ on $CH(P')$.
If the triangle
$c_ia'_{i-1}c_i$ does not intersect the interior of $CH(P')$ (see Figure \ref{ttos2}(a)), replace the triangle  $c_{i-1}a'_ic_i$ by 
$c_{i-1}a'_{i-1}c_i$ in $T$
(see Figure \ref{ttos2}(b)).  
Repeat this process for all such triangles wherever possible.
This process must terminate once the right edge of a triangle becomes the left tangent to $CH(P')$, or each vertex of 
$CH(P')$ becomes the inward vertex of only one (possibly degenerate) inward triangle.
Observe that for  all $i \le l \le i+j$, $\overrightarrow{c_la'_i}$ is the left tangent of $c_l$ to $CH(P')$. 
So, consecutive
triangles in $L(T)$ having the same inward vertex satisfy Property \ref{cons}.
Note that this step does not guarantee that all inward triangles have distinct inward vertices.
$\\ \\$
Consider the other case where every vertex of $CH(P')$ is the inward vertex of only one inward triangle in $T$.
If $c_{i-1}a'_ic_i$ is the current inward triangle and 
$\overrightarrow{c_ia'_i}$ is not the left tangent to $CH(P')$, 
then replace $c_{i-1}a'_ic_i$ by $c_{i-1}a'_{i-1}c_i$ in $T$ for all $i$ (see Figure \ref{ttos2}(c)). 
This step is required to ensure that $T$ satisfies Property \ref{cons} even after adding some degenerate inward triangles
to $T$ in the next step, making $T$ a multiset.
$ \\ \\$
In order to satisfy $|T| = |P'|$, degenerate inward triangles are added to $T$.
We find an inward triangle $c_{i-1}a'_ic_i$ in  $T$
such that $\overrightarrow{c_ia'_i}$ is a left tangent to $CH(P')$.
 Such a triangle must exist in  $T$ due to the shifting of triangles mentioned earlier.
We consider $c_ia'_i$ as a degenerate inward triangle and repeat it in 
$T$ till $|T| = |P'|$, satisfying Property 1.
Hence $T$ becomes a good set $\mathcal{S}$. $\hfill{\Box}$
$ \\ \\ $
\newpage
 \begin{algorithm}[H]
  \textbf{constructing\_good\_set(\emph{P})}
  
compute $CH(P)$ and $CH(P')$\;
 let the points of $CH(P)$ be $\{ c_1, c_2, \ldots, c_k \}$ in counterclockwise order\;
$T :=  \textbf{testing\_necessary\_conditions(\emph{P})}$\; 
\tcp{$T$ satisfies Properties 2, 3 and 6}
$a_j : = a_1$\;
\While{$i < |CH(P)|$}{
\eIf{$c_{i-1}c_{i}$ has an inward triangle in $T$  }
{
$i := i + 1$\;
}
{
scan $CH(P)$ for $c_i$ in the counterclockwise order to locate the first inward triangle $c_{j-1}a_jc_j$\;
$l := i$\;
\While{$l<i$ and $c_{l-1}a_{i}c_l$ is empty}
{
$T := T \cup c_{l-1}a_{i-1}c_l$\;
$l := l+1$\;
}
\While{$l<i$ and $c_{l-1}a_{j}c_l$ is empty}
{
$T := T \cup c_{l-1}a_{i-1}c_l$\;
$l := l+1$\;
}
}
$ i := j$\;
}
\tcp{$T$ satisfies Property 5}
 $i := 1$\;
\While{$i<|CH(P)|$}
{
\If{
inward vertices $a'_i$ and $a'_{i+1}$ of $c_{i-1}c_{i}$ and $c_{i}c_{i+1}$ are different
}
{
connect every vertex $u$ of $CH(P')$ inside $a'_ic_ia'_{i+1}$ to $c_i$ and add $c_iu$ to $T$\;
}
 $i:=i+1$\;
 }
$i := 1$\;
\tcp{$T$ satisfies Properties 4 and 8}
\While{$i<|CH(P)|$}
{
\eIf{
 $c_{i-2}a'_{i-1}c_{i-1}$, $c_{i-1}a'_{i}c_{i}$ and  $c_{i}a'_{i}c_{i+1}$ have two different inward vertices and 
 $\overrightarrow{c_ia'_i}$ is not the left tangent to $CH(P')$
 }
{
$T = T \setminus \{ c_{i-1}a'_ic_{i} \}$\;
$T = T \cup \{ c_{i-1}a'_{i-1}c_{i} \}$\;
$ i := i-1$\;
}
{
$ i := i +1$\;
}
}
\tcp{$T$ satisfies Property 7}
\While{all inward triangles have distinct inward vertices 
and the right edge $c_ia'_i$ of no inward triangle is the left tangent to $CH(P')$}
{
 $i := 1$\;
 \While{$i<|CH(P)|$}
 {
 $T = T \setminus \{ c_{i-1}a'_ic_{i} \}$\;
$T = T \cup \{ c_{i-1}a'_{i-1}c_{i} \}$\;
$i = i +1$\;
 }
 }
\tcp{inward triangles in $T$ are shifted to the left}
\If{$|T| < |P'|$}
{
\If{ a right  edge $c_ia'_i$ is a left tangent to $CH(P')$}
{
add $|P'|-|T|$ copies of $c_ia'_i$ to $T$\; 
}
%
}
\tcp{$T$ is a multiset and satisfies Property 1}
$\mathcal{S}:= T$\;
report $\mathcal{S}$\;
\end{algorithm}
$\\ \\$
The correctness of the procedure follows from Lemma \ref{cons1}. It is straight forward to show that the procedure runs 
in $O(n^2)$ time. We have the following theorem.
\begin{theorem}
 A good set $\mathcal{S}$ of $P$ can be computed in $O(n^2)$ time. 
\end{theorem}

\section{Construction of inward triangles with distinct inward vertices}
In this section, we show that $\mathcal{S}$ constructed in the previous section can be transformed 
into another good set such that no two inward triangles have the same inward vertex. The 
process of transformation is carried out by applying shift operations repeatedly. 
Suppose there exists a point $a'$ which is the inward vertex of more than one
inward triangle in $\mathcal{S}$. 
In a \emph{shift} operation, one of the inward triangles of $a'$, say, $c_{i-1}a'c_i$ 
is replaced by another
inward triangle $c_{i-1}a''c_i$ on the same edge of $CH(P)$
such that $a''$ lies to the right of $\overrightarrow{c_ia'}$.
Note that the inward triangle can be degenerate, in which case $c_{i-1}=c_i$.
Observe that $a'$ and $a''$ can be points in the interior of $CH(P')$ (denoted as $P''$) unlike in the previous section where
inward vertices are restricted to vertices of $CH(P')$.
In fact, the shift operations add points of $P''$ to the set of existing inward vertices which allows 
inward triangles in $\mathcal{S}$ to have distinct inward vertices.
Before we discuss shift operations, we state the following lemma on the properties of free points of $P'$,
which is used later in this section.
\begin{figure}[h]  
\begin{center} 
\centerline{\hbox{\psfig{figure=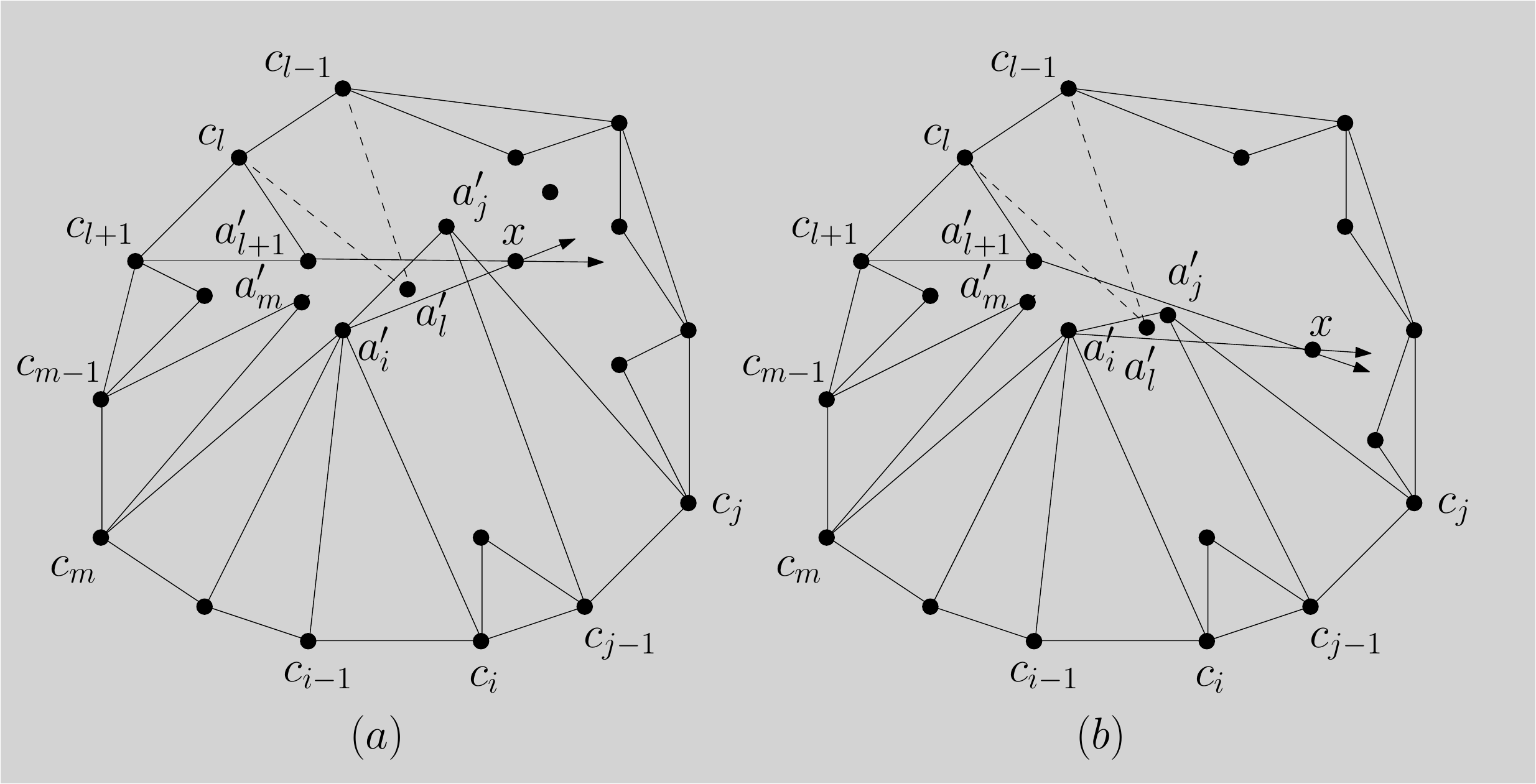,width=0.70\hsize}}}
\caption{ 
(a) The inward triangle $c_{j-1}a'_jc_j$ intersects $a'_ix$ and $a'_{l+1}x$.
(b) The inward triangle $c_{j-1}a'_jc_j$ intersects $a'_ix$ but does not intersect $a'_{l+1}x$.
}
\label{ttos4}
\end{center}
\end{figure}
\begin{lemma}
\label{left}
Let $c_{i-1}a'_ic_i$ be an inward triangle in a good set $\mathcal{S}$ and $x$ be a free point (Figure \ref{ttos4}).
If an inward triangle $c_{j-1}a'_jc_j$ in   $\mathcal{S}$ intersects $a'_ix$ then $c_j$ lies to the right of 
$\overrightarrow{a'_ix}$ and $a'_j$ lies to the left of 
$\overrightarrow{a'_ix}$.
Further, there is no inward triangle in $\mathcal{S}$ $c_{l-1}a'_lc_l$ intersecting $a'_ia'_j$ such that $c_l$ lies to the left of
$\overrightarrow{a'_ia'_j}$, and $a'_l$ lies to the right of
$\overrightarrow{a'_ia'_j}$. 
\end{lemma}
\textbf{Proof:}
Traverse $L(\mathcal{S})$ in the clockwise order from $c_i$ till an inward triangle $c_{m-1}a'_mc_m$
is reached such that $a'_m \neq a'_i$ (see Figure \ref{ttos4}). If any triangle $c_{j-1}a'_jc_j$ 
intersects $a'_ix$, then by the Property 8(b) of $c_{m-1}a'_mc_m$,
 $c_j$ lies to the right of $\overrightarrow{a'_{i}x}$, $a'_j$ lies to the left of 
$\overrightarrow{a'_{i}x}$, and the line segment
$a'_{i}a'_j$ does not intersect $c_{m-1}a'_mc_m$.
$ \\ \\$
Assume on the contrary that there exists an inward triangle $c_{l-1}a'_lc_l$ that has intersected $a'_ia'_j$
and $c_l$ lies to the left of $\overrightarrow{a'_ia'_j}$. If $c_{l-1}a'_lc_l$ is the next inward triangle 
in the clockwise order where $a'_l \neq a'_i$, then $c_{l-1}a'_lc_l$ does not intersect 
$a'_ia'_j$ by Property 8(b) of $c_{l-1}a'_lc_l$. Otherwise, there exists inward triangles between 
$c_{l-1}a'_lc_l$ and $c_{i-1}a'_ic_i$
having different vertices. Without the loss of generality, we assume that  $c_{l}a'_{l+1}c_{l+1}$ is one such inward triangle
such that $a'_l \neq a'_{l+1} \neq a'_i$ and $c_{l}a'_{l+1}c_{l+1}$ does not intersect $a'_ia'_j$. 
If $c_{j-1}a'_jc_j$ intersects $a'_{l+1}x$ (see Figure \ref{ttos4}(a)), then $c_{l-1}a'_lc_l$ cannot intersect $a'_{l+1}a'_j$
due to Property 8(b) and therefore $c_{l-1}a'_lc_l$ cannot intersect $a'_ia'_j$ without intersecting 
$a'_{l+1}a'_j$ which is a contradiction.
Again, if $c_{j-1}a'_jc_j$ does not intersect $a'_{l+1}x$ (see Figure \ref{ttos4}(b)), then $c_{l-1}a'_lc_l$ cannot intersect 
$a'_ia'_j$ without intersecting $a'_{l+1}x$ due to Property 8(a).
Hence no such triangle $c_{l-1}a'_lc_l$ in $\mathcal{S}$ intersects $a'_ia'_j$.
%
%
%
%
%
%
%
%
%
%
%
$\hfill{\Box}$ 
$\\ \\$
Let us now explain shift operations.
Let $Z = ( c_{i-1}a'_ic_i, c_{i}a'_ic_{i+1}, \ldots, c_{i+j-1}a'_ic_{i+j}  )$ be a maximal sequence   
of consecutive inward triangles in $L(\mathcal{S})$ with $a'_i$ as inward vertex.  
We call $Z$ a \emph{zone} of $a'_i$. If $a'_i$ is a vertex of $CH(P')$ (see Figure \ref{ttos3}(a)), it can have only one zone.
 Otherwise, $\mathcal{S}$ must have a forbidden triangle, violating property 2 of good sets.
 However, if $a'_i \in P''$, then $a'_i$ can have multiple zones (see Figure \ref{ttos3}(b)).
The right edge of the last triangle in a zone in counterclockwise order is called the 
right edge of the zone.
The right edges of the zones, which are all line segments joining 
$a'_i$ to some points of $CH(P)$, partition the interior of $CH(P)$ into disjoint regions. All free 
points must be contained in one region $R$, since a line segment joining two free points does not 
intersect any triangle in $\mathcal{S}$.
Note that if $a'_i$ is a vertex of $CH(P')$, then 
 there is only one region in $CH(P)$.
$ \\ \\$
Suppose $R$ is a convex region. Let $c_ia'_i$ and $c_ja'_i$ be the right edges of zones bounding 
$R$, such that $c_j$ is to the right of $\overrightarrow{c_ia'_i}$ (see Figure \ref{ttos3}(b)). 
Let $t_1 = c_{i-1}a'_ic_i$ be the last triangle in anti-clockwise
order in the zone with right edge $c_ia'_i$. 
If $t_1$ is a degenerate triangle then $t_1 = c_ia'_i$. 
Consider the case where $R$ is nonconvex.
Let $c_ia'_i$ and $c_ja'_i$ be the right edges of zones bounding $R$ 
such that $c_j$ is to the left of $\overrightarrow{c_ia'_i}$ (see Figure \ref{ttos3}(c)).
Again, let $t_1 = c_{i-1}a'_ic_i$ be the last triangle in the counterclockwise
order in the zone with right edge $c_ia'_i$. 
If $t_1$ is a degenerate triangle then $t_1 = c_ia'_i$. 
We choose the triangle $t_1$ to shift. 
The choice of $t_1$ ensures the properties stated in the following lemma.
\begin{lemma}
\label{choice}
Let $t_1 = c_{i-1}a'_ic_i$ be the triangle selected for shifting (see Figure \ref{ttos3}). Then the following properties hold:
\begin{enumerate}
\item
The triangle following $t_1$ in $L(\mathcal{S})$ has an inward vertex different from $a'_i$.
\item
Every free point is to the right of $\overrightarrow{c_{i-1}a'_i}$ and hence also of $\overrightarrow{c_ia'_i}$. 
\item
The line segment joining $c_i$ to any free point does not intersect any triangle with inward vertex $a'_i$.
\end{enumerate}
\end{lemma}
\begin{figure}[h]  
\begin{center} 
\centerline{\hbox{\psfig{figure=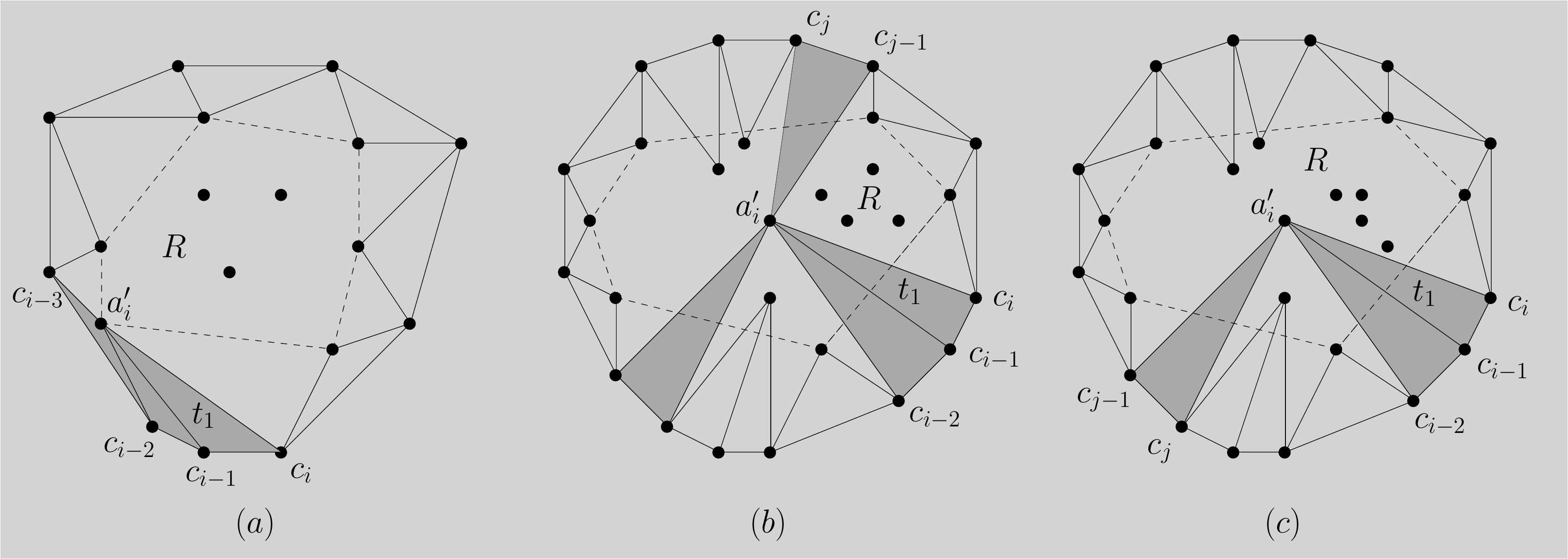,width=1.00\hsize}}}
\caption{ 
(a) Since $a'_i$ is a vertex of $CH(P')$, there can be only one zone.
(b) There are three zones with $a'_i$ as the common inward vertex and all free points lie inside the convex region $R$.
(c) All free points lie inside the nonconvex region $R$.
}
\label{ttos3}
\end{center}
\end{figure}
After $t_1$ is selected, a free point $x$ is located in $P$ 
such that all remaining free points lie to the right of $\overrightarrow{c_ix}$.
Observe that at least one such free point $x$ exists because $| \mathcal{S} | = | P' |$ and there
exist two inward triangles in $\mathcal{S}$ sharing the same inward vertex.
Let $c_ia'_{i+1}c_{i+1}$ be the next triangle of $t_1$ in $L(\mathcal{S})$ .
Assume that $c_ix$ is intersected by a set $Q$ of inward triangles in $\mathcal{S}$.
Let $c_{j-1}a'_jc_j$ be an inward triangle in $Q$ such that all inward vertices of inward
triangles in $Q$ lie to the right of $\overrightarrow{c_ia'_j}$.
We have the following four cases.
$\\ \\$
\textbf{Case 1:} The inward triangle $c_{i-1}xc_i$ is not intersected by any triangle in $\mathcal{S}$,
and $c_ixa'_{i+1}$ is empty (see Figures \ref{ttos5}(a) and \ref{ttos5}(b) ).

$ \\ $
\textbf{Case 2:} The inward triangle $c_{i-1}xc_i$ is intersected by a triangle $c_{j-1}a'_jc_j$ in $\mathcal{S}$,
and $c_ia'_ja'_{i+1}$ is empty (see Figure \ref{ttos9}(a)).

$ \\ $
\textbf{Case 3:} The inward triangle $c_{i-1}xc_i$ is not intersected by any triangle in $\mathcal{S}$,
and $c_ixa'_{i+1}$ is not empty (see Figure \ref{ttos13}(a)).

$ \\ $
\textbf{Case 4}  The inward triangle $c_{i-1}xc_i$ is intersected by a triangle $c_{j-1}a'jc_j$ in $\mathcal{S}$,
and $c_ia'_ja'_{i+1}$ is not empty (see Figure \ref{ttos17}(a)).
%
%
%
%
$\\ \\$
For the shift operation in case 1, it is sufficient to show that $\mathcal{S}$ remains a good set after $c_{i-1}a'_ic_i$
is replaced by $c_{i-1}xc_i$ to obtain $(\mathcal{S} \setminus \{ c_{i-1}a'_ic_i \} ) \cup \{ c_{i-1}xc_i \} $.
Note that the clockwise next inward triangle of $c_{i-1}a'_ic_i$ can
share the same inward vertex (see Figure \ref{ttos5}(a)) or have a different inward vertex
(see Figure \ref{ttos5}(b)).
We have the following lemma.
\begin{figure}[h]  
\begin{center} 
\centerline{\hbox{\psfig{figure=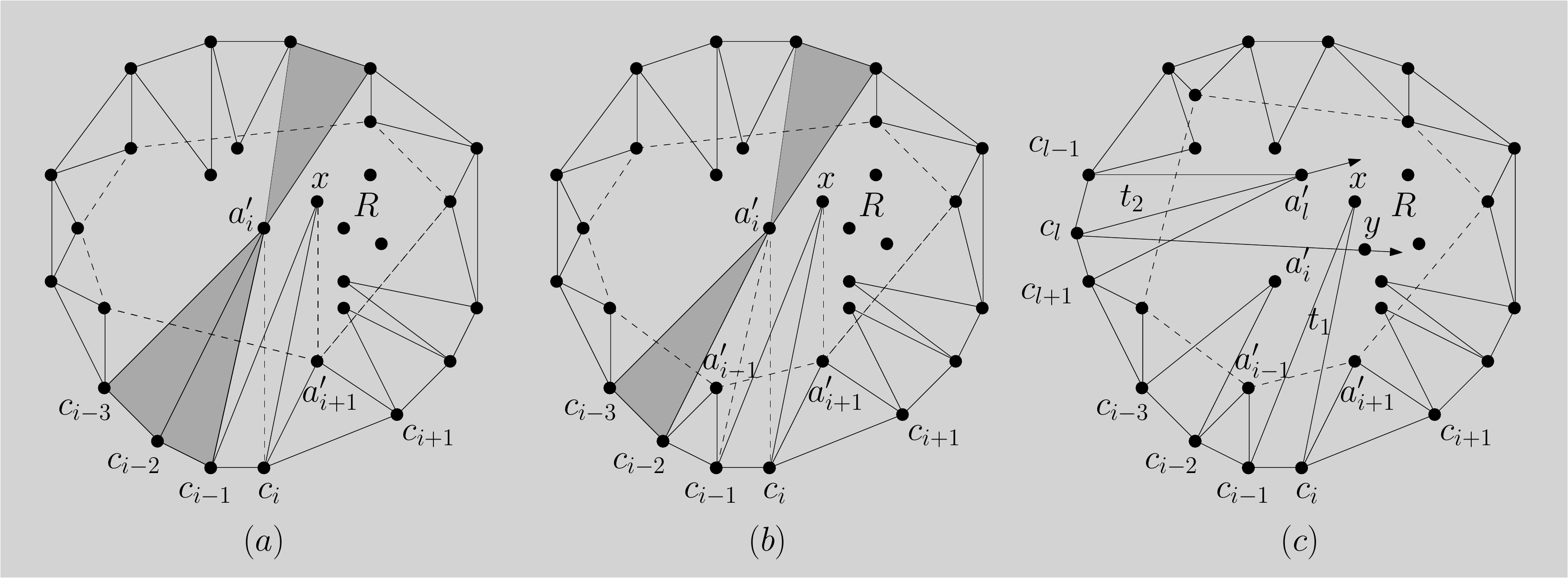,width=1.00\hsize}}}
\caption{ 
(a) The inward triangle $c_{i-1}a'_ic_i$ has been replaced by $c_{i-1}xc_i$ in $\mathcal{S}$.
(b) The inward triangle $c_{i-1}a'_ic_i$ has been replaced by $c_{i-1}xc_i$ in $\mathcal{S}$,
and $a'_{i-1} \neq a'_i$.
(c) The set  $(\mathcal{S} \setminus \{c_{i-1}a'_ic_i\} ) \cup \{c_{i-1}xc_i\}$ satisfies Property 7(b).
}
\label{ttos5}
\end{center}
\end{figure}
\begin{figure}[h]  
\begin{center} 
\centerline{\hbox{\psfig{figure=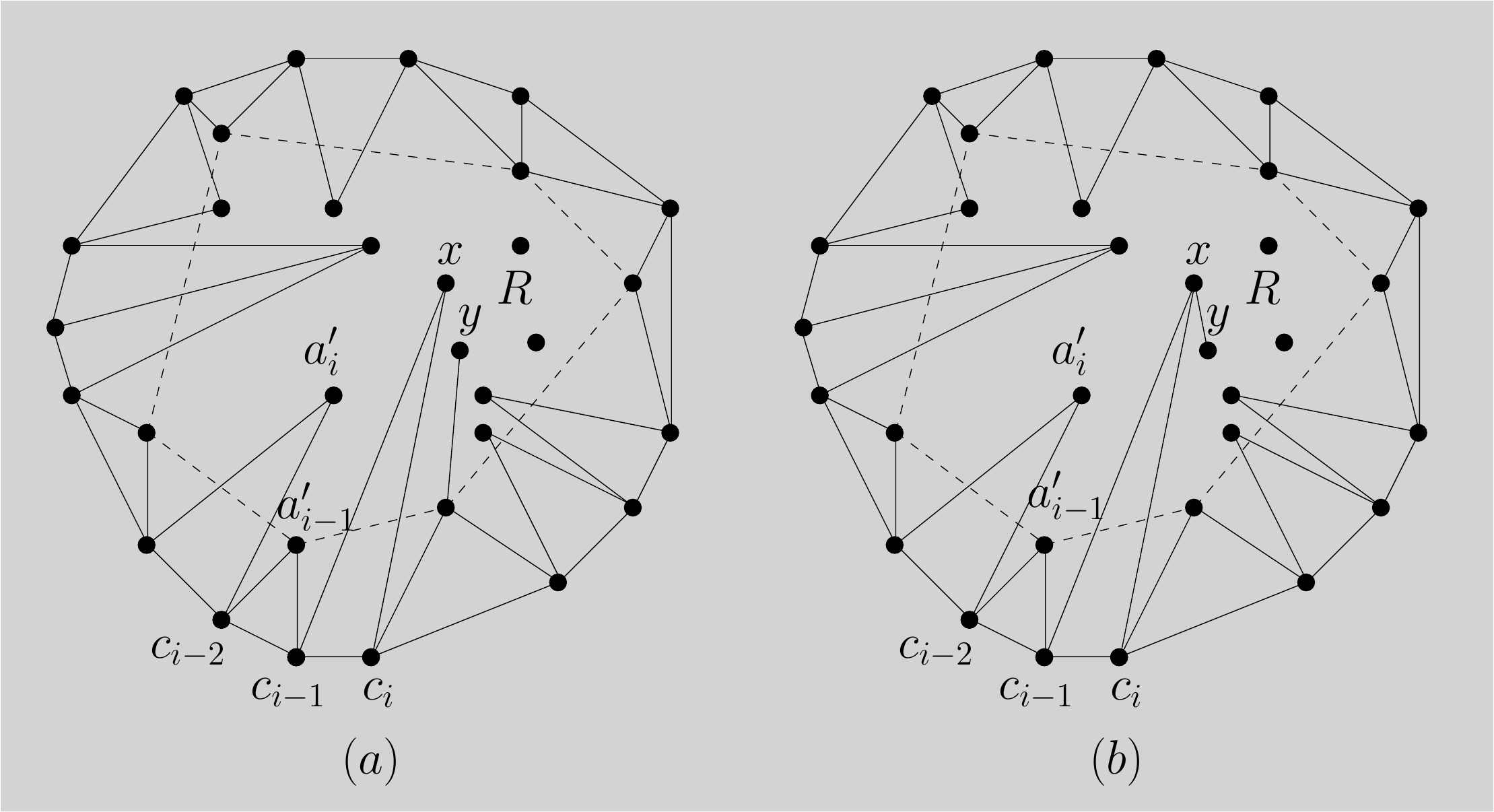,width=0.70\hsize}}}
\caption{ 
(a) The triangle $c_{i-1}xc_i$ satisfies Property 8(a) as $a'_{i+1}y$ does not intersect  $c_{i-1}xc_i$.
(b)  The triangle $c_{i-2}a'_ic_{i-1}$ satisfies Property 8(a) as $xy$ does not intersect  $c_{i-2}a'_ic_{i-1}$.
}
\label{ttos6}
\end{center}
\end{figure}
\begin{figure}[h]  
\begin{center} 
\centerline{\hbox{\psfig{figure=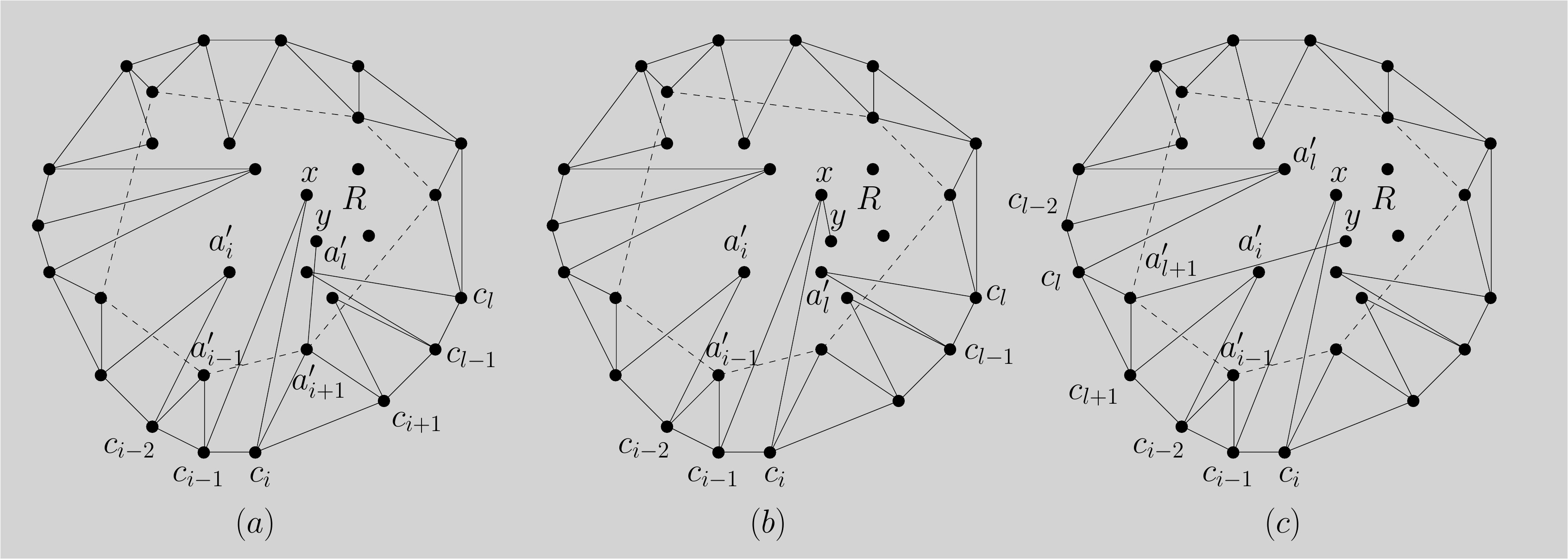,width=1.0\hsize}}}
\caption{ 
(a) The segment $a'_{i+1}y$ cannot be intersected by $c_{i-1}xc_i$.
(b) The segment $xy$ cannot be intersected by $c_{l-1}a'_lc_l$.
(c) The segment $a'_{l+1}y$ cannot be intersected by $c_{l-1}a'_lc_l$.
}
\label{ttos7}
\end{center}
\end{figure}
\begin{lemma} \label{mainlem1}
 The set $(\mathcal{S} \setminus \{ c_{i-1}a'_ic_i \} ) \cup \{ c_{i-1}xc_i \}$ is a good set
 after the shift operation in Case 1.
\end{lemma}
\textbf{Proof:}
It can be seen that $(\mathcal{S} \setminus \{c_{i-1}a'_ic_i\} ) \cup \{c_{i-1}xc_i\}$ satisfies Properties 1, 3, 4 and 5.
Property 2 is also satisfied because $x$ lies in the interior of $CH(P')$. 
The triangle $c_{i-1}xc_i$ satisfies Property 6
because all remaining free points are to the right of $\overrightarrow{c_ix}$.
Since all triangles in $\mathcal{S} \setminus \{c_{i-1}a'_ic_i\}$ satisfy Property 6,
$(\mathcal{S} \setminus \{ c_{i-1}a'_ic_i \} ) \cup \{ c_{i-1}xc_i \}$
also satisfies Property 6.
%
%

\begin{figure}[h]  
\begin{center} 
\centerline{\hbox{\psfig{figure=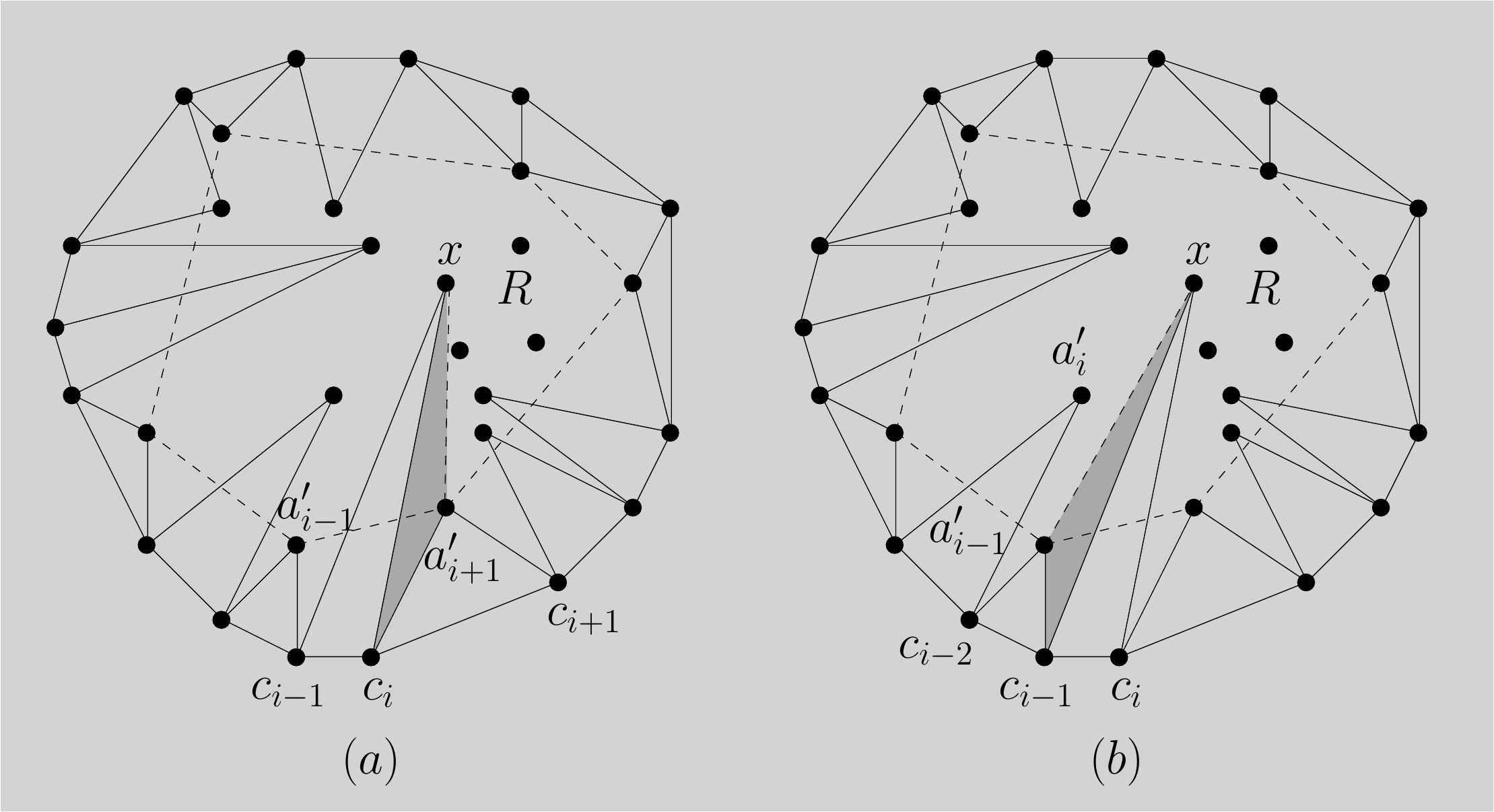,width=0.70\hsize}}}
\caption{ 
 (a) The triangle $c_ixa'_{i+1}$ is empty.
 (b) The triangle $a'_ic_{i-1}x$ is empty.
}
\label{ttos8}
\end{center}
\end{figure}

\begin{figure}[h]  
\begin{center} 
\centerline{\hbox{\psfig{figure=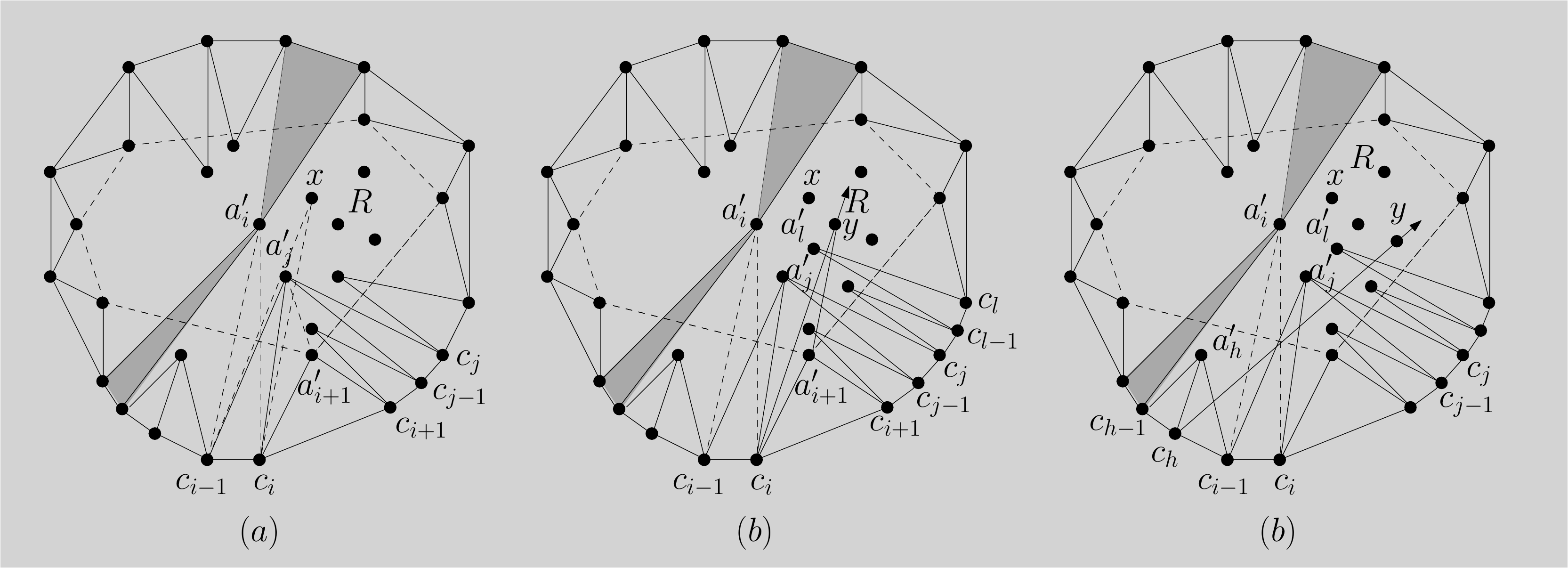,width=1.00\hsize}}}
\caption{ 
(a) The inward triangle $c_{i-1}a'_ic_i$ has been replaced by $c_{i-1}a'_jc_i$ in $\mathcal{S}$.
(b) The inward triangle  $c_{i-1}a'_jc_i$
in $(\mathcal{S} \setminus \{c_{i-1}a'_ic_i\} ) \cup \{c_{i-1}a'_jc_i\}$ satisfies Property 7(b).
(c) The inward triangle $c_{h-1}a'_hc_h$
in $(\mathcal{S} \setminus \{c_{i-1}a'_ic_i\} ) \cup \{c_{i-1}a'_jc_i\}$ satisfies Property 7(b).
}
\label{ttos9}
\end{center}
\end{figure}

$\\ \\$
Observe that $c_{i-1}xc_i$ does not satisfy the precondition of Property 7 by construction.
After the triangle replacement, it may appear that some triangle $c_{l-1}a'_lc_l$,  
that satisfies the precondition of Property 7, violates Property 7(b) due to
the intersection of 
$c_ly$ and $c_{i-1}xc_i$,
where $y$ is a free point, 
$c_{l-1}a'_lc_l$ is an inward triangle in $\mathcal{S} \setminus \{c_{i-1}a'_ic_i\}$,
and $a'_l$ is also the inward vertex of $c_{l}a'_lc_{l+1} \in \mathcal{S} \setminus \{c_{i-1}a'_ic_i\}$ (see Figure \ref{ttos5}(c)).
We know that $c_{l-1}a'_lc_l$ satisfies Property 7(a).
Since $c'_ly$ and $c_ix$ are intersecting segments, and $y$ lies to the right of $\overrightarrow{c_ix}$,
$c_i$ must lie to the right of $\overrightarrow{c_ly}$. Moreover, $x$ lies to the right of 
$\overrightarrow{c_la'_l}$ by Property 7(a). Hence, $c_{l-1}a'_lc_{l}$ satisfies Property 7(b).
$\\ \\$
%
Observe that $c_{i-1}xc_i$ satisfies the precondition of Property 8.
After the replacement, it may appear that $c_{i-1}xc_i$ violates Property 8(a) by intersecting $a'_{i+1}y$
in $(\mathcal{S} \setminus \{ c_{i-1}a'_ic_i \} ) \cup \{ c_{i-1}xc_i \}$, where 
$y$ is a free point and $c_{i}a'_{i+1}c_{i+1}$ is the next clockwise inward triangle in 
$L((\mathcal{S} \setminus \{ c_{i-1}a'_ic_i \} ) \cup \{ c_{i-1}xc_i \})$ with a different inward vertex (see Figure \ref{ttos6}(a)).
Since both $a'_{i+1}$ and $y$ lie to the right of $\overrightarrow{c_ix}$ by Property 3 and by construction,
$c_{i-1}xc_i$ cannot intersect $a'_{i+1}y$. For the counterclockwise previous inward triangle
$c_{i-2}a'_{i-1}c_{i-1}$ of  $c_{i-1}xc_i$ in $L((\mathcal{S} \setminus \{ c_{i-1}a'_ic_i \} ) \cup \{ c_{i-1}xc_i \})$,
$xy$ cannot be intersected by $c_{i-2}a'_{i-1}c_{i-1}$  due to Property 6 in $\mathcal{S}$, and therefore
$(\mathcal{S} \setminus \{ c_{i-1}a'_ic_i \} ) \cup \{ c_{i-1}xc_i \}$ satisfies Property 8(a) (see Figure \ref{ttos6}(b)).
%
%
%
%
%
%
%
%
%
$\\ \\$
After the triangle replacement, it may appear that 
$c_{i-1}xc_i$ in $(\mathcal{S} \setminus \{ c_{i-1}a'_ic_i \} ) \cup \{ c_{i-1}xc_i \}$ violates
  Property 8(b) (see Figure \ref{ttos7}(a)). Let $y$ be a free point. Consider the first sub-case where an inward triangle 
$c_{l-1}a'_lc_l$ has intersected $a'_{i+1}y$. We know that $xy$ cannot be intersected by any triangle in $\mathcal{S}$
and also in $(\mathcal{S} \setminus \{ c_{i-1}a'_ic_i \} ) \cup \{ c_{i-1}xc_i \}$ due to Property 6. So, no triangle in
$(\mathcal{S} \setminus \{ c_{i-1}a'_ic_i \} ) \cup \{ c_{i-1}xc_i \}$ can intersect $a'_{i+1}y$ by intersecting 
$xy$, and therefore, $a'_l$ and $c_l$ lie to the left and right of $\overrightarrow{a'_{i+1}y}$ respectively.
Moreover, $c_{i-1}xc_i$ cannot intersect $a'_{i+1}a'_l$ because $a'_{i+1}$ lies to the right of 
$\overrightarrow{c_ix}$ and $a'_l$ also lies to the right of $\overrightarrow{c_ix}$ 
as $c_{l-1}a'_lc_l$ does not intersect $xy$. So, $c_{i-1}xc_i$ satisfies Property 8(b). 
$\\ \\$
Suppose that $c_{i-2}a'_ic_i$  satisfies the precondition of Property 8.
Consider the second sub-case where it may appear that $c_{i-2}a'_ic_i$ in $(\mathcal{S} \setminus \{ c_{i-1}a'_ic_i \} ) \cup \{ c_{i-1}xc_i \}$
violates Property 8(b) (see Figure \ref{ttos7}(b)). We know that $xy$ cannot be intersected by any triangle in $\mathcal{S}$
and also in $(\mathcal{S} \setminus \{ c_{i-1}a'_ic_i \} ) \cup \{ c_{i-1}xc_i \}$ due to Property 6.
So, $c_{i-2}a'_{i-1}c_{i-1}$ satisfies Property 8(b).
$\\ \\$
Consider the third sub-case where it may appear that an inward triangle $c_{l-1}a'_lc_l$, 
that satisfies the precondition of Property 8 in 
$(\mathcal{S} \setminus \{ c_{i-1}a'_ic_i \} ) \cup \{ c_{i-1}xc_i \}$,
violates Property 8(b) (see Figure \ref{ttos7}(c)). Since $y$ lies to the 
right of $\overrightarrow{c_ix}$ and $c_{i-1}xc_i$ intersects $a_{l+1}y$,
then $c_i$ and $x$ must lie to the right and left of $\overrightarrow{a_{l+1}y}$ respectively. 
Moreover, $c_{l-1}a'_lc_l$ cannot intersect $a'_{l+1}x$ due to Property 8(a) of $c_{l-1}a'_lc_l$.
Therefore, $c_{l-1}a'_lc_l$ satisfies 8(b).
$\\ \\$
After the triangle replacement, it may appear that $c_{i-1}xc_i$ or $c_{i-2}a'_{i-1}c_{i-1}$ violate Property 8(c).
By the definition of Case 1, $c_ixa'_{i+1}$ is empty and therefore Property 8(c) is not violated (see Figure \ref{ttos8}(a)).
On the other hand,
$c_{i-2}a'_{i-1}c_{i-1}$ also cannot contain any free point as all free points lie to the right of $\overrightarrow{c_ix}$ 
by construction, and $x$ lies to the right of $\overrightarrow{c_ia'_i}$ by Property 3. If
$c_{i-1}a'_ix$ contains any inward vertex $a'_l$, 
it means that $a_ix$ has been intersected by the inward triangle $c_{l-1}a'_lc_l$.
Let $c_{l-1}a'_lc_l$ be the first triangle in the clockwise direction from $c_{i-1}xc_i$ on
$L((\mathcal{S} \setminus \{ c_{i-1}a'_ic_i \} ) \cup \{ c_{i-1}xc_i \})$ such that $a'_{i-1} \neq a'_i$.
Due to Property 8(b), $c_{l-1}a'_lc_l$ cannot intersect $a'_{i-1}x$ and hence $a'_l$ cannot lie inside  
$c_{i-1}a'_ix$ (see Figure \ref{ttos8}(b)). So, $c_{i-2}a'_{i-1}c_i$ also satisfies Property 8(c).
 $\hfill{\Box}$

\begin{figure}[h]  
\begin{center} 
\centerline{\hbox{\psfig{figure=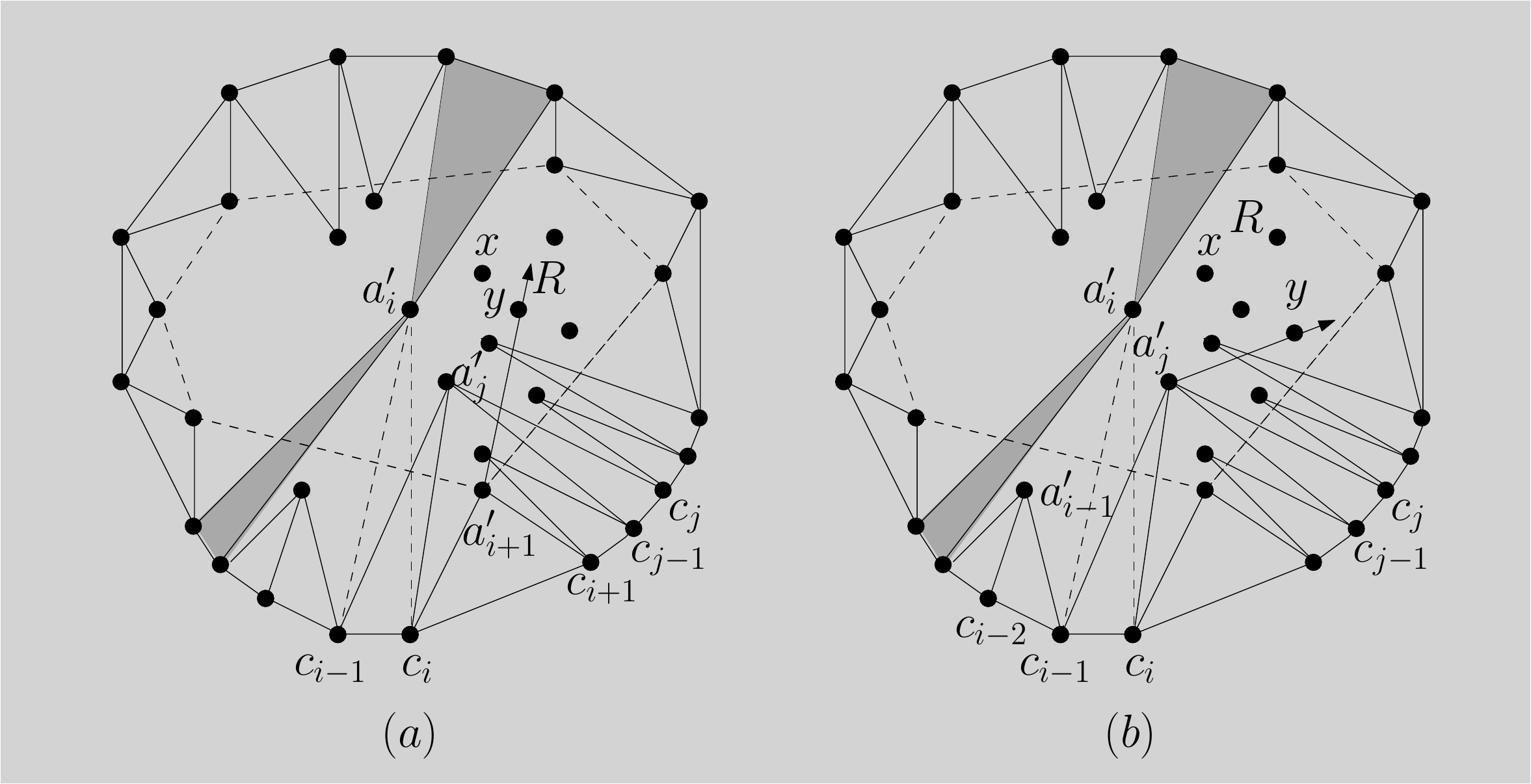,width=0.700\hsize}}}
\caption{ 
(a)  The inward triangle  $c_{i-1}a'_jc_i$
in $(\mathcal{S} \setminus \{c_{i-1}a'_ic_i\} ) \cup \{c_{i-1}a'_jc_i\}$ satisfies Property 8(a).
(b) The inward triangle  $c_{i-2}a'_{i-1}c_{i-1}$
in $(\mathcal{S} \setminus \{c_{i-1}a'_ic_i\} ) \cup \{c_{i-1}a'_jc_i\}$ satisfies Property 8(a).
}
\label{ttos10}
\end{center}
\end{figure}

\begin{figure}[h]  
\begin{center} 
\centerline{\hbox{\psfig{figure=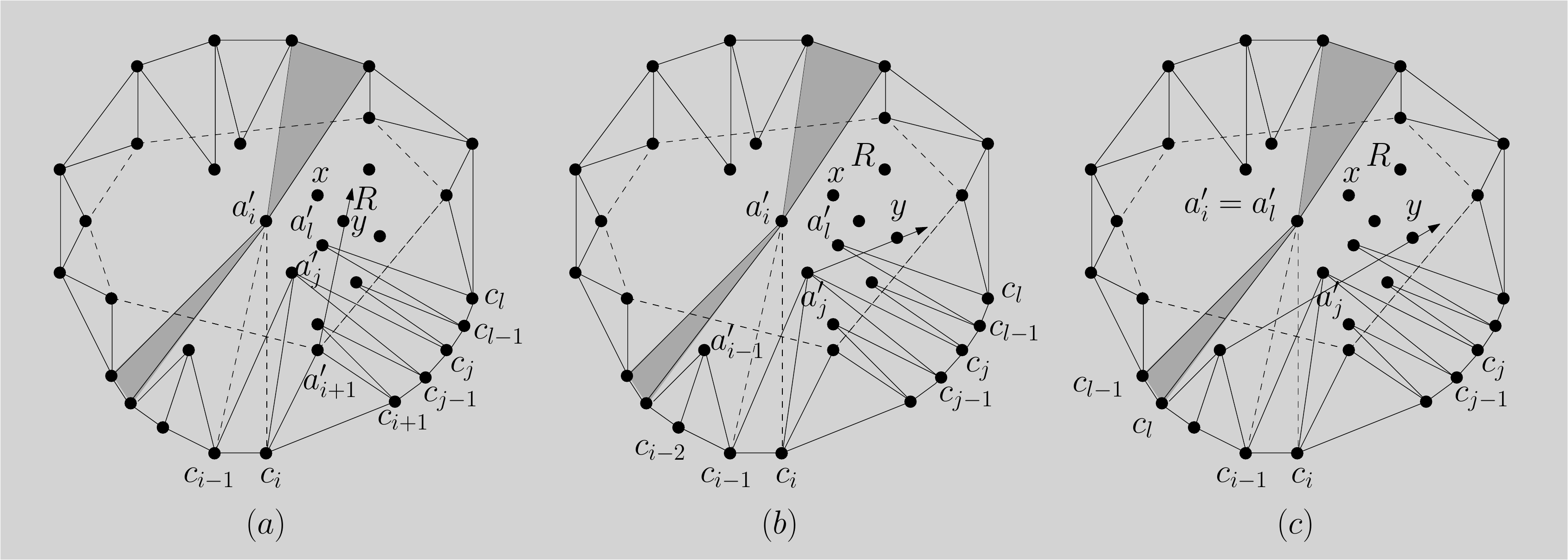,width=1.00\hsize}}}
\caption{ 
(a)  The inward triangle  $c_{i-1}a'_jc_i$
in $(\mathcal{S} \setminus \{c_{i-1}a'_ic_i\} ) \cup \{c_{i-1}a'_jc_i\}$ satisfies Property 8(b).
(b)  The inward triangle  $c_{i-2}a'_{i-1}c_{i-1}$
in $(\mathcal{S} \setminus \{c_{i-1}a'_{i}c_{i}\} ) \cup \{c_{i-1}a'_jc_i\}$ satisfies Property 8(b).
(c)  The inward triangle  $c_{l-1}a'_lc_l$
in $(\mathcal{S} \setminus \{c_{i-1}a'_ic_i\} ) \cup \{c_{i-1}a'_jc_i\}$ satisfies Property 8(b).
}
\label{ttos11}
\end{center}
\end{figure}
\begin{figure}[h]  
\begin{center} 
\centerline{\hbox{\psfig{figure=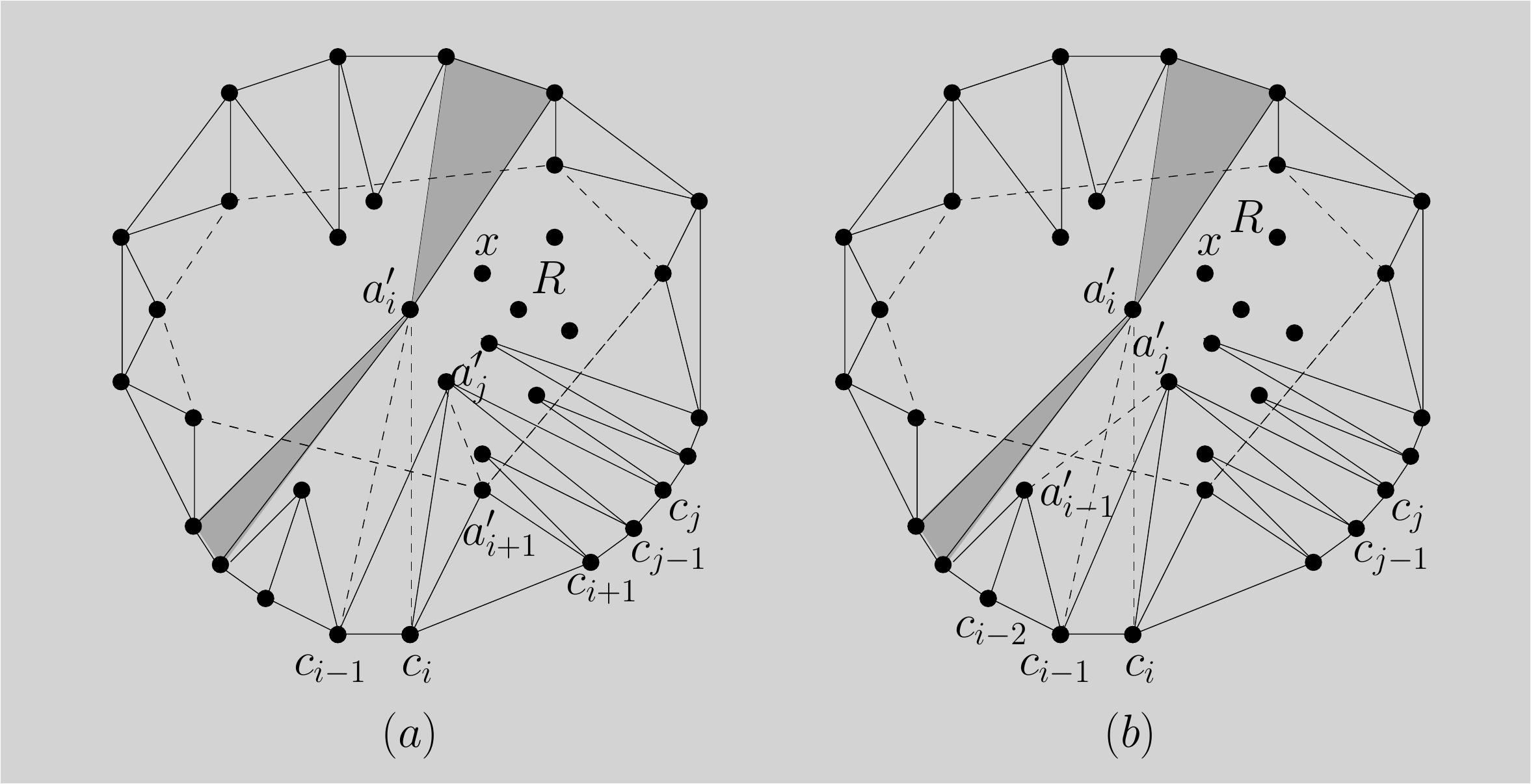,width=0.70\hsize}}}
\caption{ 
(a) The triangle $c_ia'_ja'_{i+1}$ is empty.
 (b) The triangle $a'_ic_{i-1}a'_j$ is empty.
}
\label{ttos12}
\end{center}
\end{figure}

\begin{lemma} \label{mainlem2}
 The set $(\mathcal{S} \setminus \{ c_{i-1}a'_ic_i \} ) \cup \{ c_{i-1}a'_jc_i \}$ is a good set
 after the shift operation in Case 2.
\end{lemma}
\textbf{Proof:}
It can be seen that $(\mathcal{S} \setminus \{ c_{i-1}a'_ic_i \} ) \cup \{ c_{i-1}a'_jc_i \}$
satisfies Properties 1, 4 and 5 (see Figure \ref{ttos9}(a)). 
If $a'_j$ lies in the interior of $CH(P')$, then $c_{i-1}a'_jc_i$ satisfies Property 2.
If $a'_j$ is a vertex of $CH(P')$ then $a'_j = a'_{i+1}$, where $a'_{i+1}$ is the inward vertex of 
the counterclockwise next inward triangle of $c_{i-1}a'_ic_i$ on $L(\mathcal{S})$. 
Hence, $c_{i-1}a'_jc_i$ does not intersect any edge of $CH(P')$ satisfying Property 2.
$\\ \\$
After the triangle replacement, it may appear that the triangle $c_{i-1}a'_jc_i$ violates Property 3
by intersecting an inward triangle $c_{l-1}a'_lc_l$ in $\mathcal{S}$.
By Property 7(b) of  
$c_{i-2}a'_{i-1}c_{i-1}$, $c_l$ and $a'_l$ must lie to the right
and left of $\overrightarrow{c_{i-1}x}$ respectively (see Figure \ref{ttos9}(a)). 
We know that, no inward triangle in  $\mathcal{S}$ can intersect 
$a_ix$ due to Property 8(b) of $c_{m-1}a'_{i-1}c_m$ in $\mathcal{S}$, 
where  $c_{m-1}a'_{i-1}c_m$ is the first triangle 
in the clockwise order from $c_{i-1}a'_ic_i$ on $L(\mathcal{S})$, with $a_{i-1}$ as its inward vertex.
So, no inward triangle can intersect 
 $c_{i-1}x$ without intersecting $c_ix$.
If a triangle $c_{l-1}a'_lc_l$ intersects  $c_{i-1}a'_jc_i$, then it must intersect $c_ix$.
Since $c_{l-1}a'_lc_l$ intersects both $c_{i-1}a'_jc_i$ and $c_ix$, $c_{i-1}a'_jc_i$  must intersect $c_ia'_j$.
%
However, due to our choice of the triangle $c_{j-1}a'_jc_j$,
 $a'_j$ cannot lie to the right of 
$\overrightarrow{c_ia'_l}$.
 Since no inward triangle in $(S \setminus \{ c_{i-1}a'_ic_i \}) \cup \{ c_{i-1}a'_jc_i \} $ intersects $c_{i-1}a'_jc_i$,  
 $(\mathcal{S} \setminus \{ c_{i-1}a'_ic_i \} ) \cup \{ c_{i-1}a'_jc_i \}$ satisfies Property 3.
$\\ \\$
Observe that all remaining free points
of $\mathcal{S}$ lie to the right of $\overrightarrow{c_ix}$ and  $a'_j$ lies to the 
left of $\overrightarrow{c_ix}$.
So, no line segment joining any two free points of 
$(\mathcal{S} \setminus \{ c_{i-1}a'_ic_i \} ) \cup \{ c_{i-1}a'_jc_i \}$ can intersect $c_{i-1}a'_jc_i$.
Therefore,
$(\mathcal{S} \setminus \{ c_{i-1}a'_ic_i \} ) \cup \{ c_{i-1}a'_jc_i \}$
still satisfies Property 6
even after the triangle replacement.
$\\ \\$
Clearly, all inward triangles, that satisfy the precondition of Property 7 
in $(\mathcal{S} \setminus \{ c_{i-1}a'_ic_i \} ) \cup \{ c_{i-1}a'_jc_i \}$,
satisfy Property 7(a) by construction.
Suppose that $c_{i-1}a'_jc_i$  satisfies the precondition of Property 7.
After the triangle replacement, it may appear that $c_{i-1}a'_jc_i$  violates Property 7(b) (see Figure \ref{ttos9}(b)).
Let $y$ be a free point.
Consider the first subcase where an inward triangle $c_{l-1}a'_lc_l$ 
of $\mathcal{S}$ intersects $c_iy$. If $c_{l-1}$ lies to the left of
$\overrightarrow{c_ix}$, then  $c_{l-1}a'_lc_l$ must intersect $c_iy$, contradicting 
Property 8(b) of $c_{m-1}a'_{i-1}c_m$, where  $c_{m-1}a'_{i-1}c_m$ is the first triangle 
in the clockwise order from $c_{i-1}a'_ic_i$ on $L(\mathcal{S})$, with $a_{i-1}$ as its inward vertex.
So, $c_{l-1}$ and $a'_l$ must lie to the right and left of $\overrightarrow{c_ix}$ respectively. 
If $y = x$ then by construction $a'_l$ lies to
the right of $\overrightarrow{c_ia'_j}$.  Consider the other case where $y \neq x$.
The  point $g$ lies to the right of $\overrightarrow{c_ix}$. If $a'_l$ lies to the left of
$\overrightarrow{c_ia'_j}$ then  $a'_l$ must lie to the left of $\overrightarrow{c_ix}$.
But then since $y$ lies to the right of $\overrightarrow{c_ix}$, $xy$ must intersect $c_{l-1}a'_lc_l$, which is
not possible due to Property 6 of $\mathcal{S}$.
Hence, $c_{i-1}a'_jc_i$ satisfies Property 7 (b).
$\\ \\$
Consider the second subcase where it may appear that some other triangle $c_{h-1}a'_hc_h$, that satisfies 
the precondition of Property 7, violates
Property 7(b) because $c_{i-1}a'_jc_i$ intersects $c_hy$, where $y$ is a free point (see Figure \ref{ttos9}(c)).
As shown earlier, $y$ lies to the right of $\overrightarrow{c_ia'_j}$. If $c_{i-1}a'_jc_i$ intersects 
$c_hy$, then $c_i$ must lie to the right of $\overrightarrow{c_hy}$. 
%
%
Assume $a'_l$ lies to the left of $\overrightarrow{c_ha'_h}$.
Since
no triangle intersects $a'_ix$, 
$a'_i$ or $x$ lies to the left of $\overrightarrow{c_ha'_h}$.
If $a'_i$ lies to the left of $\overrightarrow{c_ha'_h}$, then $c_{i-1}a'_ic_i$ intersects $c_hy$, contradicting
Property 7 (b) of $c_{h-1}a'_hc_h$ in $\mathcal{S}$.
If $x$ is to the left of $\overrightarrow{c_ha'_h}$, then it contradicts Property 7(a) of $c_{h-1}a'_hc_h$.
So, $c_{h-1}a'_hc_h$ satisfies Property 7(b).
%
%
%
%
%
%
%
%
%
%
%
$\\ \\$
Suppose that $c_{i-1}a'_jc_i$  satisfies the precondition of Property 8.
After the triangle replacement, it may appear that 
$c_{i-1}a'_jc_i$ in $(\mathcal{S} \setminus \{ c_{i-1}a'_ic_i \} ) \cup \{ c_{i-1}a'_jc_i \}$ violates
  Property 8(a) by intersecting $a'_{i+1}y$ (see Figure \ref{ttos10}(a)). However, this is not possible since both  
  $c_ia'_{i+1}$ and $y$ lie to the right of $\overrightarrow{c_ia'_j}$ by construction.
  Therefore, $c_{i-1}a'_{j}c_{i}$ satisfies Property 8(a).
   Let $c_{i-2}a'_{i-1}c_{i-1}$ be the clockwise next triangle of  $c_{i-1}a'_jc_i$
   on $L((\mathcal{S} \setminus \{ c_{i-1}a'_ic_i \} ) \cup \{ c_{i-1}a'_jc_i \})$ (see Figure \ref{ttos10}(b)).
   The triangle $c_{i-2}a'_{i-1}c_{i-1}$ does not intersect $a'_jy$  due to Property 8(a) of
   $c_{i-2}a'_{i-1}c_{i-1}$ in $\mathcal{S}$.
   Therefore, $c_{i-2}a'_{i-1}c_{i-1}$ also satisfies Property 8(a).
$ \\ \\$
After the triangle replacement, it may appear that 
$c_{i-1}a'_jc_i$ in $(\mathcal{S} \setminus \{ c_{i-1}a'_ic_i \} ) \cup \{ c_{i-1}a'_jc_i \}$ violates
  Property 8(b) (see Figure \ref{ttos11}(a)).
  Let $y$ be a free point.
  Consider the first sub-case where an inward triangle 
$c_{l-1}a'_lc_l$ has intersected
$a'_{i+1}y$.
The points $c_l$ and $a'_l$ lie on the right and left of $\overrightarrow{a'_{i+1}y}$ respectively, due to 
Property 8(b) of $c_{i-1}a'_ic_i$ in $\mathcal{S}$. 
If $a'_{i+1}a'_l$ intersects $c_{i-1}a'_jc_i$, then $a'_l$ must lie to the left of $\overrightarrow{c_ia'_j}$.
As $c_{l-1}a'_lc_l$ does not intersect $c_{i-1}a'_jc_i$, $a'_j$ and $y$ must lie to the left and right of
$\overrightarrow{c_la'_l}$ respectively. Thus, $x$ must also lie to the right of 
$\overrightarrow{c_la'_l}$, and $c_{l-1}a'_lc_l$ must intersect $c_{i-1}xc_i$, which is not possible due to construction.
%
%
%
%
%
%
%
%
Therefore, $c_{i-1}a'_{j}c_{i}$ satisfies Property 8(b).
%
$\\ \\$
Suppose that $c_{i-2}a'_{i-1}c_{i-1}$  satisfies the precondition of Property 8.
Consider the second sub-case where it may appear that $c_{i-2}a'_{i-1}c_{i-1}$ in 
$(\mathcal{S} \setminus \{ c_{i-1}a'_ic_i \} ) \cup \{ c_{i-1}a'_jc_i \}$
violates Property 8(b) (see Figure \ref{ttos11}(b)).
Wlog, let $c_{j-2}a'_{j-1}c_{j-1}$ be the first clockwise inward triangle on 
$L((\mathcal{S} \setminus \{ c_{i-1}a'_ic_i \} ) \cup \{ c_{i-1}a'_jc_i \})$ 
that has an inward vertex different from $a'_j$. Note that 
$c_{j-2}a'_{j-1}c_{j-1}$ and $c_{i-2}a'_{i-1}c_{i-1}$ may be the same inward 
triangle in some situations.
If any inward triangle $c_{l-1}a'_lc_l$ of $\mathcal{S}$ intersects
$a'_jy$, then 
by Property 8(b) of $c_{j-2}a'_{j-1}c_{j-1}$ in $\mathcal{S}$,
$c_l$ and $a'_l$ lie to the right and left of $\overrightarrow{a'_jy}$ respectively.
If $a'_ja'_l$ intersects $c_{i-2}a'_{i-1}c_{i-1}$, then $a'_{j-1}$ and $c_{j-1}$ lie to the left 
and right of $\overrightarrow{a'_ja'_l}$ respectively, contradicting Lemma \ref{left}.
Therefore, $c_{i-2}a'_{i-1}c_{i-1}$ satisfies Property 8(b).
$\\ \\$
Consider the third sub-case where it may appear that an inward triangle $c_{l-1}a'_lc_l$
satisfying the precondition of Property 8 in 
$(\mathcal{S} \setminus \{ c_{i-1}a'_ic_i \} ) \cup \{ c_{i-1}a'_jc_i \}$
violates Property 8(b) (see Figure \ref{ttos11}(c)). Since $y$ lies to the 
right of $\overrightarrow{c_ix}$ and $x$ lies to the right of $\overrightarrow{c_ia'_j}$, 
$y$ lies to the right of $\overrightarrow{c_ia'_j}$.
As $c_{i-1}a'_jc_i$ intersects $a_{l+1}y$,
 $c_i$ and $a'_j$ must lie to the right and left of $\overrightarrow{a_{l+1}y}$ respectively. 
Moreover, as $xy$ does not intersect any triangle of $\mathcal{S}$ due to Property 6, $a'_{l+1}y$
must intersect $c_{j-1}a'_jc_j$. Due to Property 8(b) of $c_{l-1}a'_lc_l$ in  $\mathcal{S}$,
$a'_{l+1}y$ cannot intersect $c_{l-1}a'_lc_l$.
Therefore, $c_{l-1}a'_lc_l$ satisfies Property 8(b).
%
%
$ \\ \\$
After the triangle replacement, it may appear that $c_{i-1}a'_jc_i$ or $c_{i-2}a'_{i-1}c_{i-1}$ violate Property 8(c).
By the definition of Case 2, $c_ia'_ja'_{i+1}$ is empty and therefore Property 8(c) is not violated (see Figure \ref{ttos12}(a)).
On the other hand,
$c_{i-2}a'_{i-1}c_{i-1}$ also cannot contain any 
point for the same reason as in Lemma \ref{mainlem1} (see Figure \ref{ttos12}(b)).
%
So, $c_{i-2}a'_ic_i$ also satisfies Property 8(c).
 $\hfill{\Box}$
$ \\ \\$
Now we consider Case 3
where $c_ixa'_{i+1}$ is not empty.
The inward triangle $c_{i-1}a'_ic_i$ in $\mathcal{S}$ is replaced by $c_{i-1}gc_i$
to obtain 
$(\mathcal{S} \setminus \{ c_{i-1}a'_ic_i \} ) \cup \{ c_{i-1}gc_i \}$, where (i) $g$ lies inside 
$c_ixa'_{i+1}$, 
(ii) $c_iga'_{i+1}$ is empty, and (iii)
if a point $h$ of $P'$ satisfies properties (i) and (ii),
then $h$ lies to the right of $\overrightarrow{c_ig}$ (see Figure \ref{ttos13}(a)).
Note that $g$ can be either a free point or an inward vertex.
We have the following lemma.
\begin{figure}[h]  
\begin{center} 
\centerline{\hbox{\psfig{figure=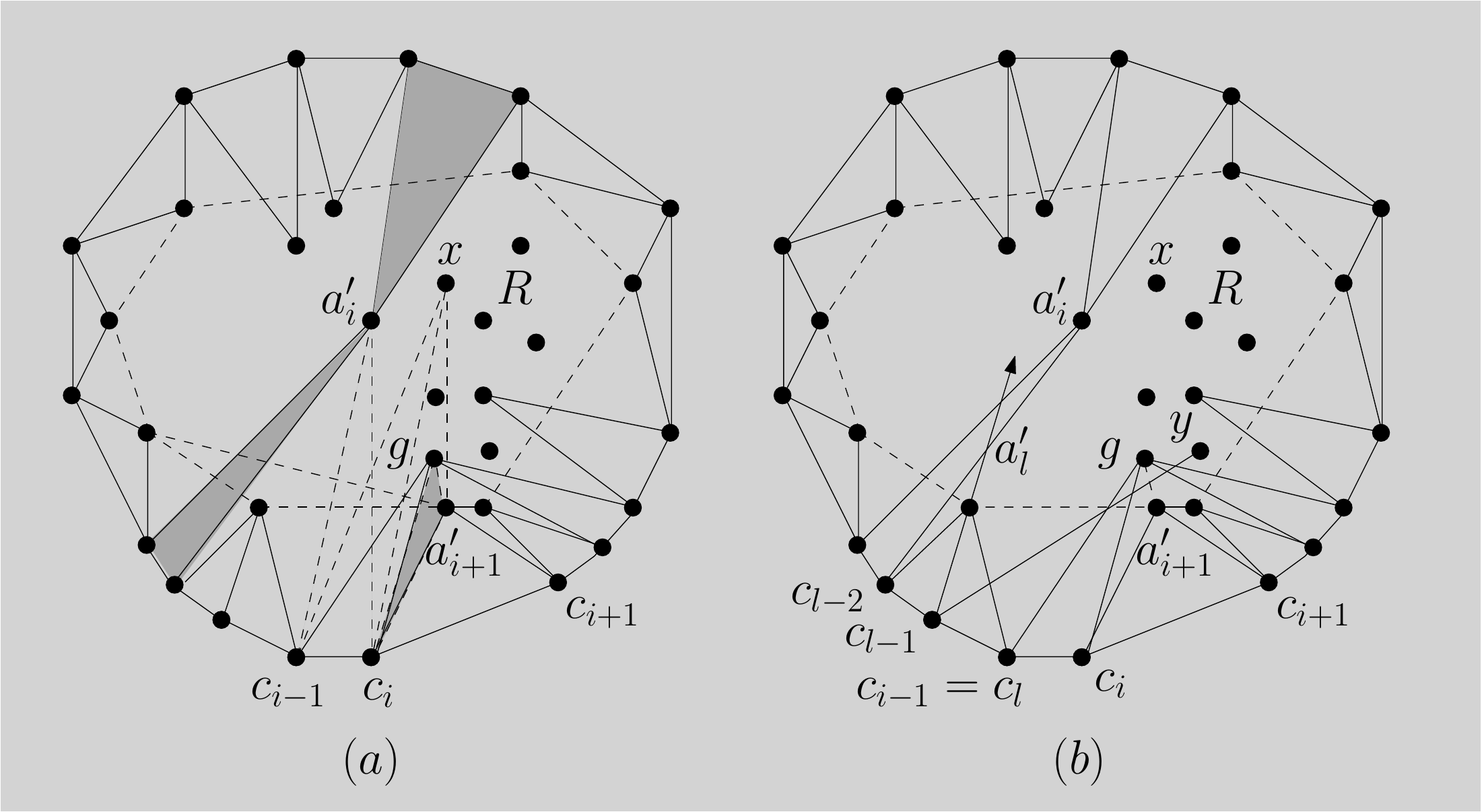,width=0.70\hsize}}}
\caption{ 
(a) The inward triangle $c_{i-1}a'_ic_i$ has been replaced by $c_{i-1}gc_i$ in $\mathcal{S}$.
(b) The inward triangle  $c_{i-1}gc_i$
in $(\mathcal{S} \setminus \{c_{i-1}a'_ic_i\} ) \cup \{c_{i-1}gc_i\}$ satisfies Property 7(b).
}
\label{ttos13}
\end{center}
\end{figure}

\clearpage

\begin{figure}[h]  
\begin{center} 
\centerline{\hbox{\psfig{figure=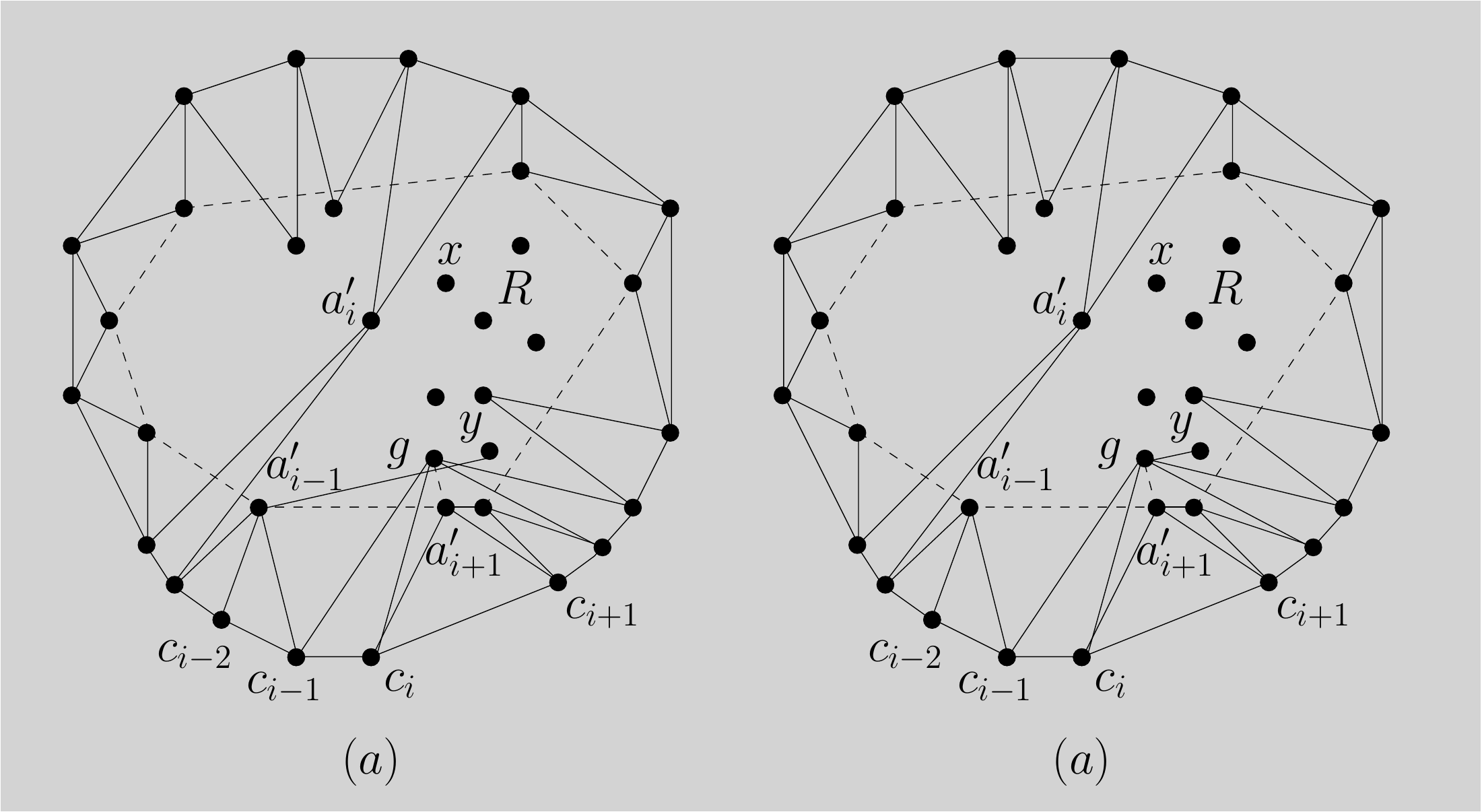,width=0.70\hsize}}}
\caption{ 
(a)  The inward triangle  $c_{i-2}a'_{i-1}c_{i-1}$
in $(\mathcal{S} \setminus \{c_{i-1}a'_ic_i\} ) \cup \{c_{i-1}gc_i\}$ satisfies Property 8(a).
(b) The inward triangle  $c_{i-1}gc_{i}$
in $(\mathcal{S} \setminus \{c_{i-1}a'_ic_i\} ) \cup \{c_{i-1}gc_i\}$ satisfies Property 8(a).
}
\label{ttos14}
\end{center}
\end{figure}
\begin{figure}[h]  
\begin{center} 
\centerline{\hbox{\psfig{figure=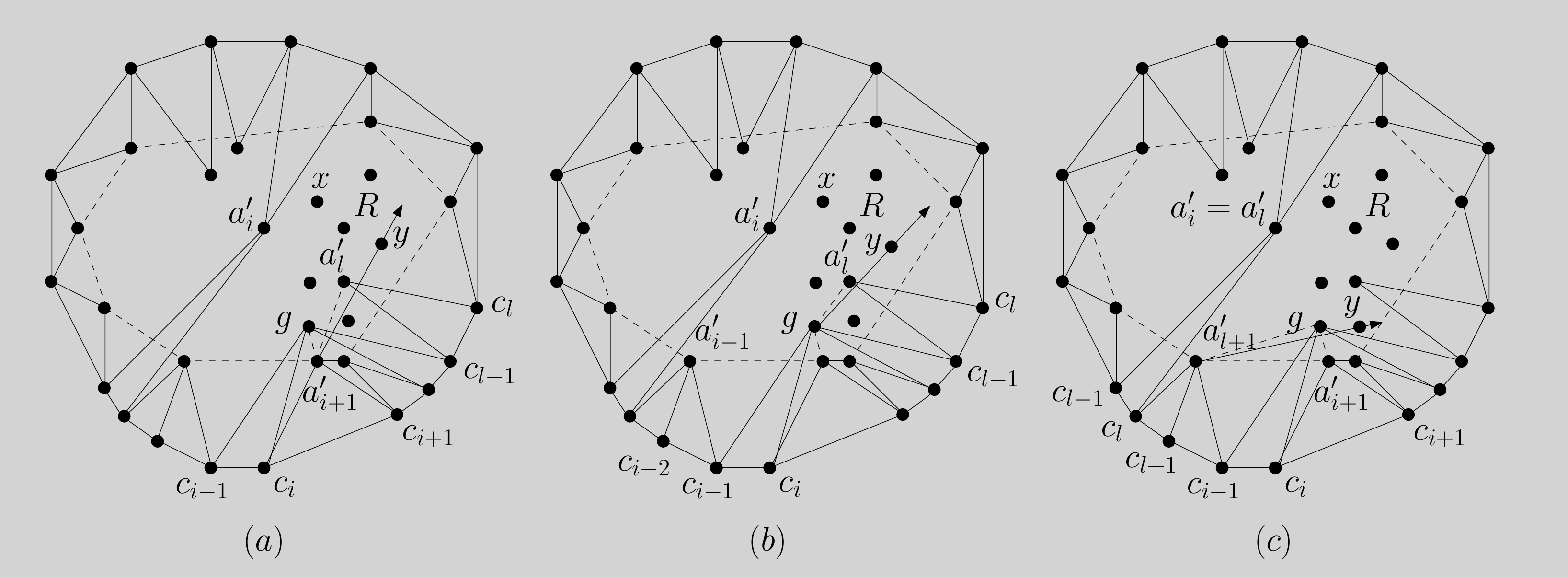,width=1.00\hsize}}}
\caption{ 
(a)  The inward triangle  $c_{i-1}gc_i$
in $(\mathcal{S} \setminus \{c_{i-1}a'_ic_i\} ) \cup \{c_{i-1}gc_i\}$ satisfies Property 8(b).
(b)  The inward triangle  $c_{i-2}a'_{i-1}c_{i-1}$
in $(\mathcal{S} \setminus \{c_{i-1}a'_{i}c_{i}\} ) \cup \{c_{i-1}gc_i\}$ satisfies Property 8(b).
(c)  The inward triangle  $c_{l-1}a'_lc_l$
in $(\mathcal{S} \setminus \{c_{i-1}a'_ic_i\} ) \cup \{c_{i-1}gc_i\}$ satisfies Property 8(b).
}
\label{ttos15}
\end{center}
\end{figure}
\begin{figure}[h]  
\begin{center} 
\centerline{\hbox{\psfig{figure=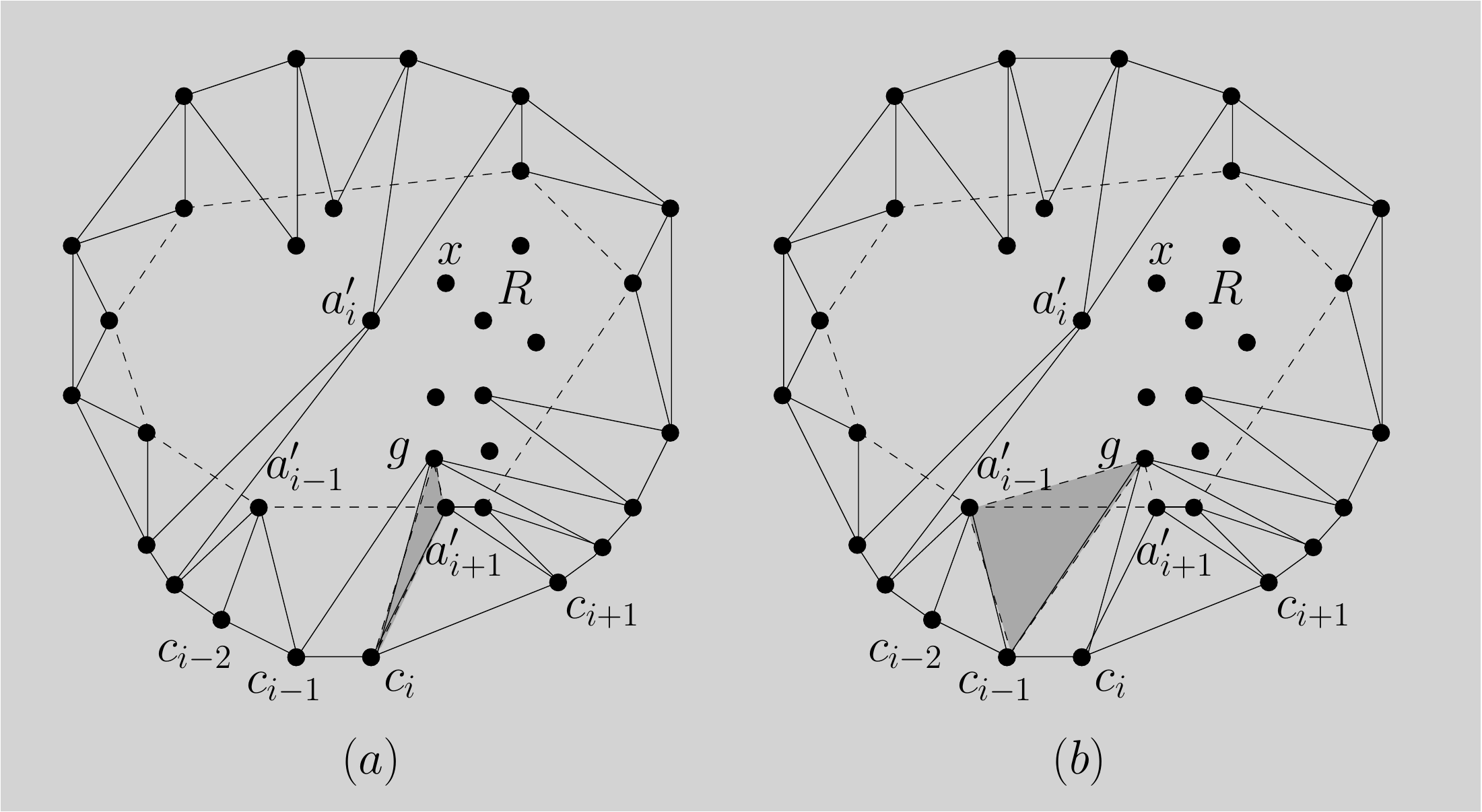,width=0.70\hsize}}}
\caption{ 
(a) The triangle $c_iga'_{i+1}$ is empty.
 (b) The triangle $a'_ic_{i-1}g$ is empty.
}
\label{ttos16}
\end{center}
\end{figure}
\begin{lemma} \label{mainlem3}
 The set $(\mathcal{S} \setminus \{ c_{i-1}a'_ic_i \} ) \cup \{ c_{i-1}gc_i \}$ is a good set
 after the shift operation in Case 3.
\end{lemma}
\textbf{Proof:}
It can be seen that the set $(\mathcal{S} \setminus \{ c_{i-1}a'_ic_i \} ) \cup \{ c_{i-1}gc_i \}$ 
satisfies Properties 1, 4 and 5. 
%
%
%
$\\ \\$
If some triangle $c_{l-1}a'_lc_l$ in $\mathcal{S} \setminus \{ c_{i-1}a'_ic_i \}$ intersects 
$c_{i-1}gc_i$, then $a'_l$ must lie to the left of $\overrightarrow{a'_{i+1}g}$. 
So, $a'_l$ must lie to the left of $\overrightarrow{c_ig}$. 
Since no triangle intersects 
$c_{i-1}xc_i$, $a'_l$ must lie to the right of $\overrightarrow{c_ix}$. Therefore, there exists a point $g'$ to the 
left of  $\overrightarrow{c_ig}$ such that $g'$ lies inside $c_ixa'_{i+1}$, and 
$c_ig'a'_{i+1}$ is empty. This contradicts our choice of $g$. Hence no inward triangle in
  $\mathcal{S} \setminus \{ c_{i-1}a'_ic_i \}$  intersects 
$c_{i-1}gc_i$, 
satisfying Property 3.
$\\ \\$
Since $g$ is an interior point of $CH(P')$, the choice of $g$ ensures that 
$c_{i-1}gc_i$ is not forbidden, and therefore $(\mathcal{S} \setminus \{ c_{i-1}a'_ic_i \} ) \cup \{ c_{i-1}gc_i \}$
satisfies Property 2.
Since $c_iga'_{i+1}$ is empty, all free points in 
 $(\mathcal{S} \setminus \{ c_{i-1}a'_ic_i \} ) \cup \{ c_{i-1}gc_i \}$ lie to the right of $\overrightarrow{c_ix}$,
and no free point can lie  to the left of $\overrightarrow{a'_{i+1}g}$
 and right of $\overrightarrow{c_ix}$ respectively.
So, no line segment joining two free points intersects  $c_{i-1}gc_i$.
%
 Hence 
 $(\mathcal{S} \setminus \{ c_{i-1}a'_ic_i \} ) \cup \{ c_{i-1}gc_i \}$
satisfies Property 6.
$ \\ \\$
Observe that $c_{i-1}gc_i$  does not satisfy the precondition of Property 7 by construction.
After the triangle replacement, it may appear that the triangle $c_{i-1}gc_i$ violates Property 7(b)
(see Figure \ref{ttos13}(b)).
Let $c_{l-2}a'_lc_{l-1}$ and $c_{i-1}a'_lc_l$ be two triangles in $\mathcal{S} \setminus \{ c_{i-1}a'_ic_i \}$
with the same inward vertex $a'_l$. Consider any free point $y$. Assume that $c_{i-1}gc_i$ intersects 
$c_{i-1}y$, and $c_i$ lies to the left of $\overrightarrow{c_{l-1}a'_l}$. Since the proof for Property
7(b) in Lemma \ref{mainlem1} still holds here, $c_{l-1}y$ cannot intersect $c_{i-1}xc_i$. Then, $y$ must lie to 
the right of $\overrightarrow{c_ix}$ and to the left of $\overrightarrow{c_ig}$ respectively,
which is not possible as shown earlier.
So, $c_i$ and $g$ must lie to the right and left of $\overrightarrow{c_{l-1}a'_l}$ respectively.
If $g$ is a free point for $\mathcal{S}$, then by Property 7(a) of $c_{i-1}a'_ic_i$ in $\mathcal{S}$, $g$ must lie 
to the right of $\overrightarrow{c_{l-1}a'_l}$. Otherwise, $g$ must be assigned as an inward vertex to some triangle in
$\mathcal{S}$.
But the quadrilateral $a'_ic_{i-1}c_ia'_{i+1}$ does not contain any points because
of Property 8(c) of $c_{i-1}a'_ic_i$ in $\mathcal{S}$. So, if $c_{i-1}xc_i$ intersects 
$c_ly$, then either $x=a'_l$ or $x$ lies to the right of $\overrightarrow{c_la'_l}$.
If $c_ly$ does not intersect $c_{i-1}xc_i$, since $c_ig$ is to the right of 
$\overrightarrow{c_ix}$, $x$ must be to the right of $\overrightarrow{c_ly}$ and hence also of $c_la'_l$.
Since $g$ is inside $xc_ia'_{i+1}$, this implies $a'_{i+1}$ is to the left of 
$\overrightarrow{c_la'_l}$. If $c_{l-2}a'_lc_{l-1}$ intersects $c_{l-1}y$, it contradicts Property 7 of 
$c_{l-2}a'_lc_{l-1}$. If $c_{l-2}a'_lc_{l-1}$ does not intersect $c_{l-1}y$,
then $y$ must be to the left of $\overrightarrow{c_ia'_{i+1}}$ and to the right of $\overrightarrow{c_ig}$. But this implies $g$ is inside the triangle
$gc_ia'_{i+1}$, contradicting the fact that it is empty (see Figure \ref{ttos13}(b)).
Therefore, $c_{i-1}gc_i$  satisfies Property 7(b).
$\\ \\$
Suppose that $c_{i-2}a'_{i-1}c_{i-1}$ or $c_{i-1}gc_i$ satisfy the precondition of Property 8.
After the triangle replacement, it may appear that the triangles $c_{i-2}a'_{i-1}c_{i-1}$ or $c_{i-1}gc_i$ violate Property 8(a).
Let $y$ be a free point. Consider the first subcase assuming that $gy$ intersects $c_{i-2}a'_{i-1}c_i$ (see Figure \ref{ttos14}(a)). 
If $a'_{i-1} \neq a'_i$, then $a'_{i-1}$ is to the left of $\overrightarrow{c_{i-1}a'_i}$. Since all free points
that are not $x$ are to the right of $\overrightarrow{c_ix}$, and $g$ is also to the right of 
$\overrightarrow{c_ix}$, any line-segment $gy$ must lie completely to the right of 
$\overrightarrow{c_ix}$ and hence cannot intersect $c_{i-2}a'_{i-1}c_{i-1}$.
Consider the second subcase assuming that $a'_{i+1}y$ intersects $c_{i-1}gc_i$
(see Figure \ref{ttos14}(b)).
Due to Lemma \ref{mainlem2}, the line-segment $a'_{i+1}y$ does not intersect the triangle $c_{i-1}xc_i$ 
for all free points $y \neq x$. The line-segment $a'_{i+1}x$ does not intersect $c_{i-1}gc_i$.
If a line-segment $a'_{i+1}y$ intersects $c_{i-1}gc_i$, then $y$ must be contained inside 
$c_{i-1}gc_i$, which is not possible. So, $(\mathcal{S} \setminus \{ c_{i-1}a'_ic_i \} ) \cup \{ c_{i-1}gc_i \}$
satisfies Property 8(a).
$\\ \\$
After the triangle replacement, it may appear that the triangle $c_{i-1}gc_i$ violates Property 8(b).
Let $y$ be a free point.
Consider the first subcase where a triangle $c_{l-1}a'_lc_l$ intersects $a'_{i+1}y$.
In that case, $a'_{i+1}y$ must intersect $c_{i-1}xc_i$ or $a'_l$ is contained in $c_{i-1}gc_i$
 (see Figure \ref{ttos15}(a)).
In the former case, there can be no point lying to the left of 
$\overrightarrow{a'_{i+1}g}$ and to the right of $\overrightarrow{c_ix}$.
The latter case is not possible because $c_{i-1}gc_i$ is empty.
Therefore, $c_{i-1}gc_i$ satisfies Property 8(b).
$\\ \\$
Consider the second subcase where an inward triangle $c_{l-1}a'_lc_l$ intersects $gy$  (see Figure \ref{ttos15}(b)).
If $g$ is a free point, then no inward triangle intersects $gy$ due to Property 6 of $\mathcal{S}$.
Consider the other situation where $g$ is the inward vertex of some triangle $c_{h-1}gc_h$. Lemma \ref{left} implies 
that if $c_{l-1}a'_lc_l$ intersects $gy$, then $c_l$ and $a'_l$ lie to the right and left of 
$\overrightarrow{gy}$ respectively. On the other hand, if $ga'_l$ intersects $c_{i-2}a'_{i-1}c_{i-1}$, then
$c_{i-1}$ is to the left of $\overrightarrow{ga'_l}$ and $a'_{i-1}$ is to the right of 
$\overrightarrow{ga'_l}$, contradicting Lemma \ref{left}.
Therefore, $c_{i-2}a'_{i-1}c_{i-1}$ satisfies Property 8(b).
$\\ \\$
Consider the third subcase where  
 $c_{l-1}a'_lc_l$ 
 and $c_la'_{l+1}c_{l+1}$
 are inward triangles in $\mathcal{S} \setminus \{ c_{i-1}a'_ic_i \}$
such that $a'_{l} \neq a'_{l+1}$, $y$ is a free point, and $c_{i-1}gc_i$ intersects
$a'_{l}y$.
Assume that $c_{l+1}y$ intersects $c_{i-1}gc_i$ where $c_i$ and $g$ lie to the left and right of $\overrightarrow{c_{l+1}y}$ respectively.
By Property 8(b) of  $c_{l-1}a'_lc_l$  in $\mathcal{S}$, $c_{l+1}y$ cannot intersect $c_{i-1}xc_i$.
Since $c_ia'_{i+1}$ lies to the right of $\overrightarrow{c_ig}$, $a'_{i+1}$ must be to the left of $\overrightarrow{c'_{l+1}y}$.
Since $g$ is contained inside the triangle $xc_ia'_{i+1}$, $x$ must be to the right of $\overrightarrow{a'_{l+1}y}$ and $y$ must be to the 
right of $\overrightarrow{c_ix}$ but to the left of $\overrightarrow{c_ig}$. However, this implies that either
$\overrightarrow{yc_ia'_{i+1}}$ is empty, or there is a point $p$
inside $yc_ia'_{i+1}$, where $pc_ia'_{i+1}$ is empty and
$p$ lies to the left of $\overrightarrow{c_ig}$. However, both 
situations
contradict the choice of $g$.
$\\ \\$ 
Consider the other situation where  $c_{l+1}y$ intersects $c_{i-1}gc_i$ such that $c_i$ and $g$ lie to the right and 
left of $\overrightarrow{a'_{l+1}y}$ respectively. Suppose $a'_{l+1}g$ intersects 
$c_{l-1}a'_lc_l$ (see Figure \ref{ttos15}(c)). 
Then $g$ must lie to the left of $\overrightarrow{a'_{l+1}a'_l}$ and $a'_l$ must lie
to the left of $\overrightarrow{a'_{l+1}y}$. 
Since $c_{i-1}gc_i$ does not intersect $c_{l-1}a'_lc_l$,
$a'_l$ must lie
to the left of $\overrightarrow{c_ig}$.
If 
 $c_{i}a'_{i+1}c_{i+1}$  intersects $a'_{l+1}y$, then by Property 8(a) of $c_{l-1}a'_lc_l$, $a'_{l+1}a'_{i+1}$
 does not intersect $c_{l-1}a'_lc_l$, and hence $a'_{i+1}$ is to the left of $\overrightarrow{a'_{l+1}y}$, but to the right of 
 $\overrightarrow{a'_{l+1}a'_l}$. Since $g$ is contained in $xc_ia'_{i+1}$, $x$ must lie to the left of 
 $\overrightarrow{c_lc_{l+1}}$. Therefore $a'_{l+1}y$ intersects 
 $c_{i-1}xc_i$ and $a'_{l+1}x$ intersects $c_{l-1}a'_lc_l$, contradicting Property 8(a) of   $c_{l-1}a'_lc_l$
 in $\mathcal{S}$.
 If $c_{i-1}xc_i$ intersects $a'_{l+1}y$, then by Property 8(b) of $c_{l-1}a'_lc_l$
 in  
 $(\mathcal{S} \setminus \{ c_{i-1}a'_ic_i \} ) \cup \{ c_{i-1}xc_i \}$,
 $x$ lies to the left of $\overrightarrow{a'_{l+1}y}$, and $a'_{l+1}x$ does not intersect  $c_{l-1}a'_lc_l$.
 This implies $x$ lies to the right of $\overrightarrow{a'_{l+1}a'_l}$. Since $g$ is contained inside  
 $xc_ia'_{i+1}$, $a'_{i+1}$ lies to the left of $\overrightarrow{a'_{l+1}a'_l}$, $c_ia'_{i+1}c_{i+1}$ 
 intersects $a'_{l+1}y$, and $a'_{l+1}a'_{i+1}$ intersects   $c_{l-1}a'_lc_l$, contradicting Property 
 8(b) of $c_{l-1}a'_lc_l$.
If $a'_{l+1}y$ does not intersect either $c_{i-1}xc_i$ or $c_ia'_{i+1}c_{i+1}$, then $y$ must be inside 
 $c_{i-1}gc_i$. However, this implies that there is a point
 $p$ to the left of $\overrightarrow{c_ig}$ and to the right of $\overrightarrow{c_ix}$
 such that $pc_ia'_{i+1}$ is empty, contradicting the choice of $g$.
 Therefore, $c_{l-1}a'_lc_l$ satisfies Property 8(b) (see Figure \ref{ttos15}(c)).
$\\ \\$
After the triangle replacement, it may appear that $c_{i-2}a'_{i-1}c_{i-1}$ or $c_{i-1}gc_i$ 
in $(\mathcal{S} \setminus \{ c_{i-1}a'_ic_i \} ) \cup \{ c_{i-1}gc_i \}$
violate Property 8(c).
But $c_iga'_{i+1}$ is empty by construction (see Figure \ref{ttos16}(a)). 
Assume that a point $p$ lies inside $a'_{i-1}c_{i-1}g$.
Suppose $a'_{i-1} = a'_i$. 
The quadrilateral $a'_ic_{i-1}c_ia'_{i+1}$ does not contain any point  
due to Property 8(c) of $c_{i-1}a'_ic_i$ in $\mathcal{S}$. So, $p$ lies to the left of $\overrightarrow{a'_ia'_{i+1}}$,
 to the left of $\overrightarrow{c_ig}$ and to the right of $\overrightarrow{c_ix}$. Moreover, there is such a point $p$ 
 to the left of $c_ig$ such that $pc_ia'_{i+1}$ is empty. This contradicts the choice of $g$.
 Consider the other situation where  $a'_{i-1} \neq a'_i$. 
By the previous arguments, the quadrilateral $a'_{i}c_{i-1}c_ix$ is empty. Also, $a'_{i-1}c_{i-1}x$ is empty
by Property 8(c) of $c_{i-2}a'_{i-1}c_{i-1}$ in $\mathcal{S}$. But $g$ must be to the right 
of $\overrightarrow{a'_{i-1}a'_i}$, and hence,
the quadrilateral
 $a'_{i-1}c_{i-1}c_ix$ must be empty. 
 Therefore, $(\mathcal{S} \setminus \{ c_{i-1}a'_ic_i \} ) \cup \{ c_{i-1}gc_i \}$ satisfies Property 8(c)
 (see Figure \ref{ttos16}(b)). $\hfill{\Box}$ 
%
%
%
$\\ \\$
Now we consider Case 4 where the triangle $c_ia'_ja'_{i+1}$ is not empty (see Figure \ref{ttos17}(a)). 
The inward triangle $c_{i-1}a'_ic_i$ in $\mathcal{S}$ is replaced by $c_{i-1}gc_i$
to obtain 
$(\mathcal{S} \setminus \{ c_{i-1}a'_ic_i \} ) \cup \{ c_{i-1}gc_i \}$, where (i) $g$ lies inside 
$c_ia'_ja'_{i+1}$, 
(ii) $c_iga'_{i+1}$ is empty, and (iii)
if a point $h$ of $P'$ satisfies properties (i) and (ii),
then $h$ lies to the right of $\overrightarrow{c_ig}$ (see Figure \ref{ttos13}(a)).
Note that $g$ can be either a free point or an inward vertex.
We have the following lemma.
\begin{figure}[h]  
\begin{center} 
\centerline{\hbox{\psfig{figure=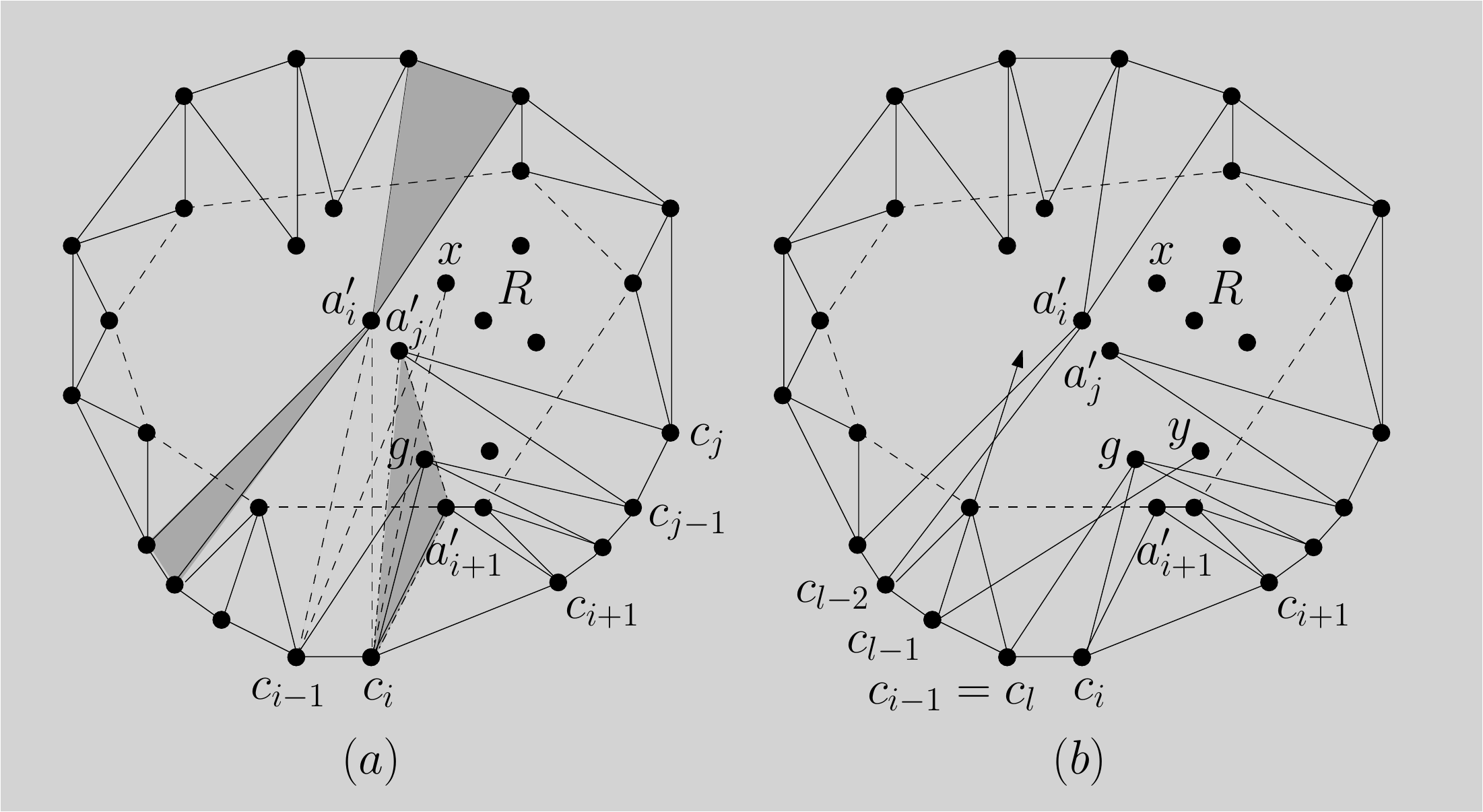,width=0.70\hsize}}}
\caption{ 
(a) The inward triangle $c_{i-1}a'_ic_i$ has been replaced by $c_{i-1}gc_i$ in $\mathcal{S}$.
(b) The inward triangle  $c_{l-2}a'_lc_{l-1}$
in $(\mathcal{S} \setminus \{c_{i-1}a'_ic_i\} ) \cup \{c_{i-1}gc_i\}$ satisfies Property 7(b).
}
\label{ttos17}
\end{center}
\vspace{-0.5cm}
\end{figure}
\begin{figure}[h]  
\begin{center} 
\centerline{\hbox{\psfig{figure=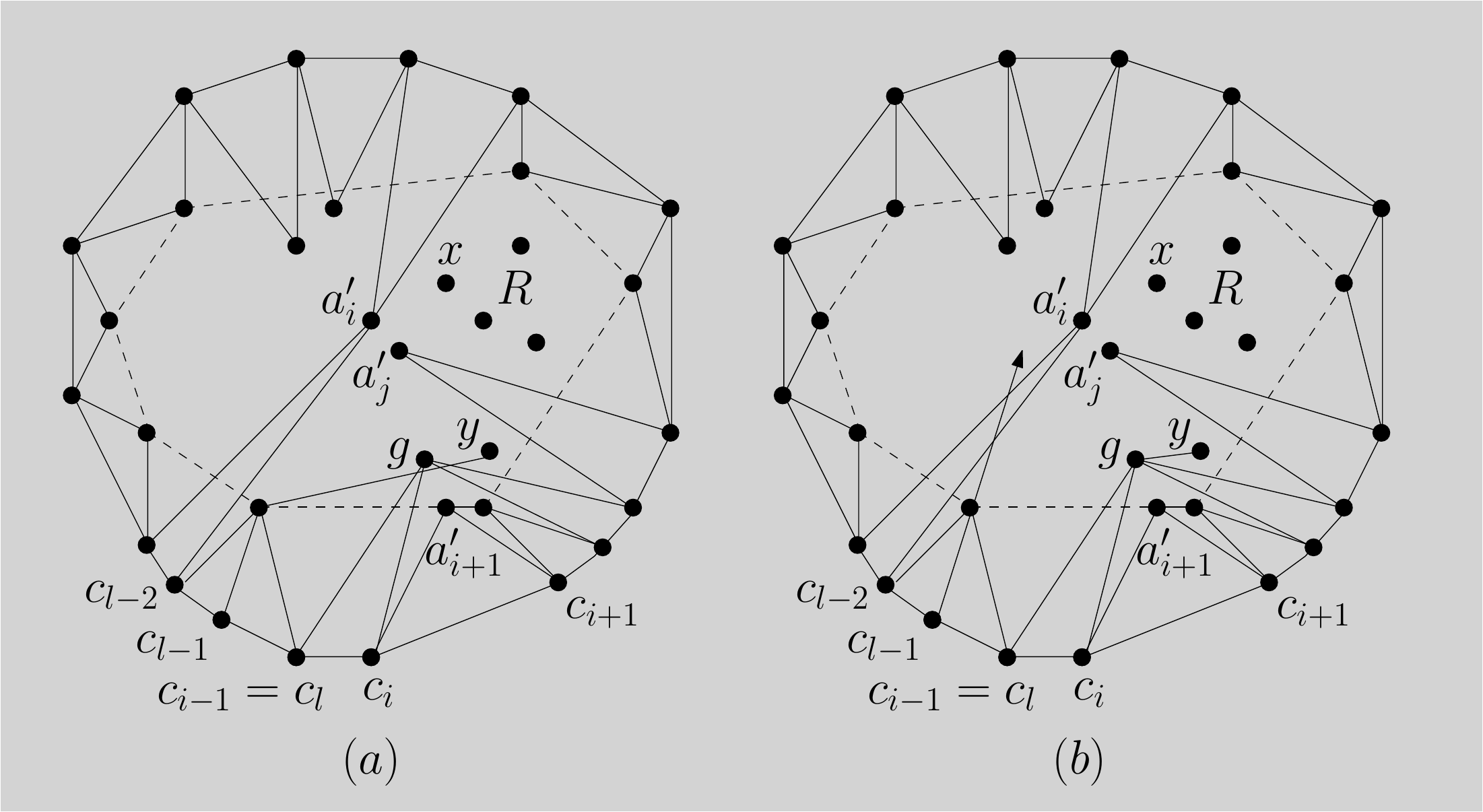,width=0.70\hsize}}}
\caption{ 
(a)  The inward triangle  $c_{i-2}a'_{i-1}c_{i-1}$
in $(\mathcal{S} \setminus \{c_{i-1}a'_ic_i\} ) \cup \{c_{i-1}gc_i\}$ satisfies Property 8(a).
(b) The inward triangle  $c_{i-1}gc_{i}$
in $(\mathcal{S} \setminus \{c_{i-1}a'_ic_i\} ) \cup \{c_{i-1}gc_i\}$ satisfies Property 8(a).
}
\label{ttos18}
\end{center}
\vspace{-0.5cm}
\end{figure}
\begin{figure}[h]  
\begin{center} 
\centerline{\hbox{\psfig{figure=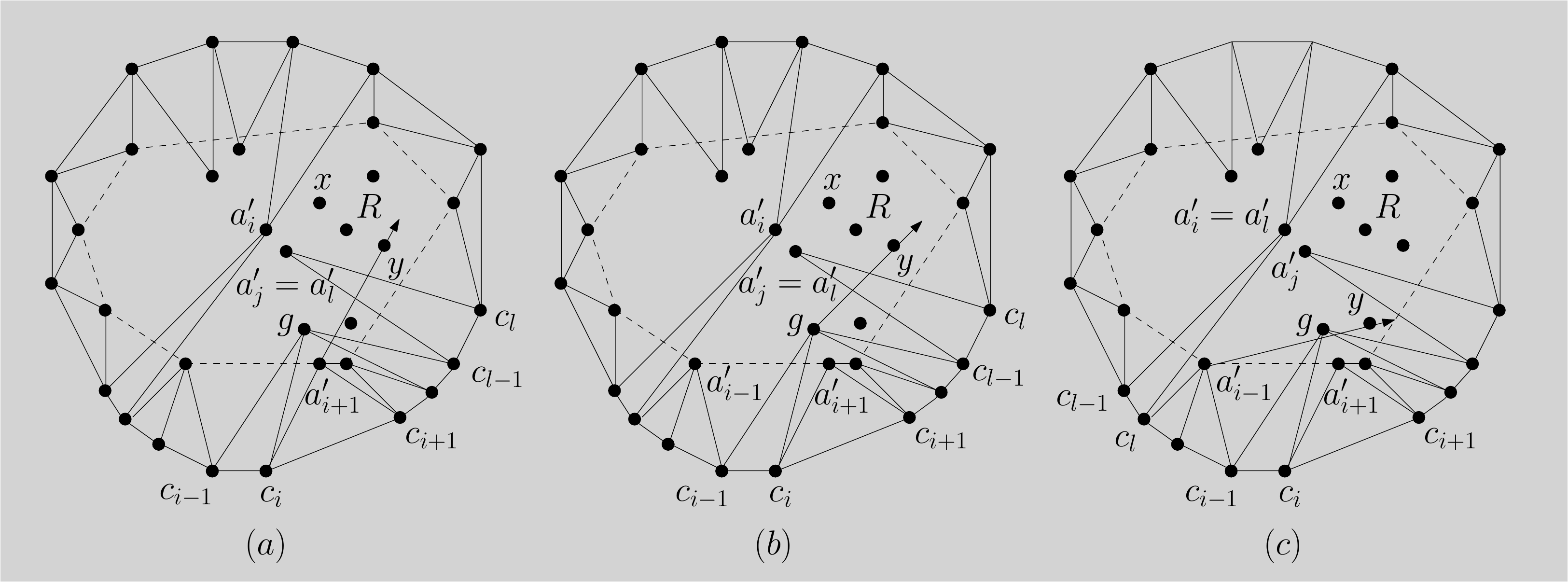,width=1.00\hsize}}}
\caption{ 
(a)  The inward triangle  $c_{i-1}gc_i$
in $(\mathcal{S} \setminus \{c_{i-1}a'_ic_i\} ) \cup \{c_{i-1}gc_i\}$ satisfies Property 8(b).
(b)  The inward triangle  $c_{i-2}a'_{i-1}c_{i-1}$
in $(\mathcal{S} \setminus \{c_{i-1}a'_{i}c_{i}\} ) \cup \{c_{i-1}gc_i\}$ satisfies Property 8(b).
(c)  The inward triangle  $c_{l-1}a'_lc_l$
in $(\mathcal{S} \setminus \{c_{i-1}a'_ic_i\} ) \cup \{c_{i-1}gc_i\}$ satisfies Property 8(b).
}
\label{ttos19}
\end{center}
\vspace{-0.5cm}
\end{figure}
\begin{figure}[h]  
\begin{center} 
\centerline{\hbox{\psfig{figure=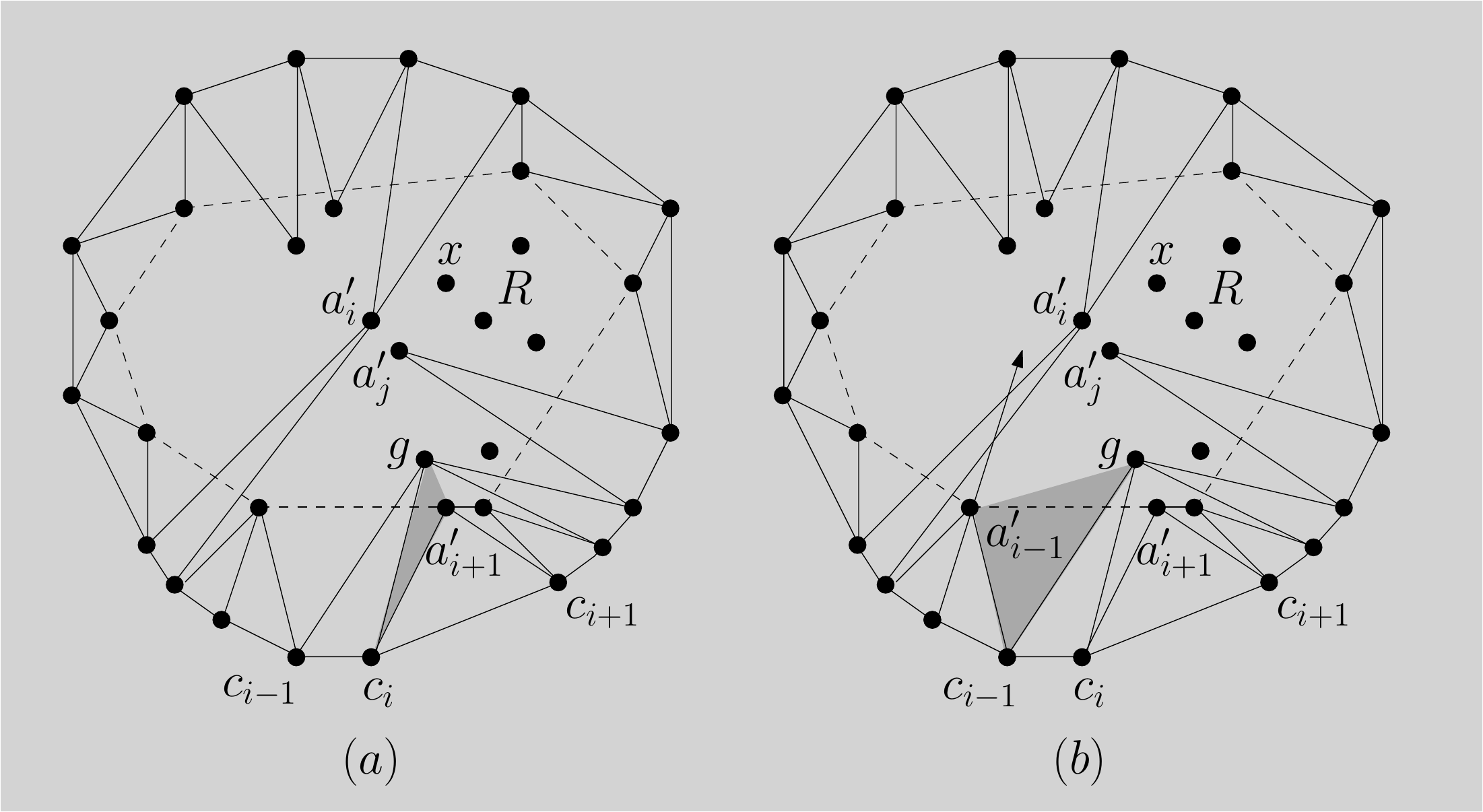,width=0.70\hsize}}}
\caption{ 
(a) The triangle $c_iga'_{i+1}$ is empty.
 (b) The triangle $a'_ic_{i-1}a'_j$ is empty.
}
\label{ttos20}
\end{center}
\vspace{-0.5cm}
\end{figure}
\begin{lemma} \label{mainlem4}
 The set $(\mathcal{S} \setminus \{ c_{i-1}a'_ic_i \} ) \cup \{ c_{i-1}gc_i \}$ is a good set
 after the shift operation in Case 4.
\end{lemma}
\textbf{Proof:}
It can be seen that the set $(\mathcal{S} \setminus \{ c_{i-1}a'_ic_i \} ) \cup \{ c_{i-1}gc_i \}$ satisfies Properties 1, 4 and 5.
$\\ \\$
The inward triangle $c_{i-1}gc_i$ is not forbidden and does not intersect the interior of any other triangle
of $(\mathcal{S} \setminus \{c_{i-1}a'_ic_i\} ) \cup \{c_{i-1}gc_i\}$ as shown in the proof of 
Properties 2 and 3 in Lemma \ref{mainlem3}. Therefore, $(\mathcal{S} \setminus \{c_{i-1}a'_ic_i\} )\cup \{c_{i-1}gc_i\}$ 
satisfies Properties 2 and 3. Moreover, no line segment joining two free points in 
$(\mathcal{S} \setminus \{c_{i-1}a'_ic_i\} )\cup \{c_{i-1}gc_i\}$ intersects 
$c_{i-1}gc_i$ as shown in the proof of Property 6 in Lemma \ref{mainlem3}.
Therefore, $(\mathcal{S} \setminus \{c_{i-1}a'_ic_i\} )\cup \{c_{i-1}gc_i\}$ 
satisfies Property 6 (see Figure \ref{ttos17}(a)).
$\\ \\$
After the triangle replacement, if $c_{i-2}a'_{i-1}c_{i-1}$ satisfies the precondition of Property 7, then
$c_{i-2}a'_{i-1}c_{i-1}$
satisfies Property 7 as shown in the proof of Property 7 in Lemma \ref{mainlem3},
where $c_{i-2}a'_{i-1}c_{i-1}$ is the clockwise next triangle of $c_{i-1}gc_i$
on $L((\mathcal{S} \setminus \{c_{i-1}a'_ic_i\} )\cup \{c_{i-1}gc_i\})$
(see Figure \ref{ttos17}(b)).
The other inward triangles in $(\mathcal{S} \setminus \{c_{i-1}a'_ic_i\} )\cup \{c_{i-1}gc_i\}$, 
that satisfy the precondition of Property 7,
also satisfy Property 7 as shown in the proof of Property 7 in Lemma \ref{mainlem3}.
Therefore, $(\mathcal{S} \setminus \{c_{i-1}a'_ic_i\} )\cup \{c_{i-1}gc_i\}$ 
satisfies Property 7.
$\\ \\$
After the triangle replacement, the inward triangles $c_{i-2}a'_{i-1}c_{i-1}$
or $c_{i-1}gc_i$, that satisfy the precondition of Property 8, also
satisfy Property 8(a) as shown in the proof of Property 8(a) in Lemma \ref{mainlem3}
(see Figures \ref{ttos18}(a) and  \ref{ttos18}(b)).
After the triangle replacement, the inward triangles $c_{i-2}a'_{i-1}c_{i-1}$, $c_{i-1}gc_i$
and any other inward triangle $c_{l-1}a'_lc_l$, that satisfy the precondition of Property 8 in 
$(\mathcal{S} \setminus \{c_{i-1}a'_ic_i\} )\cup \{c_{i-1}gc_i\}$, satisfy 
Property 8(b) as shown in the proof of Property 8(b) in Lemma \ref{mainlem3}
(see Figures \ref{ttos19}(a),  \ref{ttos19}(b) and  \ref{ttos19}(c)).
After the triangle replacement, the inward triangles
 $c_{i-1}gc_i$ and  $c_{i-2}a'_{i-1}c_{i-1}$ satisfy Property 8(c) as shown in the proof of Property 8(c) in Lemma \ref{mainlem3}
(see Figures \ref{ttos20}(a) and  \ref{ttos20}(b)).
Therefore, $(\mathcal{S} \setminus \{c_{i-1}a'_ic_i\} )\cup \{c_{i-1}gc_i\}$ 
satisfies Property 8. $\hfill{\Box}$ 
\newpage
 \begin{algorithm}[H]
  \textbf{transforming\_good\_set(\emph{P})}
  
compute $CH(P)$ and $CH(P')$\;
 let the points of $CH(P)$ be $\{ c_1, c_2, \ldots, c_k \}$ in counterclockwise order\;
$\mathcal{S} :=  \textbf{constructing\_good\_set(\emph{P})}$\; 
\tcp{$\mathcal{S}$ is a good set}
compute $L(\mathcal{S})$\;
compute all zones and regions of all inward vertices in $\mathcal{S}$ and store in a set $Z$\;
\While{$\mathcal{S}$ has multiple copies of inward triangles or has inward triangles having the same 
inward vertex
}
{
locate an inward vertex $a'_i$ assigned to multiple inward triangles of $\mathcal{S}$\;
locate the region $R$ of $a'_i$ containing all free points\;
locate the right edges $c_ja'_i$ and $c_ka'_i$ of zones bounding $R$ with 
$c_k$ lying to the right of $\overrightarrow{c_ja'_i}$\;
\eIf{$R$ is convex}
{
assign $c_{j-1}a'_ic_j$ to $c_{i-1}a'_ic_i$\; 
}
{
assign $c_{k-1}a'_ic_k$ to $c_{i-1}a'_ic_i$\; 
}
\tcp{$c_{i-1}a'_ic_i$ is the triangle to be shifted}
identify the set $F$ of free points\;
scan $F$ and locate a vertex $x$ of $CH(F)$ with $\overrightarrow{c_ix}$ being the left tangent 
to $CH(F)$\;
\eIf{no inward triangle in $\mathcal{S}$ intersects $c_{i-1}a'_ic_i$}
{
\eIf{$c_ixa'_{i+1}$ is empty}
{
$\mathcal{S} = ( \mathcal{S} \setminus \{ c_{i-1}a'_ic_i \} ) \cup \{ c_{i-1}xc_i \}$\;
\tcp{case 1}
}
{
compute the set $Q'$ of every point $q$ of $P'$ lying inside $c_ixa'_{i+1}$
with $c_iqa'_{i+1}$ being empty\;
scan $Q'$ till a vertex $g$ of $CH(Q')$ is located with $\overrightarrow{c_ig}$ being the left tangent 
to $CH(Q')$\;
$\mathcal{S} = ( \mathcal{S} \setminus \{ c_{i-1}a'_ic_i \} ) \cup \{ c_{i-1}gc_i \}$\;
\tcp{case 3}
}
}
{
compute the set $Q$ of inward vertices of all inward triangles intersecting $c_{i-1}a'_ic_i$\;
scan $Q$ till a vertex $a'_j$ of $CH(Q)$ is located with $\overrightarrow{c_ia'_j}$ being the left tangent 
to $CH(Q)$\;
\eIf{$c_ia'_ja'_{i+1}$ is empty}
{
$\mathcal{S} = ( \mathcal{S} \setminus \{ c_{i-1}a'_ic_i \} ) \cup \{ c_{i-1}a'_jc_i \}$\;
\tcp{case 2}
}
{
compute the set $Q'$ of every point $q$ of $P'$ lying inside $c_ia'_ja'_{i+1}$
with $c_iqa'_{i+1}$ being empty\;
scan $Q'$ till a vertex $g$ of $CH(Q')$ is located with $\overrightarrow{c_ig}$ being the left tangent 
to $CH(Q')$\;
$\mathcal{S} = ( \mathcal{S} \setminus \{ c_{i-1}a'_ic_i \} ) \cup \{ c_{i-1}gc_i \}$\;
\tcp{case 4}
}
}
update $L(\mathcal{S})$\;
update $Z$\;
}
\tcp{$\mathcal{S}$ is a good set with all points of $P'$ assigned as inward vertices
of distinct inward triangles}
report $\mathcal{S}$\;
\end{algorithm}
\begin{lemma}\label{complxlemgs}
Given a good set $\mathcal{S}$, the procedure \textbf{transforming\_good\_set(\emph{P})} transforms $\mathcal{S}$ such that 
every inward triangle in $\mathcal{S}$ has a distinct inward vertex and the transformation can be 
carried out in $O(n^3)$ time.
\end{lemma}
\textbf{Proof:}
The correctness of the procedure follows from Lemmas \ref{mainlem1}, \ref{mainlem2}, \ref{mainlem3} and
\ref{mainlem4}. 
Computation of $L(\mathcal{S})$ takes $O(n^2)$ time.
Computation of $Z$ takes $O(n^3)$ time.
Since there are less $n^2$ non-forbidden triangles,
there are at most $n^2$ modifications of $\mathcal{S}$.
Since each modification can take $O(n)$ time, for computing $Q$ or $Q'$, the overall 
time complexity of the procedure is $O(n^3)$.
\section{A 4-connected triangulation}
\begin{figure}[h]  
\begin{center} 
\centerline{\hbox{\psfig{figure=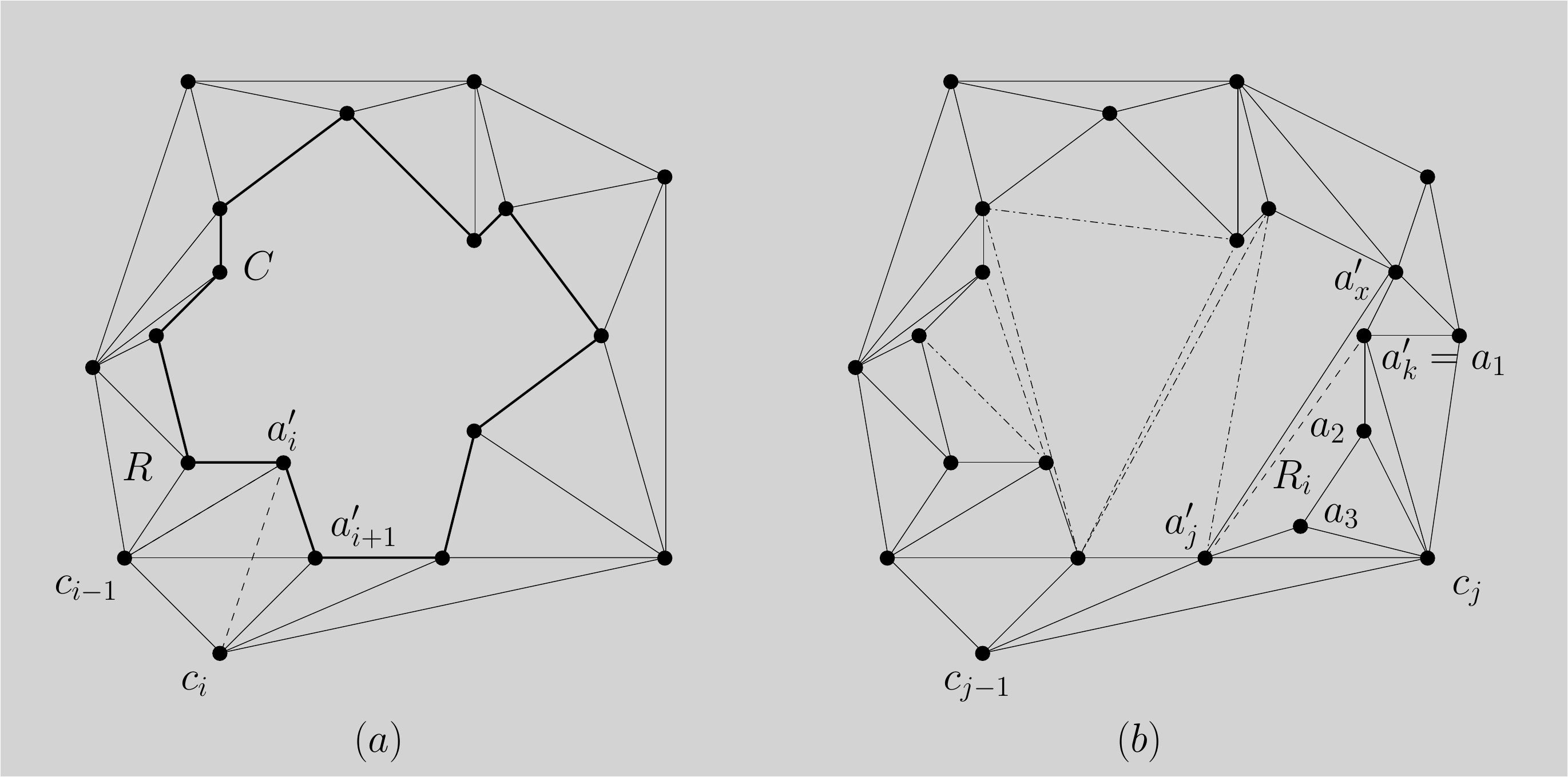,width=0.80\hsize}}}
\caption{ 
(a) The inward vertex $a'_i$ of $c_{i-1}a'_ic_i$ is reflex in $C$.
The inward triangle $c_{i-1}a'_ic_i$ and the degenerate triangle $c_ia'_{i+1}$ are replaced in $\mathcal{S}$
by the degenerate inward triangle $c_{i-1}a'_i$ and the inward triangle $c_{i-1}a'_{i+1}c_i$, 
such that the degree of $c_i$ becomes four.
(b) The region $R_i = a'_xa'_ka_2a_3a'_k$ is a bad region and $a'_ja'_k$ is a 
diagonal of $R_i$.
}
\label{lastlem1}
\end{center}
\vspace{-0.5cm}
\end{figure}

In this section, we show that $\mathcal{S}$ constructed in the previous section can be transformed into a 4-connected
triangulation of $P$.
An inner cycle $C$ is constructed
by connecting $a'_{i}$ and $a'_{i+1}$ for all $i$, where $a'_{i}$ and $a'_{i+1}$ are the inward 
vertices of two consecutive inward triangles $c_{i-1}a'_ic_i$ and $c_ia'_{i+1}c_{i+1}$ in $L(\mathcal{S})$
(see Figure \ref{lastlem1}(a)).
Let $R$ be the annular region enclosed by $C$ and $CH(P)$.
Observe that by Property 8(c) of good sets, $C$ is non self-intersecting,
and no inward triangle of $\mathcal{S}$ intersects $C$. 
Note that $C$ contains all points of $P'$.
Starting from $C$, 
a 4-connected triangulation of $P$ is constructed as follows.
$\\ \\$
Let $T$ be the triangulation of $R$ 
formed by the inward triangles in $\mathcal{S}$.  
Let $D$ be a maximal set of pairwise non-intersecting diagonals of $C$ such  
that there is no complex triangle in $T \cup C \cup D$. 
We show that
there exists a triangulation 
$T$ of $R$ and a collection $D$ of pairwise  
non-intersecting diagonals of $C$ satisfying the following properties: 
\begin{enumerate} 
\item 
$T$ does not contain a chord of $CH(P)$. 
\item 
No two inward triangles in $T$ have the same inward vertex. 
\item 
There is no complex triangle in $T \cup C \cup D$. 
\item 
Let $R_1,R_2, \ldots,R_m$ be the interior regions of $C$ partitioned by $D$ (see Figure \ref{lastlem1}(b)).
If $|R_i| = 3$ for all $i$ then it is a 4-connected triangulation. 
So we assume that $|R_i| \ge 4$.  Then the points on 
the boundary of $R_i$ can be labeled in clockwise or counterclockwise order as  
$a_0, a_1, a_2, \ldots, a_l$ such that (i) $a_1, a_2,\ldots,a_l$ are consecutive  
points of $C$, 
(ii) $a_1, a_2,\ldots,a_l$ are adjacent in $T$ to a point $c_j$ in $CH(P)$,   
(iii) $a_1a_l$ is a diagonal of $R_i$, 
(iv) for all $1 < m < l$,  
$a_m$ is contained in the interior of the triangle $a_1a_lc_j$, and
(v) $a_0a_m$ is not a diagonal of $R_i$. 
\end{enumerate} 
   $\\ $
We call a triple $(T,C,D)$ satisfying the above properties as 
a \emph{consistent} triple. We have the following lemmas.
\begin{lemma} \label{triple}
If $P$ has a good set $\mathcal{S}$ with all inward triangles of $\mathcal{S}$ having distinct inward vertices,
then $P$ has a consistent triple $(T,C,D)$.
%
\end{lemma} 
%
 \textbf{Proof:}
Observe that $(T,C,D)$ satisfies Properties 1 and 2 since $\mathcal{S}$ is a good set.
Also, $(T,C,D)$ satisfies Property 3 by the construction of $D$.
$\\ \\$
Consider any region $R_i$ such that $|R_i| \ge 4$. We call such a region a \emph{bad region} (see Figure \ref{lastlem1}(b)).
 Let $a'_ja'_k$ be a diagonal of $R_i$. By the maximality of $D$, adding  
$a'_ja'_k$ to $D$ must create a complex triangle. 
In that case, there exists a vertex $c_j$ of $CH(P)$ adjacent to  both $a'_j$ and $a'_k$ in $T$.
Since every  
vertex in $C$ is adjacent in $T$ to some vertex of $CH(P)$, the vertices of  
$R_i$ that are contained in the interior of the triangle $a'_jc_ja'_k$ must all be  
adjacent to $c_j$.
Choose a diagonal of $R_i$ (say, $a'_ja'_k$) such that the number of 
vertices in the interior of $a'_jc_ja'_k$ 
 is maximum.  
  $\\ \\$
Consider an arbitrary triangulation $T'_i$ of $R_i$ which includes the diagonal $a'_ja'_k$.
Let $a'_x$ be the vertex of $R_i$ outside the triangle $a'_jc_ja'_k$ such that $a'_ja'_ka'_x$ is a 
triangle in $T'_i$ (see Figure \ref{lastlem1}(b)).  
If $a'_x$ is adjacent to $c_j$ in $T$, then either $a'_k$ is contained in the 
interior of  $a'_jc_ja'_x$ or $a'_j$ is contained in the interior of  
$a'_kc_ja'_x$. 
If the former condition holds,  
then $a'_ja'_x$ is a  
diagonal of $R_i$ for which the number of vertices in the interior of   
$a'_jc_ja'_x$  is greater than the number of vertices in the interior of $a'_jc_ja'_k$.  
This contradicts the choice of the diagonal $a'_ja'_k$. 
Similar arguments hold for the latter case also.
Thus we assume $a'_x$ is  
not adjacent to $c_j$ in $T$. 
  $\\ \\$
  Before we consider the cases depending on edges of $R_i$ connecting $a'_x$, we present a procedure to ensure
that if the inward vertex $a'_i$ of some inward triangle $c_{i-1}a'_ic_{i}$
is reflex in $C$, then $c_i$ has degree four in $T$. 
If
the degree of $c_i$ is
five or more, and there is a degenerate inward triangle
$c_{i}a'_{i+1}$ in $\mathcal{S}$,
replace the
inward triangles $c_{i-1}a'_ic_{i}$ and $c_{i}a'_{i+1}$ in $\mathcal{S}$
by the inward triangles $c_{i-1}a'_i$ and $c_{i-1}a'_{i+1}c_{i}$ respectively,
to get a new triangulation of $R$ (see Figure \ref{lastlem1}(a)).
Note that the new $(T,C,D)$ satisfies the first three properties.
Since 
this operation shifts the inward vertex on some edge $c_{i-1}c_i$ to the next counterclockwise vertex of $C$,
the degree of all such $c_i$ becomes four by repeating this operation at most $n$ times.
%
%
%
  $\\ \\$
Consider the first case where both
 $a'_ja'_x$ and $a'_ka'_x$ are edges of $R_i$. Therefore, $c_j$ is adjacent to all vertices 
of $R_i$ except $a'_x$. Thus, $R_i$ satisfies Property 4(ii). The edges of $R_i$ that are contained inside the 
triangle $a'_jc_ja'_k$ must be edges of $C$, otherwise we have a complex triangle 
in $T \cup C \cup D$. Thus we can label $a'_x$ as $a_0$ and the vertices 
of $C$ from $a'_j$ to $a'_k$ that are contained in the triangle $a'_jc_ja'_k$ as $a'_j = a_1, 
a_2, \ldots,a_l = a'_k$, satisfying Properties 4(i) and 4(iii). The maximality of $D$ implies that there is no diagonal 
of $R_i$ incident with $a'_x$. Thus, $R_i$ satisfies Property 4(iv). 
  $\\ \\$
%
%
%
  Consider the other case where 
$a'_x$ is not adjacent to $c_j$, and at least one of $a'_ja'_x$ and $a'_ka'_x$,
(say, $a'_ja'_x$)
is a diagonal of $R_i$.
There must be a vertex $c_{j-1} \neq c_j$ of $CH(P)$  
such that both $a_j$ and $a_x$ are adjacent to $c_{j-1}$ in $T$. This implies $a'_j$ 
is the inward vertex of $c_{j-1}a'_jc_j$ in $T$. Observe that both 
$c_j$ and $c_{j-1}$ must have degree at least five in $T$ because they must be adjacent 
to some vertex of $R_i$ in the interiors of triangles $a'_jc_ja'_k$ and $a'_jc_{j-1}a'_x$ 
respectively. 
%
%
Since the degrees of $c_{j-1}$ and $c_j$ are more than four, $a'_j$ cannot be reflex in $C$ as shown earlier.
So, $a'_j$ is a convex vertex of $C$ and hence of $R_i$. 
$\\ \\$
We show that another diagonal can be added to $D$, contradicting its maximality.
  Since both $a'_ja'_k$ and $a'_ja'_x$ are diagonals of $R_i$,
  there must be two triangles $a'_ja'_ka'_q$ and $a'_ja'_xa'_r$ in the triangulation of $R_i$,
  where $a'_ja'_ka'_q$, $a'_ja'_xa'_r$ and $a'_ja'_ka'_x$ are all distinct.
So, $a'_q$ is contained in the interior of triangle $a'_jc_ja'_k$ and it must be  
adjacent to only $c_j$ among all vertices of $CH(P)$.
Similarly, $a'_r$ is contained in the interior of triangle $a'_jc_{j-1}a'_x$, and is adjacent  
to only $c_{j-1}$ among all vertices of $CH(P)$.
If 
quadrilaterals
$a'_ja'_qa'_ka'_x$ or $a'_ja'_ka'_xa'_r$ is convex, then either $a'_qa'_x$ 
or $a'_ka'_r$ 
can be added as a diagonal to $D$ without creating a complex triangle,
contradicting the maximality of $D$. 
$\\ \\$
Recall that $a'_j$ is 
a convex vertex of $R_i$, and $a'_ja'_k$ and $a'_ja'_x$ are diagonals of $R_i$, 
So, $a'_k$ must be a reflex vertex of $a'_ja'_qa'_ka'_x$ and $a'_x$ must be a reflex vertex of 
$a'_ja'_ka'_xa'_r$. Thus $a'_k$ is contained in the interior of triangle $a'_ja'_qa'_x$ and $a'_x$ is  
contained in the interior of triangle $a'_ja'_ka'_r$. However, $a'_q$ is contained in  
the interior of triangle $a'_jc_ja'_k$ and $a'_r$ is contained in the interior of 
triangle $a'_jc_{j-1}a'_x$. Therefore, the triangle $c_{j-1}a'_jc_j$ contains the points  
$a'_q,a'_k,a'_x,a'_r$ in its interior, contradicting the 
fact that it is an inward triangle of $T$. 
Therefore, all bad regions $R_i$  must satisfy property 4  
and hence $(T,C,D)$ is a consistent triple. 
  $\hfill{\Box}$
 \begin{figure}[h]  
\begin{center} 
\centerline{\hbox{\psfig{figure=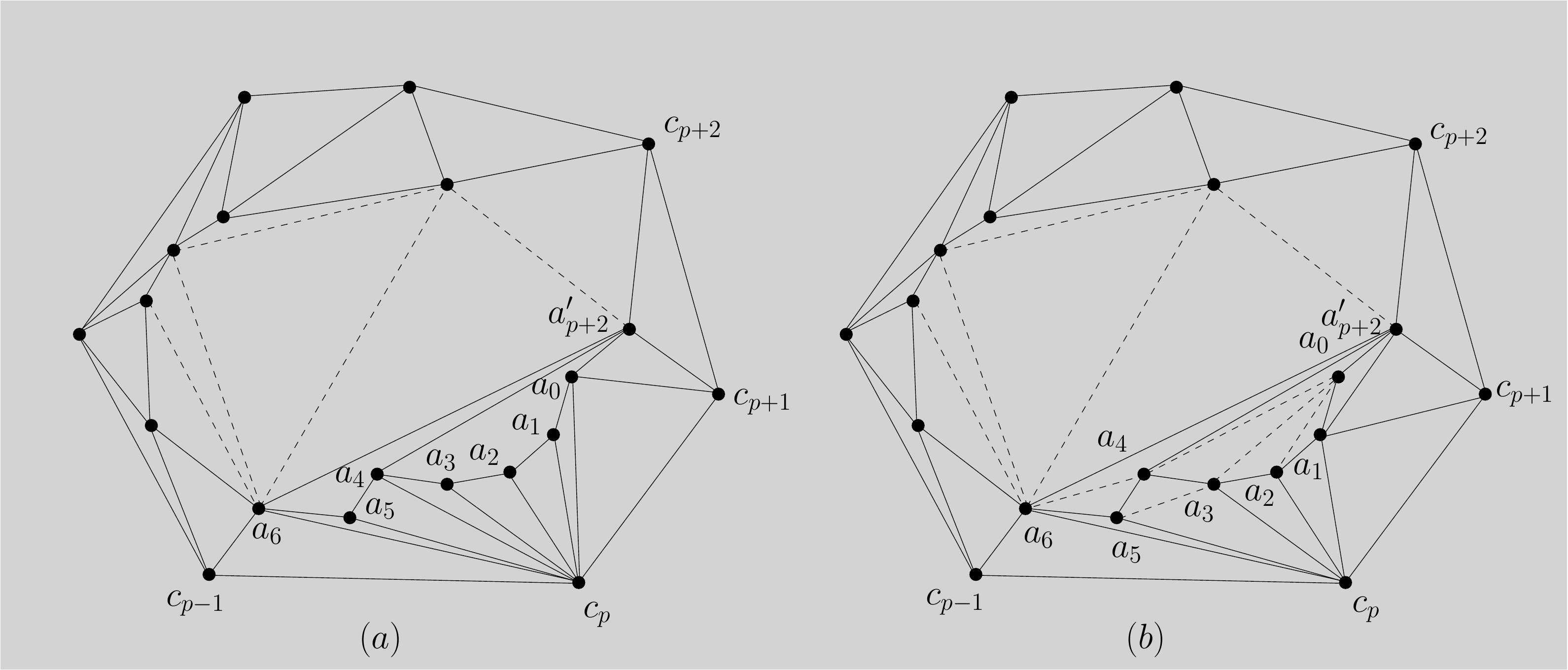,width=0.80\hsize}}}
\caption{ 
(a) The regions $a'_{p+2}a_6a_5a_4a'_{p+2}$ and $a'_{p+2}a_4a_3a_2a_1a_0a'_{p+2}$ 
are bad regions in the neighbourhood of $c_p$.
(b) $C$ and $R$ are re-triangulated by replacing $a_4c_p$, $a_0c_p$ and $a_0c_{p+1}$ with
$a_5a_3$, $a_1c_{p+1}$ and $a_1a'_{p+2}$ respectively.
}
\label{lastlem2}
\end{center}
\vspace{-0.5cm}
\end{figure}

\begin{lemma} \label{triple2}
If $P$ has a consistent triple $(T,C,D)$,
then $P$ admits a 4-connected triangulation. 
\end{lemma} 
%
\textbf{Proof:}
Consider a consistent triple $(T,C,D)$ such that $|D|$ is maximum.
If $D$ triangulates $C$, then the first three properties 
of a consistent triple ensure that 
$T \cup C \cup D$ is a 4-connected triangulation of $P$.  
If $D$ does not triangulate $C$, we show that 
another consistent  
triple  $(T',C, D')$ 
can be located such that $|D'| > |D|$ as follows. 
  $\\ \\$
  Let $R_i$ be a bad region of $C$ due to $(T,C,D)$.
So, by Property 4 of a consistent triple,
there exists a unique vertex $c_p$ in $CH(P)$ that is adjacent to 
all but one vertex of $R_i$ (see Figure \ref{lastlem2}(a)). 
Moreover, the vertices of $R_i$ that are adjacent to 
$c_p$ are consecutive vertices of $C$.
We say that the region $R_i$ is in 
the neighbourhood of $c_p$. 
If
the vertices of another bad region $R_j$ due to $(T,C,D)$ are also neighbours of $c_p$,
then $R_j$ is also in the neighbourhood of $c_p$.
So the neighbourhood of
$c_p$ may contain several such bad regions. 
By Property 4 of $(T,C,D)$,
the vertices of $R_i$ 
can be labelled as $a_0,a_1,a_2,\ldots,a_l$  such that 
$a_1a_l$ is a diagonal of $R_i$,  $a_0a_j$ is not a 
diagonal for all $1 < j < l$, and the vertices $a_2,\ldots,a_{l-1}$ are 
contained in the interior of $a_1c_pa_l$. This implies that 
either $a_1$ or $a_l$ 
is a reflex vertex of $R_i$, and hence also of $C$. 
  $\\ \\$
Consider a vertex in $CH(P)$ (say, $c_p$) such that the neighbourhood of $c_p$ contains 
at least one bad region $B$. Let the neighbours of $c_p$ in $C$ be 
$a_0,a_1,a_2,\ldots,a_r$ in clockwise order (see Figure \ref{lastlem2}(a)). Let $c_{p+1}$ denote the 
vertex of $CH(P)$ such that $c_{p+1}c_pa_0$ is an inward triangle in $T$. 
Similarly, let $c_{p-1}$ be the vertex of $CH(P)$ such that $c_pc_{p-1}a_r$ 
is an inward triangle in $T$.
So, $B$
must contain a consecutive subsequence of these vertices, 
say, $a_i,a_{i+1},\ldots,a_j$, $0 \le i < i+2 \le j \le r$, such that either 
$a_i$ or $a_j$ is a reflex vertex in $B$. We call $B$ as 
a \emph{left} bad region (or, \emph{right} bad region) if $a_i$ (respectively, $a_j$) is the reflex vertex in $B$.
  $\\ \\$
Let $i$ be 
the smallest index such that there exists a left bad region $R'$ of $c_p$ containing points 
$a_i,a_{i+1},\ldots,a_j$ and having $a_i$ as the reflex vertex (see Figure \ref{lastlem2}(a)).
If $a_i$ belongs to any other bad region $R''$, then 
it must be the rightmost vertex in $R''$, and it cannot be a 
reflex vertex in $R''$ 
and $R'$.
Therefore,
$a_i$ belongs to $R'$ only due to the choice of $i$.
%
However, if $a_i = a_0$, $a_0$ may belong to another bad region in 
the neighbourhood of some other vertex of $CH(P)$. 
$ \\ \\$
Now 
$T$ and $D$ can be modified as follows. 
Consider the first case where $a_i = a_0$ and $a_0$ is reflex. 
The next counterclockwise vertex of $a_0$ on $C$ must be $a'_{p+2}$, where $a'_{p+2}$
is the inward vertex of the inward triangle $c_{p+1}a'_{p+2}c_{p+2}$ in $\mathcal{S}$.
Observe that the pentagon $c_pc_{p+1}a'_{p+2}a_0a_1$ must be convex. Replace 
the diagonals $c_pa_0$ and $c_{p+1}a_0$ by $c_{p+1}a_1$ and $a'_{p+2}a_1$.
Consider the other case where $a_i \neq a_0$.
Since $a_i$ is a reflex vertex in $C$, 
the quadrilateral $a_{i-1}a_ia_{i+1}c_p$ is a convex quadrilateral. Replace   
 $a_ic_p$ by $a_{i-1}a_{i+1}$ in $T$. 
 These replacements do not create a complex 
triangle.
For both cases, 
all possible diagonals of  $R'$ that are 
incident with $a_i$ are added to $D$. In particular, $a_ia_j$ can always be added. 
$\\ \\$
We show
that the new triple $(T',C,D')$ is a consistent triple with $|D'| > |D|$. 
We know that all bad regions other than $R'$ satisfy all four properties of consistent triples.
On the other hand,
$R'$ is
broken into smaller bad regions,
and we show that each region satisfies the same properties.
Suppose, $a_{i_1}, a_{i_2}, \ldots, a_{i_s}$, 
$i+2 \le i_1 < i_2 < \cdots <  i_s = j$ are the vertices of $R'$ such 
that the diagonals $a_ia_{i_t}$ are added, for $1 \le t \le s$. If $i_{t+1} 
> i_t+1$, new bad region is obtained with the labelling $a_i, a_{i_t}, a_{i_t+1}, 
\ldots, a_{i_{t+1}}$, for $1 \le t < s$. Similarly, if $i_1 > 2$, 
then the region $a_i, a_1,a_2,\ldots,a_{i_1}$ is a bad region.
All these bad regions continue to 
satisfy property 4 
of consistent triples
(see Figure \ref{lastlem2}(b)).  
Analogous arguments hold for right bad regions in the neighbourhood of $c_p$.
Thus the size of $D$ can be increased in a consistent 
triple $(T,C,D)$ if $D$ does not triangulate $C$. 
 $\hfill{\Box}$
 \begin{algorithm}[H]
  \textbf{four\_connected\_triangulation(\emph{P})} 
  
  compute $CH(P)$ and $CH(P')$\;
 let $\{ c_1, c_2, \ldots, c_k \}$ be the vertices of $CH(P)$ in the counterclockwise order\;
  $\mathcal{S} := \textbf{transforming\_good\_set(\emph{P})}$\;
  \tcp{$\mathcal{S}$ is a good set with all points of $P'$ assigned as distinct inward vertices}
  compute $L(\mathcal{S})$\;
  $C := \phi$, $i:=1$\;
  \While{$i \leq |CH(P)|$}
  {
  locate $c_{i-1}a'_{i}c_{i}$ and $c_{i}a'_{i+1}c_{i+1}$ in $\mathcal{S}$\;
  \tcp{$c_{i-1}a'_{i}c_{i}$ or $c_{i}a'_{i+1}c_{i+1}$ can be degenerate inward triangles}
  $C = C \cup \{ a'_ia'_{i+1}\}$\;
  $i = i + 1$\;
  }
  \tcp{$C$ is the inner cycle of $P$ corresponding to $\mathcal{S}$
  (see Figure \ref{lastlem1}(a))}
  $i:=1$\;
  \While{$i \leq|P'|$}
  {
  \If{$a'_i$ is reflex in $C$ and $a'_i$ is the inward vertex of non-degenerate inward triangle $c_{i-1}a'_iC_i$
  and $c_i$ has degree at least five}
  {
  $\mathcal{S} = (\mathcal{S} \setminus \{c_{i-1}a'_ic_{i}, c_{i}a'_{i+1}\}) 
  \cup \{c_{i-1}a'_i, c_{i-1}a'_{i+1}c_{i}\}$\;
  }
  $i = i +1$\;
  }
  \tcp{
  if the inward vertex $a'_i$ of some inward triangle $c_{i-1}a'_ic_{i}$
       is reflex in $C$, 
       $\mathcal{S}$ is transformed to ensure that $c_i$ has degree four
       (see Figure \ref{lastlem1}(b))}
  $D := \phi$, $i:=1$\;
  \While{$i \leq|P'|$}
  {
  $j:=i+1$\;
  \While{$j \leq|P'|$}
  {
  \If{$a'_ia'_j$ does not intersect $C$ and $a'_ia'_j$ does not intersect any edge in $D$}
  {
   \If{$D \cup \{ a'_ia'_j \}$ does not create any complex triangle}
   {
   $D = D \cup \{ a'_ia'_j \}$\;
   }
   }
   $j= j +1$\;
  }
  $i= i +1$\;
  }
  \tcp{$D$ is a maximal set of diagonals of $C$}
  compute annular region $R$ from $C$\;
  compute triangulation $T$ of $R$ from $\mathcal{S}$\;
  \tcp{$(T,C,D)$ is a consistent triple due to Lemma \ref{triple2}}
\end{algorithm}
\newpage
\begin{algorithm}[H]
  $i : = 1$\;
  \While{$i \leq |CH(P)|$}
  {
  \While{the neighbourhood of $c_i$ contains a bad region}
  {
  assign the neighbours of $c_i$ in the clockwise order on $C$ to $\{ a_0, a_1, \ldots, a_{deg(c_i)-3}\}$\;
  \eIf{the neighbourhood of $c_i$ contains a left bad region}
  {
 scan $\{ a_0, a_1, \ldots, a_{deg(c_i)-3}\}$ from $a_0$ and locate the first reflex vertex $a_j$
 of a bad region $R_x$ contained in the neighbourhood of $c_i$\;
  \eIf{$a_j=a_0$}
  {
  locate the inward vertex $a'_{i+2}$ of $c_{i+1}a'_{i+1}c_{i+2}$ in $\mathcal{S}$\;  
  $T = (T \setminus \{c_ia_0, c_{i+1}a_0 \}) \cup \{ c_{i+1}a_1, a'_{i+1}a_1\}$\;
  add all possible diagonals of $R_x$ incident on $a_j$ to $D$\;
  }
  {
  $T = (T \setminus \{c_ia_j \}) \cup \{ a_{j-1}a_{j+1} \}$\;
  add all possible diagonals of $R_x$ incident on $a_j$ to $D$\;
  }
  update $\mathcal{S}$\;
  }
  {
scan $\{ a_0, a_1, \ldots, a_{deg(c_i)-3}\}$ from $a_0$ and locate the last reflex vertex $a_j$
 of a bad region $R_x$ contained in the neighbourhood of $c_i$\;
  {
  $T = (T \setminus \{c_ia_j \}) \cup \{ a_{j-1}a_{j+1} \}$\;
  add all possible diagonals of $R_x$ incident on $a_j$ to $D$\;
  }
  update $\mathcal{S}$\;
    
  }
  }
  \tcp{modified $\mathcal{S}$ may not remain a good set but modified (T,C,D) is a consistent triple
   due to Lemma \ref{finlem}}
  $ i = i + 1$\;
  }
  \tcp{all bad regions in the neighbourhood of every vertex of $CH(P)$ are eliminated}
  $T = T \cup  C \cup D$\;
  \tcp{the resulting triangulation is a  4-connected triangulation of $P$}
  report $T$\;
  \end{algorithm}
  \begin{lemma}\label{finlem}
   The procedure {four\_connected\_triangulation(\emph{P})}
   computes a 4-connected triangulation of $P$ in $O(n^3)$ time.
  \end{lemma}
  \textbf{Proof:} 
  The correctness of the procedure follows from Lemmas \ref{triple2} and \ref{finlem}.
  Let us analyse the time complexity of the procedure.
  Computing $\mathcal{S}$ takes $O(n^3)$ time due to Lemma \ref{complxlemgs}.
  Constructing a consistent triple $T,C,D)$ takes $O(n^3)$ time.
  Since the diagonals in $D$ are non-intersecting, the total number of bad regions is $O(n)$. 
  So, $D$ of maximum size can be constructed in $O(n^2)$ time.
 Thus, the overall time complexity of the procedure is $O(n^3)$.   $\hfill{\Box}$
 $\\ \\$
We have the following theorems.
  \begin{theorem}
   A given set of points $P$ admits a 4-connected triangulation if and only if $P$ satisfies Necessary Condition \ref{nc3}.
  \end{theorem}
\begin{theorem}
 A 4-connected triangulation of a point set $P$ (if it exists) can be constructed in $O(n^3)$ time.
\end{theorem}

\section{Concluding remarks}
\begin{figure}[h]  
\begin{center} 
\centerline{\hbox{\psfig{figure=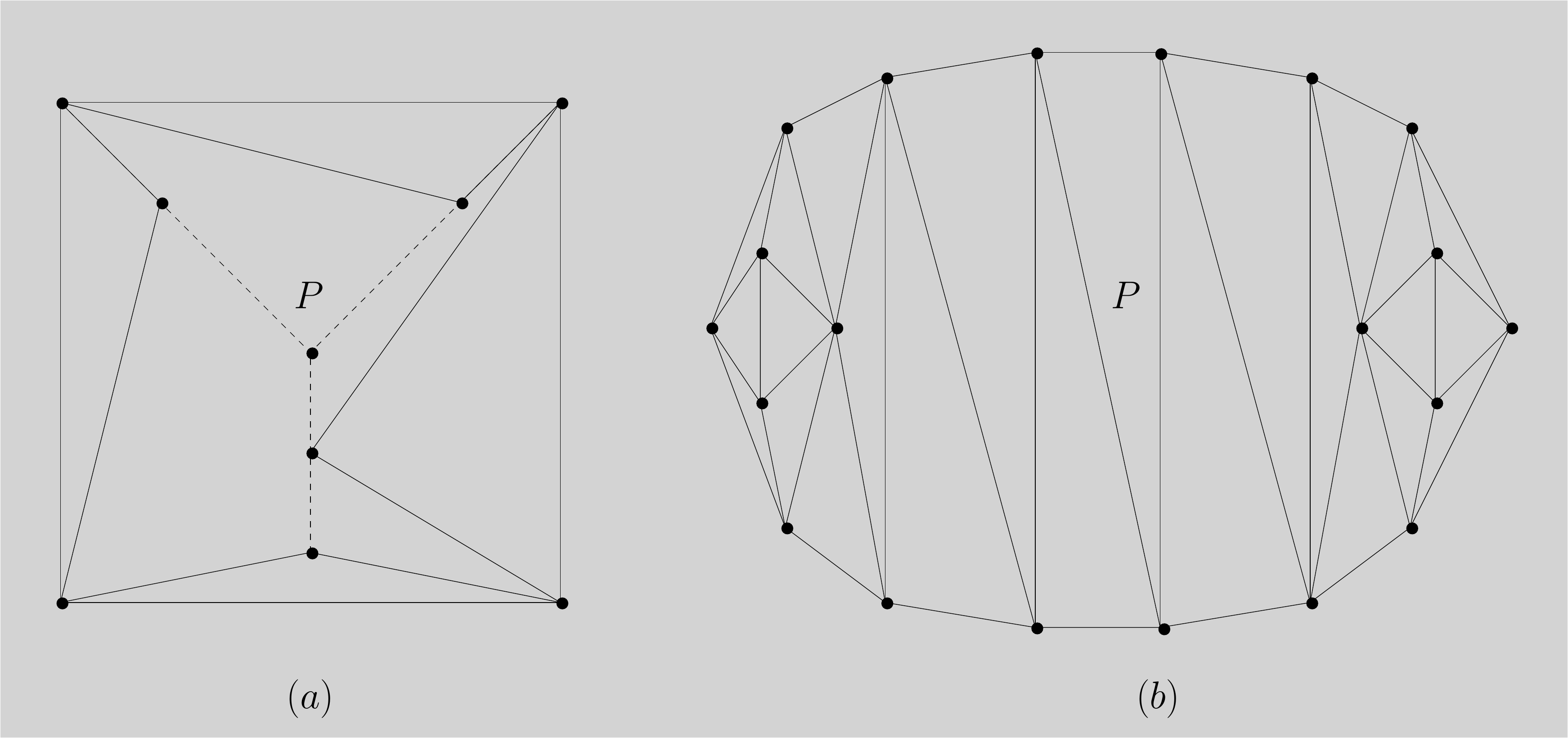,width=0.90\hsize}}}
\caption{ 
(a) The point set $P$ satisfies the third necessary condition but it does not admit a 4-connected triangulation.
(b) The point set $P$ admits a 4-degree triangulation but not a 4-connected triangulation as $|P'|=6$ and $|CH(P)|=14$. 
}
\label{concrem1}
\end{center}
\end{figure}

In this paper, we have characterized point sets $P$ that admit 4-connected triangulation.
Furthermore, we have presented an $O(n^3)$ time algorithm for constructing a 4-connected triangulation of $P$.
%
Observe that the third necessary condition is sufficient for characterizing $P$ only under the assumption
that no three points of $P$ are collinear.
If $P$ contains collinear points, then the third necessary condition is no longer sufficient
as shown in Figure \ref{concrem1}(a).
$\\ \\$
Consider a triangulation $T$ of $P$ such that at least four edges of $T$ are incident on every point of $P$.
We call such a triangulation as a \emph{4-degree} triangulation of $P$.
Observe that a 4-connected triangulation of $P$ is always a 4-degree triangulation of $P$ but a 4-degree triangulation 
of $P$ may not be a 4-connected triangulation of $P$ (see Figure \ref{concrem1}(b)).
%
Thus, the problem of characterizing point sets 
that admit 4-degree triangulation remains open.
$\\ \\$
Consider the problem of characterizing point sets that admit 5-connected triangulation.
Our method for constructing 4-connected triangulation does not generalize to 
the problem of 5-connected triangulation.
It will be interesting to see if a new method can be developed for constructing a 5-connected triangulation of $P$.
%
%
%
Also, the problem of characterizing point sets 
that admit 5-degree triangulation remains open.
 \bibliographystyle{plain}
 \bibliography{vis}
\end{document}